# All-optical Lithography for Spatiotemporal Patterning of High-Modulation Azopolymer Microreliefs

I Komang Januariyasa,[1] Francesco Reda,[1] Nikolai Liubimtsev,[2] Achu Purushothaman,[1] Marina Saphiannikova,[2,3] Fabio Borbone,[4] Marcella Salvatore,[1,*] and Stefano Luigi Oscurato[1,*]

[1]Physics Department "E. Pancini", University of Naples Federico II, Complesso Universitario di Monte Sant'Angelo, via Cinthia, 80126, Naples, Italy.

[2]Division Theory of Polymers, Leibniz Institute of Polymer Research Dresden, 01069 Dresden, Germany.

[3]Faculty of Mechanical Science and Engineering, Dresden University of Technology, 01062 Dresden, Germany.

[4]Department of Chemical Sciences, University of Naples "Federico II", Complesso Universitario di Monte Sant'Angelo, Via Cintia, 80126 Naples, Italy.

[*]Stefano Luigi Oscurato, e-mail: stefanoluigi.oscurato@unina.it

[*]Marcella Salvatore, e-mail: marcella.salvatore@unina.it

## Abstract

Microstructured surfaces are central to photonics, biointerfaces, and coating technologies, but they are typically fabricated through multistep workflows involving masks, masters, development, and etching. Azopolymers offer a direct light-driven route to surface structuring, although holographic photopatterning of flat films has largely remained limited to smooth, shallow, and engraving-like reliefs with limited vertical modulation. Here we show that computer-generated holograms with co-designed bright and dark regions can spatially confine inward mass transport and directly generate isolated protruding microstructures with surface modulations of several micrometers from pristine flat azopolymer films. Single exposures produce individual protrusions and arrays, whereas sequential tailored exposures generate flat-topped and hierarchical microreliefs. The same workflow produces connected freeform structures and enables optical erasure and rewriting within the same surface region. We further transfer the optically programmed azopolymer morphologies into another transparent polymer and demonstrate replicated microlenses operating at multiple visible wavelengths, with focal response controlled through the master-writing parameters. These results establish an all-optical framework for generating and reconfiguring complex, high-modulation azopolymer microreliefs from flat films through combined control of holographic illumination geometry and exposure history, with replication extending their use to materials selected for their final functional properties.

# 1. Introduction

Microstructured surfaces enable a wide range of technologies, including photonic components, biointerfaces, and functional coatings.[1–6] In many of current lithographic workflows, surface microreliefs are fabricated through multistep routes that rely on physical masks or master molds, dedicated processing facilities, and post-processing steps such as development, etching, or replication.[1,5,7,8] Advanced photolithographic schemes,[5] such as projection photolithography, direct laser writing, two-photon polymerization, and holographic lithography[9] can eliminate the need for a physical mask by properly engineering the illumination, but they still typically remain coupled to resist chemistry, material removal, or development steps.[5,8] These operations separate optical pattern definition from actual surface formation and often lock the final morphology, limiting post-fabrication reconfiguration. When the structured surface serves as a master for subsequent replication,[1,8] this constraint also limits the ability to refine or update the geometry transferred to other materials.

All-optical patterning approaches such as laser-induced periodic surface structures (LIPSS)[10] and laser-driven thin film flow or dewetting[11,12] can directly convert optical exposure into surface topography without post-processing. However, these methods rely on ablation, melting, or irreversible material reorganization, which naturally favor periodic or quasi-periodic morphologies while constraining geometrical flexibility and post-fabrication reconfigurability. Relying on reversible and non-destructive mass transport, all-optical patterning of azobenzene-containing polymers[13–20] has the potential to overcome these limitations. In these materials, both the direction and the magnitude of the mass transport are governed by the illuminating optical field, and the resulting surface relief pattern can be spatially engineered through intensity and polarization gradients.[21–25] This distinctive vectorial response has motivated holographic lithography strategies[26,27] in which the vectorial optical field is digitally structured to prescribe the surface topography. Current implementations of these approaches use digital polarization holograms[28,18] or intensity computer generated holograms (CGHs),[29] with the latter offering broader design freedom for complex patterns and multilevel reliefs with diffraction-limited spatial resolution, tunable modulation depths, and reconfigurable relief morphologies.[29–32] These studies, including previous work from our group, established digital holographic lithography in azopolymers as a versatile route to light-driven surface patterning. Because the relief forms during illumination without chemical development, its evolution can be monitored optically, as demonstrated by real-time holographic microscopy and integrated optical writing and characterization.[33,34] However, the accessible topographies have mainly remained in a regime in which the relief follows mainly the geometry of the bright regions of the holograms, yielding smooth and engraved surface patterns with typical modulations below one micron.[29–31,35] Accordingly, the direct all-optical formation of isolated protruding microstructures with several micrometers of modulation and hierarchical morphologies from a flat azopolymer film has remained challenging and has not been demonstrated so far.

High-modulation microstructures with amplitudes of several micrometers, such as micropillar arrays, have instead generally required prior lithographic processing, typically through replication from masters fabricated via conventional lithography.[36] In that case, digital holographic illumination can still reshape or reconfigure a pre-patterned template,[37–42] but key geometrical parameters remain defined by the initial master template, including the micropillar shape, array arrangement and periodicity, or microchannel morphology. The primary geometry is therefore inherited from the pre-patterned surface rather than being defined directly by the holographic illumination. Bringing this primary geometry under optical control would allow protrusion formation and subsequent reshaping to be programmed within the same digital workflow.

A distinct patterning regime that overcomes these limitations can emerge when not only bright but also dark regions are co-designed to control the spatial confinement of mass transport during exposure of a flat azopolymer film. In this configuration, illuminated and dark regions play complementary roles, with the illuminated regions driving material redistribution and the spatially confined dark regions defining where this redistribution accumulates. The dark features therefore become active geometrical elements of the surface design.

In this work, we introduce a holographic design strategy that uses co-engineered bright and dark regions of CGHs to spatially confine azopolymer mass transport through tailored intensity gradients. In this framework, dark regions act as boundaries that localize inward mass transport and enable the direct all-optical formation of isolated protruding microstructures with high modulation from pristine flat films. We first investigate this concept experimentally and through finite-element modelling of viscoplastic deformation for single microstructures and for arrays, establishing how the illumination geometry can even control the primary morphology of three-dimensional microreliefs. We then extend the approach to sequential exposures, exploiting the reconfigurability of azopolymer mass migration[13,14,43,33] together with digitally reconfigured illumination to reshape and progressively add features and complexity to the final relief morphology over time. Spatial and temporal degrees of freedom are thus combined within the same digital holographic framework, with successive optical fields determining the formation, reshaping and hierarchical evolution of the surface topography. Using this strategy, we fabricate and reshape microrelief arrays, continuous freeform structures, and hierarchical microreliefs, which can also be transformed one into another through multiple write-erase-rewrite cycles in the same surface regions.

We further demonstrate that this morphological programmability can be translated into functional use. Representative topographies are transferred through an intermediate PDMS mold into polycarbonate, demonstrating the use of the fabricated azopolymer surface as an optically programmable primary master when a different final material is required. As a quantitative functional demonstration, replicated rounded microreliefs are used as microlenses operating at multiple visible wavelengths. Their focal response is controlled through the optically programmed morphology of the azopolymer master and characterized through focal-distance and focal-intensity measurements, with comparison to estimates derived from the measured microlens morphology.

By integrating direct high-modulation patterning from flat films with dynamic optical reconfiguration, our approach opens a route to programmable three-dimensional surface architectures whose geometry can be all-optically defined, reshaped, and rewritten, with replication enabling transfer into materials selected for their final function.

# 2. Results

## 2.1 Isolated protruding microreliefs from bright-dark illumination engineering

To establish the physical basis of our approach, we first compare the response of azopolymers to a circularly polarized bright spot with that obtained when bright and dark regions are deliberately co-designed within the same illumination field. As illustrated in **Figure 1**a and **1**b, illumination with a bright spot, used here as a reference case, produces the conventional outward mass transport from high to low intensity along the radial gradient.[23,44] This light-driven migration leads to the formation of the characteristic circular micro-well surface relief sketched in the inset of **Figure 1**b, which corresponds in lithographic terms to engraved or depressed reliefs. This response already highlights the limitation of bright-spot illumination for generating protruding microstructures, since outward transport prevents material accumulation in the center of the bright region. To overcome this limitation, the illumination field can be engineered to include a dark region within the bright pattern, introducing additional intensity gradients that further redirect and confine the mass transport. **Figure 1**c and **1**d illustrate this concept using a doughnut-shaped intensity profile. In this configuration, the inner boundary of the bright ring promotes inward material accumulation in the central dark core of the doughnut, while outward redistribution also produces a peripheral rim. The bright ring and the dark core therefore play complementary roles in directing the deformation and defining the region of material accumulation. By properly engineering the doughnut geometry, this configuration can be used to generate isolated protruding microstructures.

We first tested this design principle through finite-element simulations based on the Viscoplastic PhotoAlignment (VPA) model.[24,42,45] In this model, the azopolymer is treated as an incompressible viscoplastic material in which optical irradiation induces photoinduced molecular reorientation, thereby generating stresses that drive surface deformations. Under circularly polarized illumination, the irradiated material tends to expand radially within the polarization plane. For doughnut-shaped illumination, this tendency promotes both inward and outward material redistribution, while the surrounding non-irradiated film mechanically constrains the deformation. The resulting central protrusion and peripheral rim therefore arise from spatially distributed deformation of the film under the constraint of volume conservation. The VPA calculations are used here to provide a physically grounded interpretation of the principal deformation regimes and their dependence on illumination geometry. The modeling assumptions and parameter selection are detailed in the "Methods" section. **Figure 1**e presents a three-dimensional view and a cross-sectional profile of the modeled microstructure produced by a binary doughnut-shaped intensity pattern, as described in the "Methods" section. The simulation supports the design strategy, revealing the formation of a localized surface protrusion in the center of the region of interest defined by the doughnut. By contrast, when the central dark region is removed from an otherwise equivalent illumination pattern, no central protrusion is observed, and the expected conventional micro-well relief is recovered (see Supplementary Fig. 1).

To implement this design principle experimentally, we used a CGH scheme (**Figure 2**a) based on a phase-only spatial light modulator (SLM), which can be programmed to project tailored circularly polarized intensity distributions at 491 nm onto azopolymer films with micrometer-scale spatial resolution.[17,29,31,40] The chemical structure and the UV/vis absorption spectrum of the azopolymer used here are shown in Supplementary Fig. 2. Following the conceptual picture above, we employed doughnut intensity patterns defined by two length scales, the outer radius $R$ of the bright ring and the dark core radius $r$, as indicated in the inset of **Figure 2**a. **Figure 2**b shows a representative time-averaged doughnut intensity pattern produced by our CGH setup at the azopolymer film plane. Details of the optical system, hologram design, computation and characterization, sample synthesis and preparation, and irradiation conditions are provided in the "Methods" section.

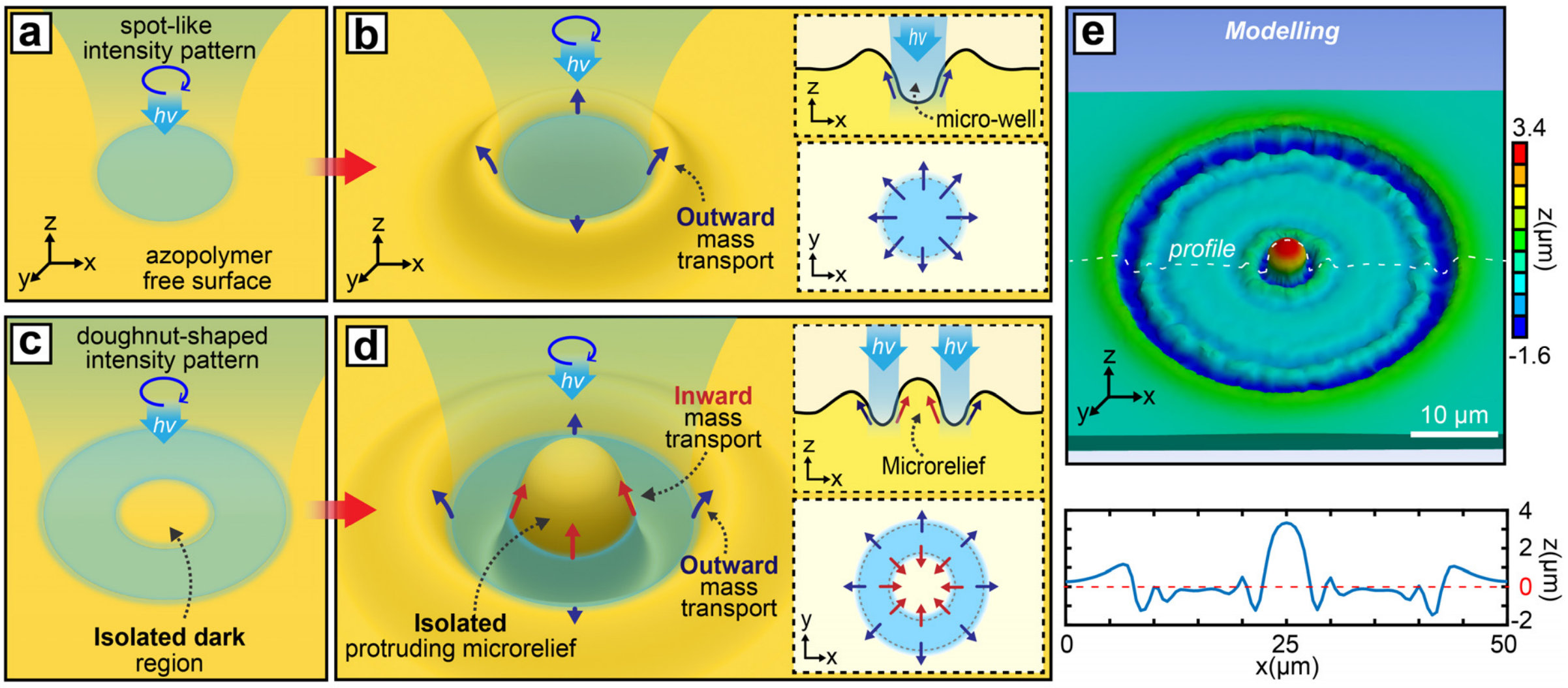

**Figure 1**. **Concept of inward azopolymer mass transport through dark-bright illumination co-designing**. **a–b)** Conceptual illustration of a circularly polarized bright spot producing depressed relief on the flat surface of the azopolymer driven by the outward radial intensity gradient. **c–d)** Illustration of a doughnut-shaped, circularly polarized intensity pattern, which introduces an inward intensity gradient toward the central dark region. **e)** VPA-model prediction of the surface deformation under binary doughnut illumination, shown as a three-dimensional view and a cross-sectional profile along the indicated line.

All experiments in this section were performed in the presence of a uniform assisting beam at $\lambda = 405$ nm, incident from the substrate side simultaneously with the writing holograms (**Figure 2**b). In this configuration, the structured 491 nm beam defines the spatial pattern, whereas the uniform 405 nm beam enhances the relief formation dynamics, as demonstrated in previous works,[29,46] and suppresses the roughness that would otherwise develop within the bright regions of the holograms. Additional details about its experimental implementation, physical role, and phenomenological representation in the VPA model are described in the "Methods" section.

For reference, **Figure 2**c shows the atomic force microscopy (AFM) image of the conventional micro-well relief produced by a spot-like hologram with only an outward radial intensity gradient (see "Methods" section for AFM details). To test the concept of inward mass transport, the bright-dark area ratio was varied by increasing the outer radius $R$ of the bright doughnut ring from 3.9 μm to 17.6 μm while keeping $r$ fixed at approximately 1.95 μm (see Supplementary Fig. 3). In this illumination design, increasing R enlarges the illuminated annular area and hence the volume of material in the film potentially available for redistribution. Across this design range, the doughnut holograms retained an average dark-core to bright-ring contrast of 0.45 ± 0.04 (see Supplementary Fig. 4), maintaining an inward intensity gradient that is sufficient to achieve efficient mass transport despite the residual background in the dark core.[17] As $R$ increases, the AFM topography of the patterned region evolves from a depression-dominated microrelief to a morphology with a well-defined central protrusion (**Figure 2**d). As summarized in **Figure 2**e, the protrusion modulation $h$, increases with both $R$ and exposure time. The modulation $h$ is defined as the vertical distance between the protrusion peak and the local baseline estimated from the immediately surrounding depleted surface. It therefore includes the contribution of the surrounding depletion and should be distinguished from the protrusion height above the original unexposed film surface (See Supplementary Fig. 5). The observed trend is consistent with the volume-conserving redistribution of material during relief formation (see Supplementary Notes 1). For the geometry at fixed $R$ examined in Supplementary Fig. 5, $h$ increases approximately linearly at short exposure time before reaching saturation at longer times.

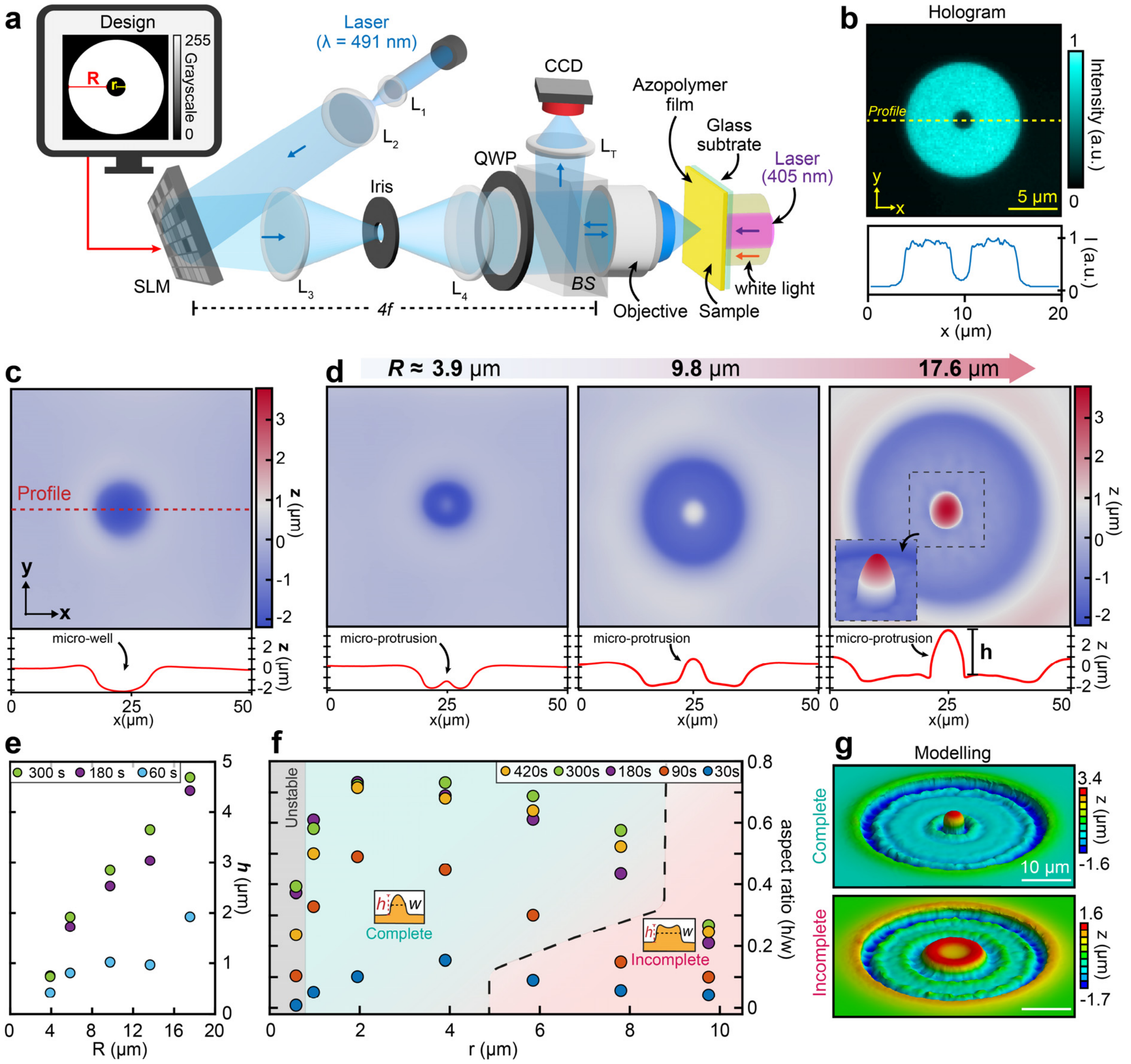

**Figure 2. Isolated protruding microreliefs. a)** CGH setup. The laser beam is first expanded using lenses $L_1$ and $L_2$, and then impinges on a reflective SLM. The lenses $L_3$ and $L_4$ are arranged in the 4f configuration, and together with the objective, projects the hologram onto the sample plane. An imaging arm, implemented with a beam splitter (BS), collects the light reflected from the sample, which is then relayed by lens $L_T$ in the CCD. $R$ and $r$ in the grayscale design represent the outer and inner radii of the doughnut pattern, respectively. **b)** Representative doughnut intensity pattern produced by the CGH setup at the sample plane recorded with the CCD. **c), d)** AFM images of the micro-well and of isolated protruding microreliefs produced by doughnut holograms with different $R$, after 180 s of exposure. **e)** Isolated micro-protrusion height $h$ versus $R$ at three different time exposures. The intensity of hologram for experiments shown in c), d), and e) is ~ 63 $Wcm^{-2}$. **f)** Evolution of the aspect ratio $h/w$ of the fabricated isolated protruding microstructure for increasing $r$ values at different exposure time. The intensity of hologram for experiments shown in f) is ~ 38 $Wcm^{-2}$. **g)** The VPA modelling results showing the two different regimes (complete and incomplete) of isolated protruding microstructures formation.

Having established the bright-dark co-design principle as the mechanism enabling protrusion formation, we next investigated how the size of the central dark region affects the morphology of the resulting structures. The design principle is based on controlling the balance between the lateral extent of the accumulation region, set by the dark-core radius $r$, and the lateral spatial scale of the illuminated material available for redistribution, set by $R$-$r$. Elucidating this interplay provides practical design guidelines required for obtaining well-defined protrusions, thereby laying the foundation for generalizing the approach to the wider family of microstructures demonstrated in the following sections.

To investigate this interplay, we systematically varied $r$ at fixed $R$ for a series of exposure times, thereby tuning both the transport length scale and the amount of illuminated material available for transport and accumulation. We extracted the modulation $h$ and the full width at half maximum $w$ of the protrusions from the AFM profiles (see Supplementary Fig. 6-7). For well-defined single protrusions, $w$ scales approximately linearly with $r$ and depends only weakly on exposure time, whereas $h$ shows a non-monotonic dependence on $r$, first increasing and then decreasing at larger $r$.

These combined effects are captured by the aspect ratio $h/w$ of the microstructures, reported in **Figure 2**f, which also exhibit a similar non-monotonic dependence of $h(r)$. As $r$ is varied from ~0.6 µm to ~13.6 µm at fixed $R$ ~19.5 µm, three qualitative regimes of relief morphology emerge, highlighted by different background colors in **Figure 2**f. For very small $r$, unstable reliefs were obtained under our conditions (gray region), with modulation comparable to the roughness of the depleted region (see Supplementary Fig. 7). By increasing $r$, a threshold separates a regime where a single, complete and well-developed protrusion forms (green region) from a regime where accumulation in the central region becomes incomplete and the topography is characterized by a double-peak cross-sectional profile (red region). This behavior is consistent with a finite transport length of the material from the depletion region of the bright hologram toward the center along the inward intensity gradients. For smaller $r$, the inward deformation produces a single central protrusion, whereas for larger $r$ the accumulation remains spatially separated around the dark region. In the design used here, this effect is further enhanced by the reduced material volume available at larger $r$ (i.e. small $R$-$r$ values). The transition threshold between complete and incomplete formation depends on exposure time, with longer exposures enabling greater accumulation and shifting the boundary for complete protrusion formation toward larger $r$.

Notably, the VPA model qualitatively reproduces complete and incomplete protrusion formation, as illustrated by representative simulated morphologies in **Figure 2**g. The fine-scale roughness visible in the simulated depleted regions is attributed to the tetrahedral finite-element discretization and is not interpreted as a prediction of physical surface features. The simulations further indicate that increasing $r$ while maintaining a constant $R/r$ ratio can introduce a buckling-like deformation within the protrusion. Within the VPA framework, this behavior is interpreted in terms of reduced light-driven stress in the central region of the feature and the onset of a mechanical instability during the surface deformation process.

An important outcome of this study is that, for the geometries and exposure times examined, the process approaches a saturation regime in which the changes in protrusion morphology becomes smaller at longer exposure times (Supplementary Fig. 7). This reduced sensitivity to further exposure provides a useful operating regime for obtaining well-developed protrusions. For a suitable hologram design, increasing the exposure mainly drives the system toward this regime, with progressively smaller changes in the final relief morphology. In this limit, the hologram geometry, and in particular the size of the dark region, sets the main lateral scale for material accumulation and thus becomes the primary parameter controlling the morphology of the resulting protruding microstructures within the region of interest defined by the bright pattern. The use of CGHs makes this geometrical control digitally programmable, allowing the size, shape, and arrangement of bright and dark regions to be prescribed through the illumination design.

## 2.2 From single protrusions to ordered arrays

Building on the single-feature results above, we extended the holographic bright-dark co-design to multiple dark regions embedded within a shared bright pattern, with the aim of producing ordered arrays of isolated protruding microreliefs in a single exposure from the pristine flat film.

As a first step, we used a two-core benchmark illumination pattern to analyze how neighboring protrusions evolve from a single feature to distinct and progressively separated structures as the spacing between dark regions is increased. As schematized in **Figure 3**a and **3**b, two identical dark circles of radius $r \approx 1.95$ µm were arranged along a line and their center-to-center spacing $L$ was progressively increased starting from $L = 0$, which corresponds to a single circle. The corresponding holographic intensity images are shown in Supplementary Fig. 8. The measured topographies in **Figure 3**c evolve from a single protrusion at $L = 0$, to overlapping relief features at intermediate spacing, and finally to two discrete and distinct protruding microreliefs at larger $L$, as also shown in Supplementary Fig. 9. This transition is shown in **Figure 3**d through the evolution of the normalized peak-to-peak distance $P/w$ of the protruding microstructures as a function of the normalized spacing $L/w_0$ of the dark circles in the holograms. Here, $P$ and $w$ are extracted from the two-peak fit of the measured profiles, and $w_0$ is the FWHM of the reference protrusion produced for $L = 0$ (see "Methods" section for details). The analysis distinguishes protrusions that are resolved at half maximum from those that are separated at their bases. For example, Supplementary Fig. 9 shows two peaks that are resolved at half maximum at $L/w_0 \approx 1.2$, whereas a transition to base-separated protrusions is observed at $L/w_0 \approx 2.1$. In this regime, which is of main interest for this work, the separation between the two isolated microstructures can be deliberately controlled within the entire shared bright area (**Figure 3**c). It is important to note that the these numerical thresholds should be regarded as practical design values for this two-feature benchmark. The single-feature footprint depends indeed on the dark-core size, illumination geometry, and exposure conditions, as demonstrated in Supplementary Fig.7 and 8, and the separation criterion therefore cannot be transferred unchanged to other designs. Nevertheless, the analysis establishes how the spacing of dark features can be used to control the separation of neighboring protrusions and provides a basis for extending the approach to the formation of multiple microstructures within a shared illuminated region.

To fabricate a microprotrusion array using this principle, we designed a holographic pattern consisting of a 6 × 6 square array of dark circles with $r \approx 4.9$ µm and $L \approx 19.5$ µm ($L/w_0 \sim 2.3$) within a shared square bright illumination field of ~120 × 120 µm$^2$, as shown in **Figure 3**e. As shown in **Figure 3**f, VPA modeling supports the formation of an array of protruding microstructures under this illumination geometry. **Figure 3**g shows the AFM morphology of the azopolymer surface after 300 s of illumination at a local writing intensity of approximately 27 W cm$^{-2}$, with simultaneous 405 nm assisting illumination as described in the "Methods" section. The analyzed array exhibits an average local modulation $h = 5.4 \pm 0.3$ µm from the depleted baseline, an average FWHM $w = 8.6 \pm 0.2$ µm, and a periodicity $P = 19.6 \pm 0.3$ µm, where the uncertainties are expressed as standard deviation estimated across the array (see also Supplementary Fig. 10). The measured periodicity is in good agreement with the spacing prescribed by design in the hologram. Complementary scanning electron microscopy (SEM, see "Methods" section) images provide independent confirmation of the array morphology. The top view in Supplementary Fig. 10, shows the regular arrangement and lateral separation of the protrusions, while the oblique view in **Figure 3**h reveals their three-dimensional geometry. The higher-magnification image in the inset of **Figure 3**h shows the rounded apex and curved sidewalls of an individual protrusion, consistent with the overall morphology observed by AFM, supporting the formation of discrete protruding microstructures directly from the initially flat film.

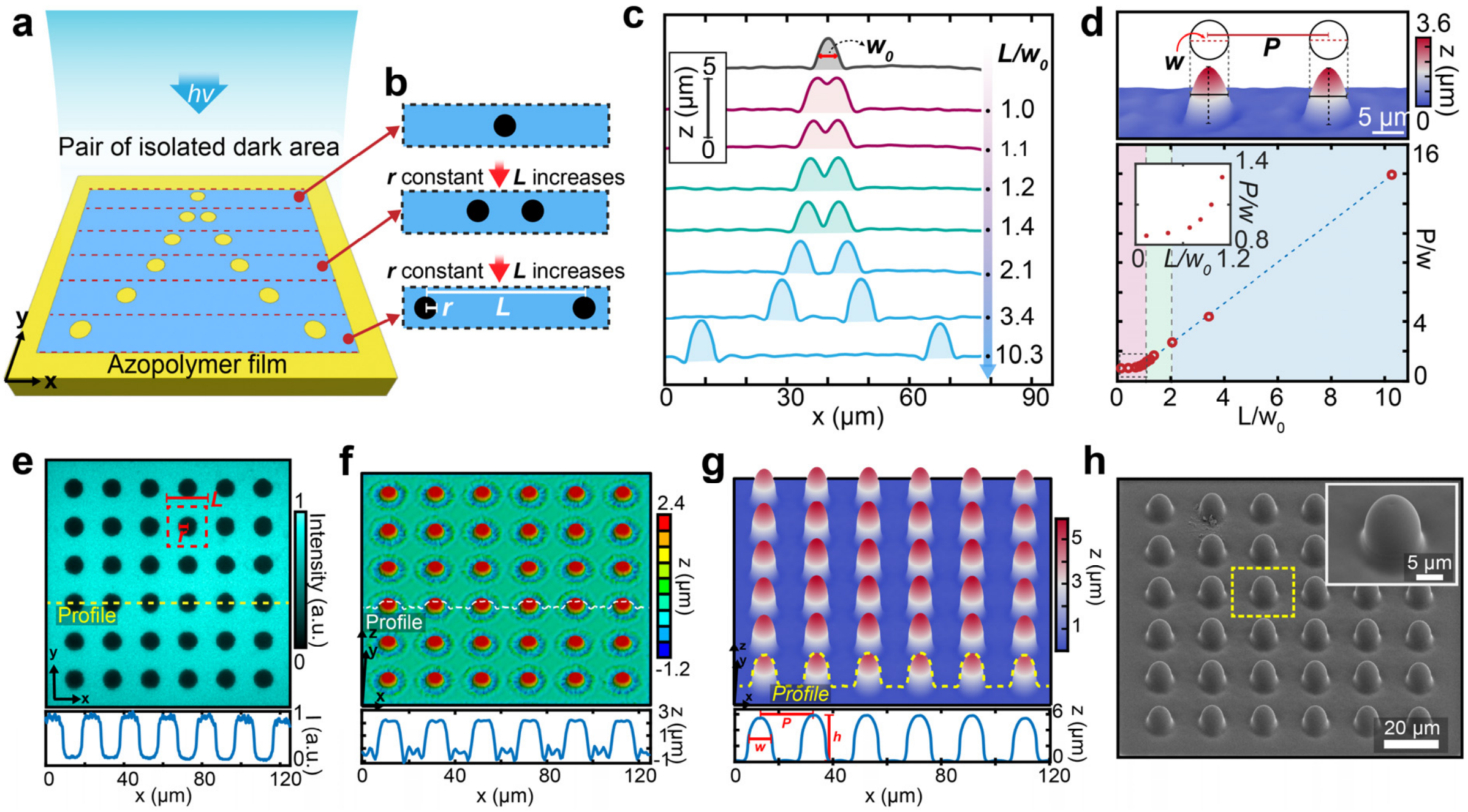

**Figure 3. Direct formation of spatially isolated protruding microrelief protrusion arrays. a)** Schematic of the two-core illumination study used to investigate the separation of neighboring protrusions. **b)** Illustration of two isolated dark circles with constant radius $r \approx 1.95$ µm and gradually increasing spacing center-to-center $L$. **c)** Evolution of the measured surface profiles for different values of $L/w_0$. The gray profile corresponds to $L=0$ and defines the reference FWHM $w_0$ of the single protruding microrelief. **d)** (top) Definition of the FWHM $w$ and the peak-to-peak distance $P$ for the pair of protrusions and (bottom) $P/w$ plotted as function of $L/w_0$. Red, green, and blue shaded areas indicates protrusions that are unresolved at half maximum, resolved at half-maximum but not separated at their basis, and base-separated, respectively. The inset shows enlarged view for data at small $L/w_0$. A linear fit to data in the green and blue regions gives a slope of ~ 1.4. **e)** Holographic intensity pattern comprising a 6 × 6 array of dark circles with radius $r \approx 4.9$ µm and periodicity $L \approx 19.5$ µm, with relative intensity profile traced along the yellow line. **f)** VPA-simulated surface topography for an exposure of the equivalent intensity pattern as shown in e), with relative profile along the with line. **g)** AFM topography of the resulting protrusion array and the profile along the yellow line. produced by the illumination with the hologram shown in e) for 300 s and intensity of ~ 27 Wcm$^{-2}$. **h)** Oblique SEM view of the array, with a higher magnification of an individual protrusion in the inset. Experimental data shown in c), d, and g) were obtained after 300 s of exposure at writing intensity of ~ 27 Wcm$^{-2}$, with simultaneous assisting beam illumination.

Supplementary Fig. 10, shows an enlarged view of the VPA map and top SEM image to highlight the effect of the peripheral relief arising at the boundary of the patterned region. The peripheral relief follows the square outline of the illuminated field and is consistent with outward material redistribution at its external edge, as discussed for the single-feature geometry

in Section 2.1. The AFM boundary profile in Supplementary Fig. 10 further shows that the local array baseline lies below the original unexposed surface, while the protrusion apices extend above it. In lithographic terms, the array therefore forms a pattern of discrete, well-isolated positive microstructures in the targeted illuminated area, accompanied by a peripheral boundary relief.

## 2.3 Temporal control of protruding microrelief morphology

Given the capacity of azopolymer microreliefs to undergo further light-induced reshaping, temporal control can be integrated into the holographic workflow.[13,14,20,29] Here we use this capability to structure the process into successive illumination steps that add an additional level of geometrical control to the all-optically generated protruding microreliefs.

**Figure 4**a illustrates the concept through a sequential double exposure designed to obtain a flat-topped protruding microrelief from a pristine flat azopolymer surface. The first exposure uses a hologram containing a dark circle that produces an isolated protruding microstructure, as described above. The second exposure then applies an inverted hologram design that places a bright disk at the same position as the previous dark region. This second step reconfigures the existing microstructure, preferentially redistributing the material near the upper part due to the absorption gradient.[37,39,41,45]

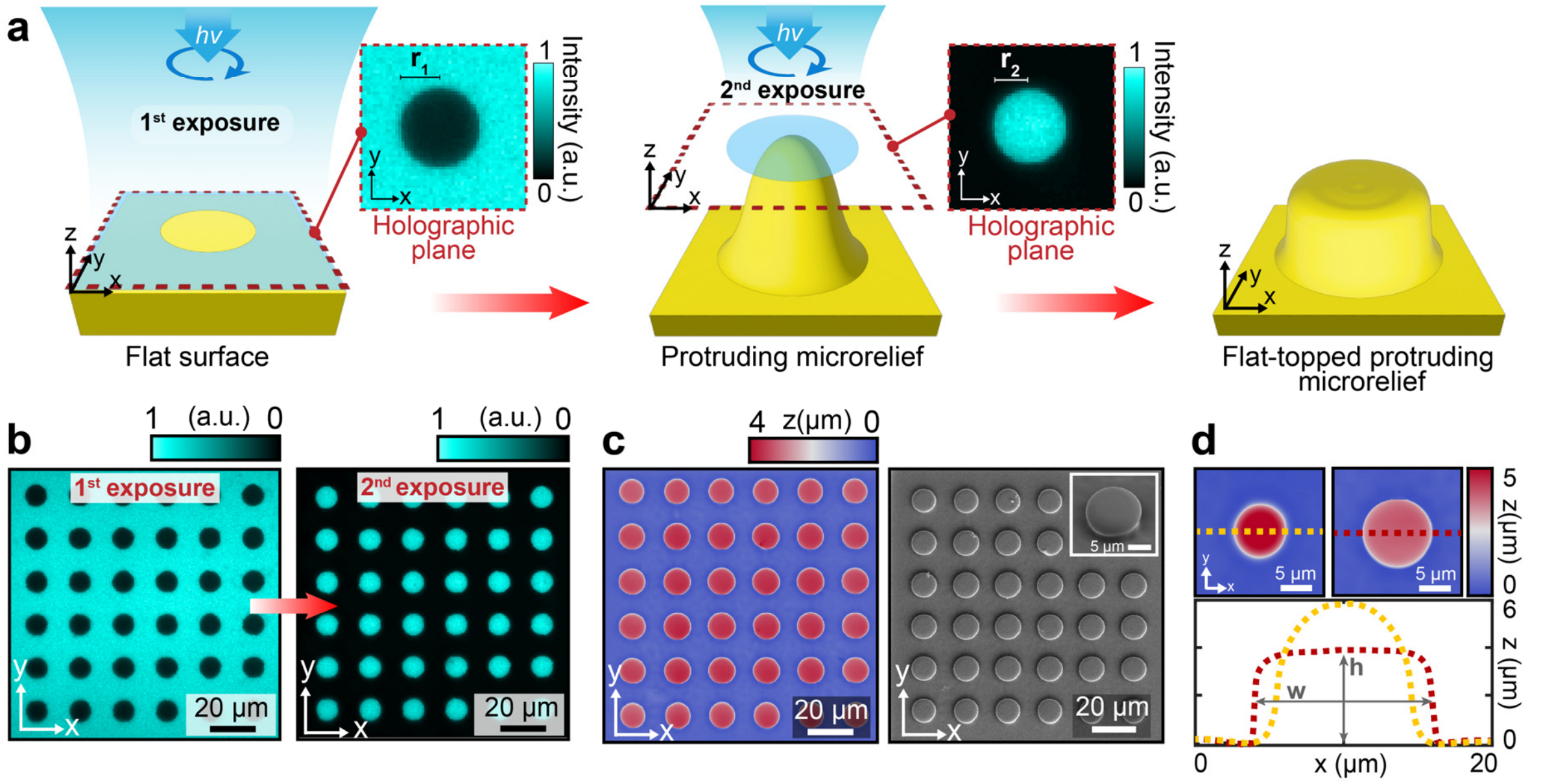

**Figure 4. Double-step exposure for controlling the three-dimensional architecture of the microstructures. a)** Schematic of the two-step exposure sequence. The 1st exposure uses a dark circle embedded within a bright area to generate the protruding microrelief from the flat surface. The 2nd exposure places a bright disk at the position of the former dark circle, reshaping the protrusion into a flat-top protruding microrelief. **b)** Images of the holographic intensity patterns used for 1st and 2nd ($r_1 = r_2 \approx 4.9$ µm). **c)** AFM topography (left) and SEM image (right) of the array after the two-step sequence. The inset shows a higher-magnification SEM view of an individual structure. **d)** Representative AFM topographies after first (top left) and second (top right) exposures, with the corresponding profiles. Yellow and red dotted lines identify the profiles after the first and second exposures, respectively. The local writing intensities and exposure times for the 1st and 2nd steps were ~ 27 $Wcm^{-2}$ for 300 s and ~ 133 $Wcm^{-2}$ for 30 s, respectively.

The resulting structure develops a broader and flatter upper region while retaining rounded transitions toward the surrounding film. The final profile can also be further tuned through the relative size of the bright disk and the exposure dose (Supplementary Fig. 11).
We experimentally implemented this illumination strategy for a square array of flat-topped protruding microreliefs using the hologram sequence shown in **Figure 4**b. The first exposure followed the same writing conditions adopted for the array in **Figure 3**g and produced a periodic array of protruding microreliefs from the flat film. The relative size of the bright disks in the second exposure was optimized at fixed exposure time by minimizing the deviation of the measured upper profiles from fitted rectangular plateaus, following the procedure described in Supplementary Fig. 11. The selected geometry used equal radii $r_1 = r_2 \approx 4.9$ µm for the dark and bright disks. The optimized two-step illumination sequence yielded the array characterized by AFM and complementary SEM imaging in **Figure 4**c. The comparison of representative AFM topographies and profiles in after the first and second exposures in **Figure 4**d shows the transition from an initially rounded protrusion to a broader, lower structure with a substantially flatter top. The resulting array is characterized by an average modulation $h = 3.5 \pm 0.1$ µm relative to the local baseline and an average fitted width $w = 11.4 \pm 0.4$ µm. Here, $w$ is the width parameter obtained from the rectangular-pulse fit to the AFM profiles (Supplementary Fig. 11) and provides an effective measure of the lateral extent of the reshaped structures, even though not sensitive to possible reentrant portions of the sidewalls.
The combined spatial and temporal control enabled by the holographic multi-step exposure can be generalized from the circular patterns to a broad class of shapes, arrays, and freeform designs, as schematized in **Figure 5**a. In this workflow, the initial lateral geometry of each microstructure is defined by the shape of the dark feature embedded within the shared bright field used for the first exposure. The second exposure then refines the topography through spatially tailored holograms, whose pattern can be arbitrary designed, for example with a complimentary pattern of the first exposure to further flatten the upper surface.
Combining different feature shapes and positions withing the illumination sequence enables programmable arrays and assemblies of microstructures. As representative examples, **Figure 5**b shows an irregular array of circular structures, while **Figure 5**c demonstrates an assembly containing mixed feature geometries. AFM and SEM images confirm the spatial arrangement and overall morphology of these structures in agreement with the corresponding designed illumination sequences (Supplementary Fig. 12). **Figure 5**d provides magnified views of some of the microreliefs in the array, further supporting design flexibility and morphological control. Notably, the same design framework can extend beyond discrete microstructures to continuous periodic or freeform protruding motifs exploiting the digital flexibility of CGH illumination (see also Supplementary Fig. 13). **Figure 5**e shows the two-step formation and reshaping of a branched relief. The first exposure generates a connected protruding structure from a continuous dark motif embedded within a shared bright field. A second exposure with the complementary illumination pattern broadens and flattens the upper region, following the same morphological evolution observed for the discrete structures, as also shown by the corresponding AFM profiles. **Figure 5**f demonstrates the design flexibility of this process through a different branched geometry, produced by modifying the holographic patterns without changing the general exposure workflow. The paired AFM and SEM images confirm the continuity and overall morphology of the resulting relief. The hologram sequences used to produce the patterns in **Figure 5** are shown in Supplementary Fig. 12, together with additional examples of programmable shapes and assemblies of protruding microstructures produced from a flat film using light alone.

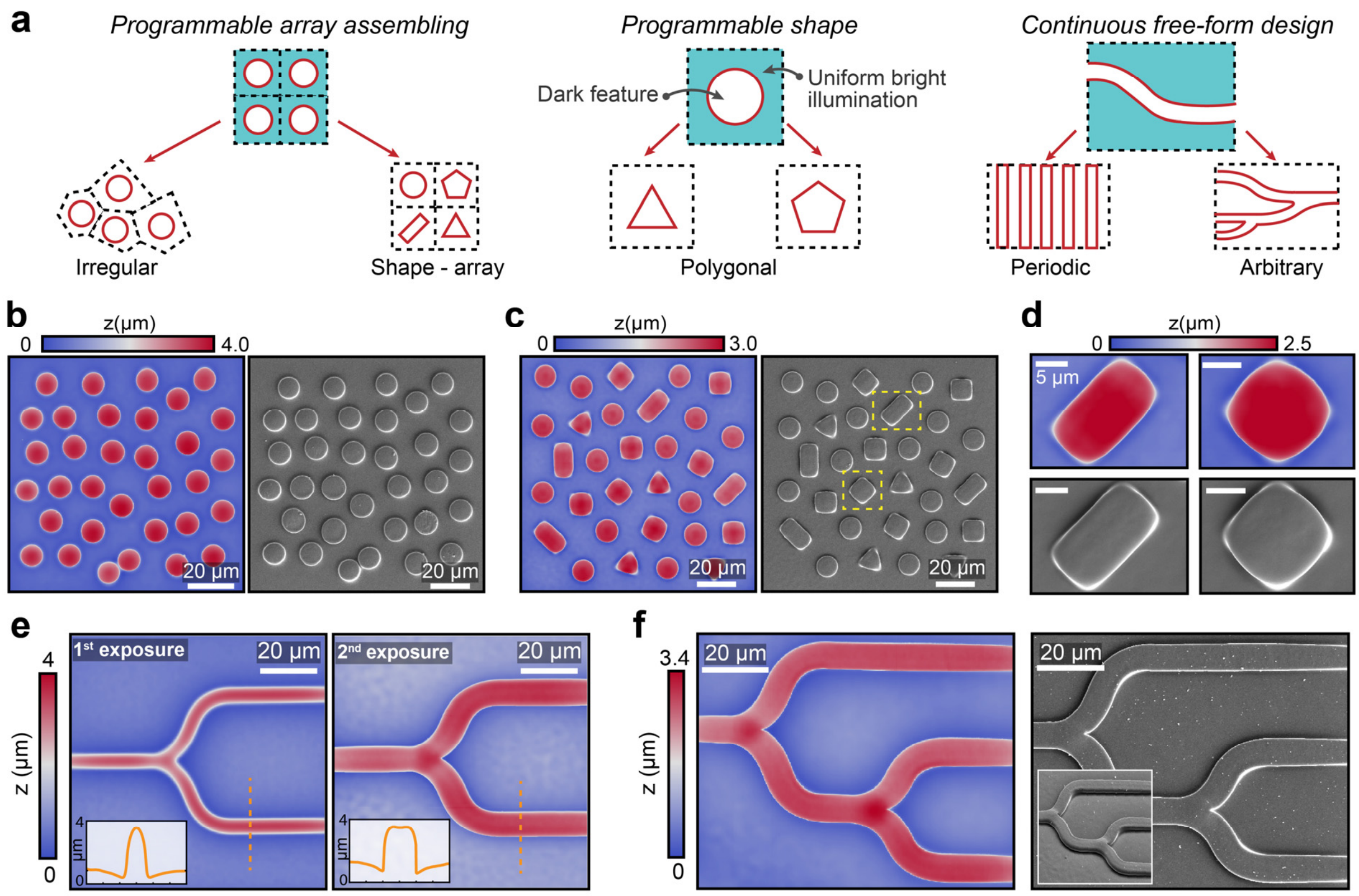

**Figure 5. Programmable arrays, feature geometries, and continuous freeform reliefs produced by sequential optical patterning. a)** Schematics of the design possibilities enabled by controlling the shape and spatial arrangement of dark features within a shared bright illumination field. **b) -c)** AFM topographies (left) and SEM images (right) of an irregular array of circular protruding microreliefs and assembly containing mixed feature geometries, respectively, after two-step exposure. **d)** Magnified AFM topographies (top) and SEM images (bottom) of representative polygonal structures. **e)** AFM topographies of a branched relief after the first exposure (left) and the second exposure (right). The inset profiles, extracted along the dashed lines, show the change from a rounded cross-section to a broader profile with a flattened upper region. **f)** AFM topography (left) and SEM image (right) of a tri-branched freeform relief. The oblique SEM inset provides an additional view of the connected structure and its three-dimensional morphology. For b) the first and second exposures were 300 s at ~30 W cm$^{-2}$ and 30 s at ~ 153 W cm$^{-2}$, respectively. For c), the corresponding conditions were 300 s at ~ 28 W cm$^{-2}$ and 30 s at ~ 142 W cm$^{-2}$. For e) the first and second exposures were 360 s at ~38 W cm$^{-2}$ and 15 s at ~ 329 W cm$^{-2}$, respectively. For f), they were 300 s at ~ 40 W cm$^{-2}$ and 20 s at ~ 230 W cm$^{-2}$.

## 2.4 Hierarchical and reprogrammable microstructures

While the double-step scheme above already demonstrates morphological shaping through combined spatial and temporal control of the holographic illumination, additional architectural complexity can be obtained by extending the workflow to multi-step exposure sequences. Here, we use three illumination steps to produce hierarchical microstructures with digitally defined primary geometries and secondary surface features.

**Figure 6**a illustrates our three-step illumination sequence. The first exposure generates an array of protruding microstructures using dark features embedded within a shared bright field. The second exposure applies an inverted bright holographic pattern to reshape the array in flattened-top microstructures, defining the primary morphological level. A third, independently designed

hologram is then used to introduce secondary structural features on top of each element of the array, yielding the hierarchical morphology schematized in **Figure 6**b.

**Figure 6**c demonstrates this concept for an array combining different primary shapes with secondary features that vary in geometry and orientation. The corresponding SEM image in **Figure 6**d and the magnified oblique views in **Figure 6**e independently corroborate the hierarchical morphology observed by AFM as the result of the three-step illumination sequence (the AFM images and profiles of the surfaces after each step of illumination are reported in Supplementary Fig. 14). Additional examples obtained using either binary fringe pattern to generate secondary grooves (see also Movie S1) or a combination of different illumination motifs are presented Supplementary Fig. 15, together with their holographic illumination sequences and those for the structures in **Figure 6**. Using smooth grayscale intensity gradients during the third exposure can also expand this flexibility, enabling also secondary features with gradually varying heights (Supplementary Fig. 16).

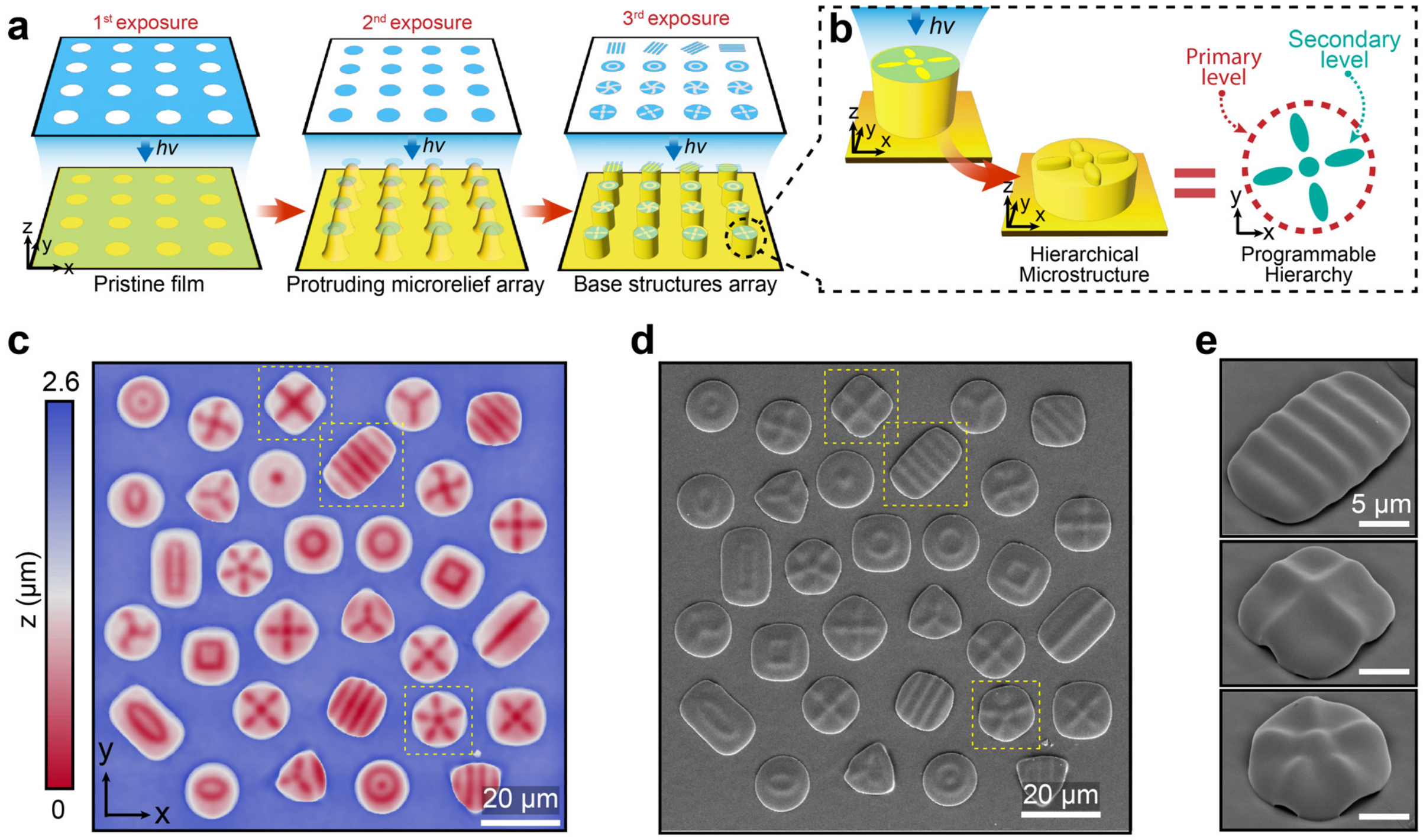

**Figure 6. Three-step optical patterning of hierarchical microstructures. a)** Illumination sequence used to generate protruding microreliefs from a pristine film, reshape them into flat-topped primary structures, and introduce secondary surface features. **b)** Schematic illustrating the formation of the secondary level of a hierarchical microstructure through a spatially tailored third exposure. **c)** AFM topography of a hierarchical array of microstructures combining different primary shapes and secondary motifs. **d)** Corresponding SEM image. **e)** Oblique SEM views of the structures identified by the yellow boxes in d) showing higher magnification secondary features on individual microreliefs. The first, second, and third exposures were performed for 300 s, 30 s, and 45 s at local writing intensities of approximately 28, 142, and 90 W cm$^{-2}$, respectively.

As a further demonstration of the spatiotemporal tuning capabilities arising from the combination of the reconfigurable nature of azopolymers and our all-optical holographic patterning approach, we show optical erasure and rewriting of different protruding microstructures within the same sample area. Optical erasure can be achieved through irradiation at a wavelength strongly absorbed by the azopolymer, as demonstrated in our previous works.[17,29,35]

**Figure 7** demonstrates this concept through five successive writing and erasing cycles. In each cycle, the desired microstructure topography was first inscribed using the holographic workflow described above. Between successive writing operations, erasure was implemented by increasing the intensity of the 405 nm assisting beam to 0.6 Wcm$^{-2}$ for 60 s, as detailed in the "Methods" section. The AFM images in **Figure 7** show that this erasing step restores a nearly flat surface after each exposure, allowing a different geometry to be written in the same region (see also Movie S2).

Additional AFM analysis in Supplementary Fig. 17 shows that erasure reduces the initially micrometre-high reliefs to surfaces with an RMS roughness (Rq) of approximately 0.11-0.19 µm obtained form the analysis of the 100 ×100 µm$^2$ area. The residual topography is dominated by shallow, broad height variations associated with the overall square illumination footprint, while the individual features of the previously written patterns are largely suppressed. Despite this residual background, the substantial reduction in surface modulation enables successive writing of distinct microstructure geometries within the same film area without observable crosstalk.

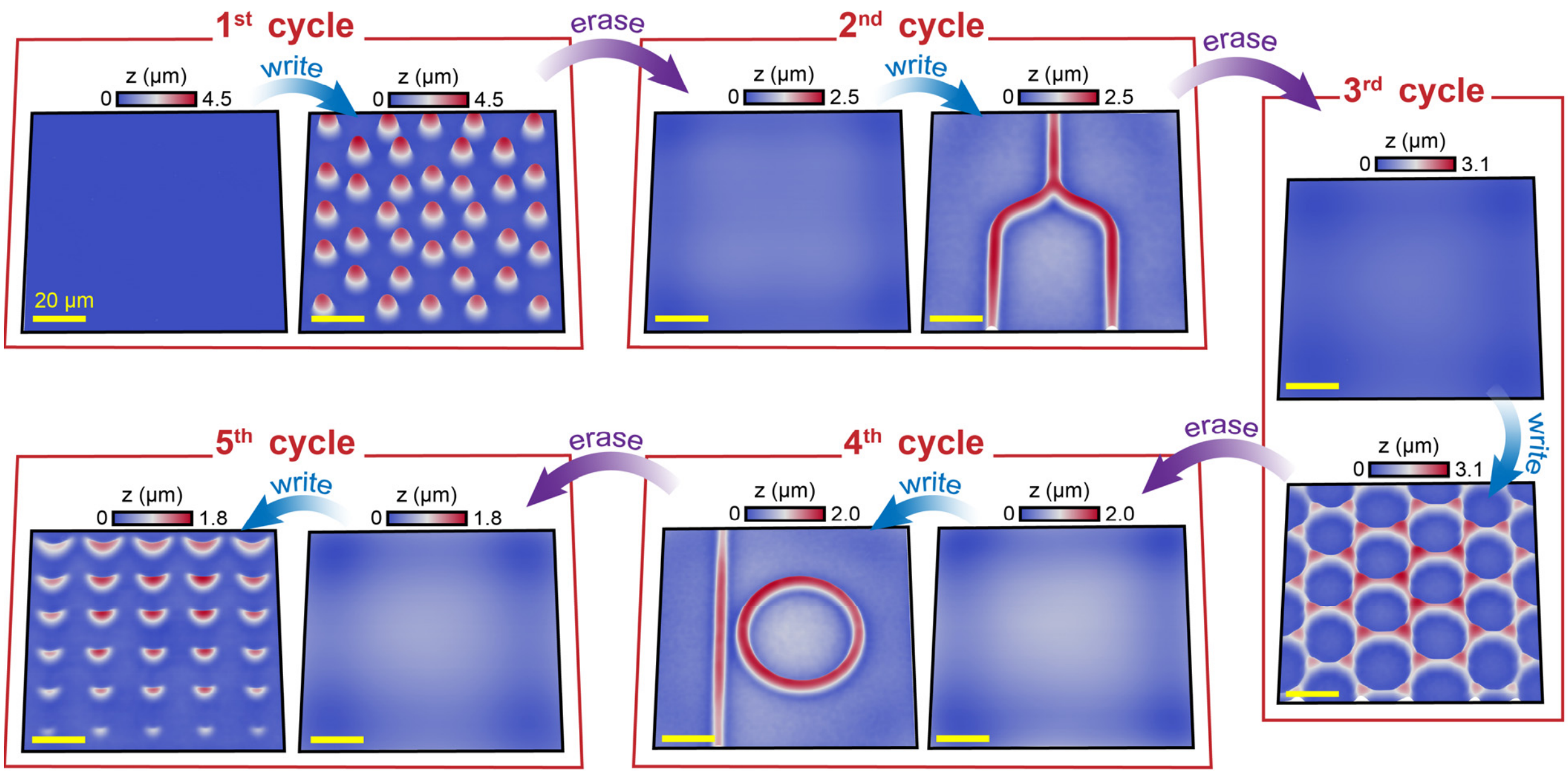

**Figure 7. Repeated optical writing and erasure in the same film region.** The 3D AFM images showing five successive writings-erasing cycles. Each pair shows the surface before and after holographic writing, following the direction of the arrows. Writing was performed using the holographic illumination with simultaneous 405 nm assisting illumination. Erasure was performed by increasing the 405 nm intensity to ~0.6 W cm$^{-2}$ for 60 s. The exposure time was 300 s for all the patterns and the writing intensities from first to the fifth pattern were ~ 38 - 47 Wcm$^{-2}$.

## 2.5 Replication and optical functionality of programmed microreliefs

The programmability and versatility demonstrated above arise directly from the photoinduced deformation of the azopolymer, which enables successive exposures to refine microrelief morphology, introduce hierarchical features, or erase and rewrite different structures within the same film region. For some applications, however, it may be desirable to retain the programmed surface geometry while benefiting from the chemical and physical properties of a different material. Building on the established use of structured azopolymer surfaces as replication masters, here, we demonstrate that the high-modulation, hierarchical protruding microreliefs generated directly from initially flat films through co-engineered bright-dark illumination and sequential optical reshaping can serve as optically programmable masters for morphology transfer into another polymer, without requiring a pre-patterned azopolymer surface.

To demonstrate this capability, we replicated the hierarchical array introduced in **Figure 6**c into polycarbonate (PC) using an intermediate polydimethylsiloxane (PDMS) mold, following the replication procedure schematized in **Figure 8**a, detailed in the "Methods" section.

The AFM topography in **Figure 8**b shows that the PC replica reproduces the spatial arrangement of the array and its combination of different primary shapes and secondary motifs. **Figure 8**c compares three representative structures on the azopolymer master and the PC replica, showing the transfer of both the primary geometry and the secondary features. These comparisons show that the transfer preserves recognizable features at both levels of the hierarchy. The corresponding profiles in **Figure 8**d retain the main lateral features and secondary modulations, while showing a reduction ~13-18% in vertical relief amplitude in the PC replica (see also Supplementary Fig. 18 for the size reduction study).

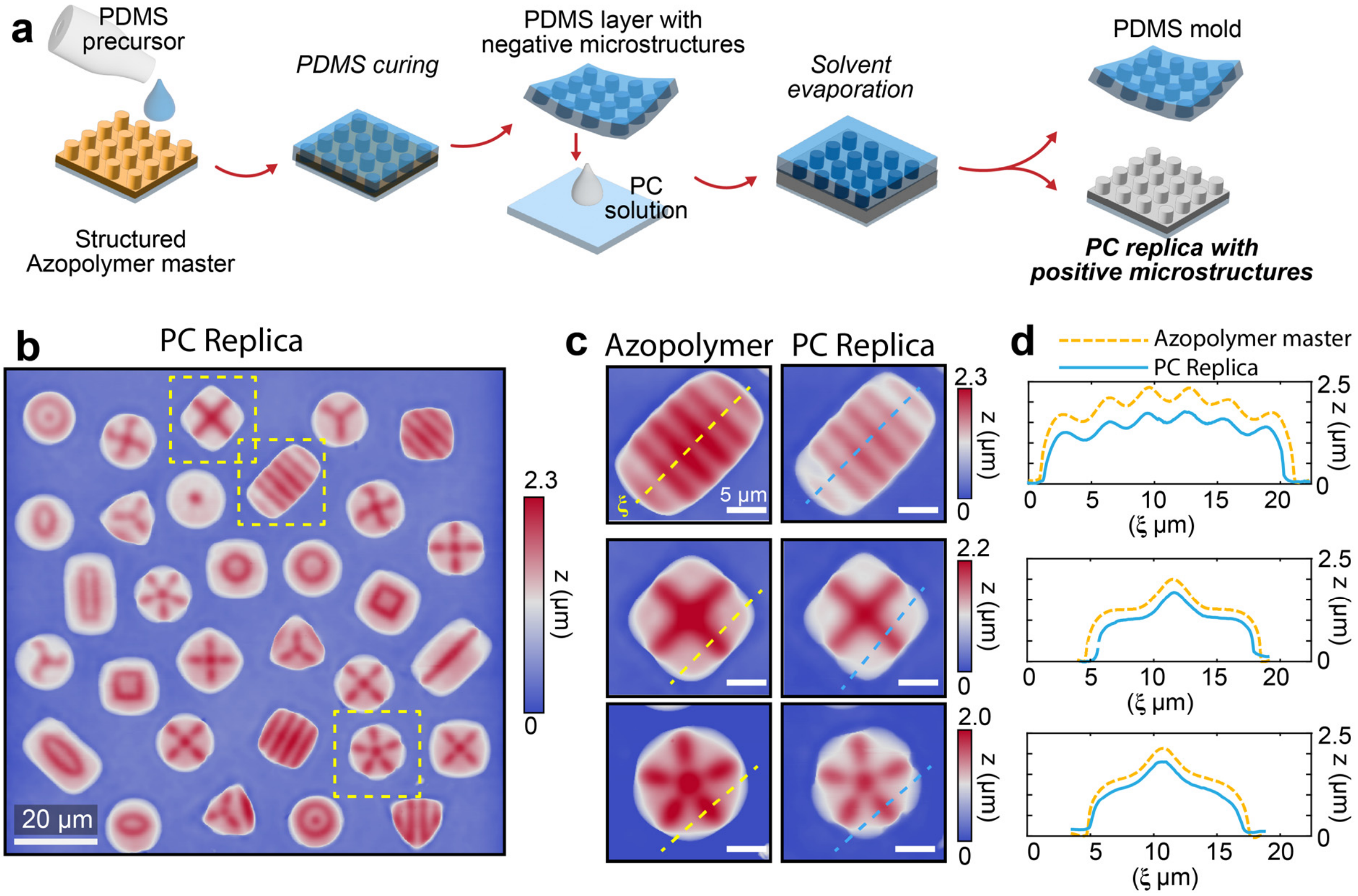

**Figure 8. Transfer of optically programmed hierarchical microreliefs into polycarbonate. a)** Schematic of the replication procedure. The structured azopolymer master is used to form a negative PDMS mold by casting and curing a PDMS precursor. The mold is brought in contact with a PC solution on a glass substrate. Solvent evaporation and demolding yield the final positive PC replica. **b)** AFM topography of the replicated hierarchical array in PC. Yellow dashed boxes identify the representative structures compared in c). **c)** Magnified AFM topographies of corresponding structures on the azopolymer master (left) and the PC replica (right). A common height scale is used within each pair. **d)** Comparison of the profiles extracted along the colored dashed lines in c), with the azopolymer master shown by orange dashed curves and the PC replica by blue solid curves. The coordinate ξ denotes distance along the indicated profile direction. The profiles show transfer of the primary geometry and secondary surface features, accompanied by reduced vertical modulation in the replica.

This transfer provides a route to combine optical programming of the master geometry with material selection for the resulting structures. We demonstrate its functional relevance by replicating the rounded protrusions generated in a single exposure to obtain PC microlenses.[47–49] **Figure 9**a compares the measured absorbance spectra of the two materials. The azopolymer exhibits appreciable absorption at shorter visible wavelengths, whereas the PC shows negligible absorbance over the displayed spectral range. **Figure 9**b compares AFM images and cross-sectional profiles of a representative azopolymer structure and its PC replica, showing the transfer of the curved surface profile that provides the microlens geometry.

**Figure 9**c summarizes the fitted height $h$ and radius of curvature $R_c$ for both the masters and replicas as functions of writing time and dark-circle radius, with the spherical-cap geometry illustrated in the inset. The illumination patterns are provided in Supplementary Fig. 19, while the fitting procedure and residuals are reported separately for the azopolymer masters and PC replicas in Supplementary Fig. 20 and 21. These measurements characterize both the morphology defined by optical writing and the changes introduced during replication.

At fixed dark-circle radius, increasing the writing time increases the relief height and decreases the fitted radius of curvature for both materials.[50] The replicas generally exhibit lower heights, while retaining a similar dependence of curvature on writing time. At fixed exposure time, varying the dark-circle radius provides a further means of adjusting the profile, with the fitted radius of curvature increasing more markedly for the larger dark features. These results show that the geometrical control achieved during master fabrication is retained after transfer, although replication modifies the resulting surface profile.

**Figure 9**d compares longitudinal maps of the point spread function (PSF) in the y-z plane and focal-plane images measured for a microlens fabricated with $r = 5.9$ μm and $t = 120$ s in the azopolymer master and its corresponding PC replica at wavelengths of 450, 550, 600, and 650 nm (optical setup shown in Supplementary Fig. 22). These values denote the central wavelengths of the 10 nm spectral bands selected from a supercontinuum source using bandpass filters (see "Methods" section). The images show localized focusing at the four wavelengths, with a particularly weak signal from the azopolymer master at 450 nm. For each material, the camera settings were selected to avoid saturation at 600 nm and kept unchanged at the other wavelengths, and all focal signals were normalized to the respective peak at 600 nm. The display factors indicated in **Figure 9**d allow visualization of the weaker signals, while the normalized focal peaks are compared directly in Supplementary Fig. 23. The PC replica exhibits a higher relative focal signal at 450nm, consistent with the lower absorbance of PC measured at this wavelength (**Figure 9**a), and a smaller increase at 550 nm. Because this comparison retains the spectral weighting of the illumination and detection system, it accurately describes the relative focal response of the microlenses but cannot be directly used for absolute transmission or focusing efficiency estimations.

The focal response of the microlenses can also be adjusted through the illumination parameters used to fabricate the master microstructures (see Supplementary Fig. 24). **Figure 9**e shows the measured focal distances at a probe wavelength of 633 nm for PC microlenses replicated from the azopolymer master obtained at a fixed dark-circle radius $r = 6.8$ μm and writing time ranging from 30 to 180s. The plot shows a decrease of the focal distance from approximately 65 to 20 μm, accompanied by a reduction in the transverse and axial widths of the focal intensity distribution (Supplementary Fig. 25). The morphological analysis in **Figure 9**c shows increasing relief height and decreasing fitted radius of curvature with writing time, consistent with the stronger focusing observed for the resulting microlenses. The smaller changes in focal distance at longer exposure times indicate reduced sensitivity of this optical parameter to further increases in writing time within the investigated range, consistent with the morphological analysis for this type of microstructures in Section 2.1. PSF-derived MTF estimates provide complementary characterization of the focal response, with the corresponding analysis reported in Supplementary Fig. 25.

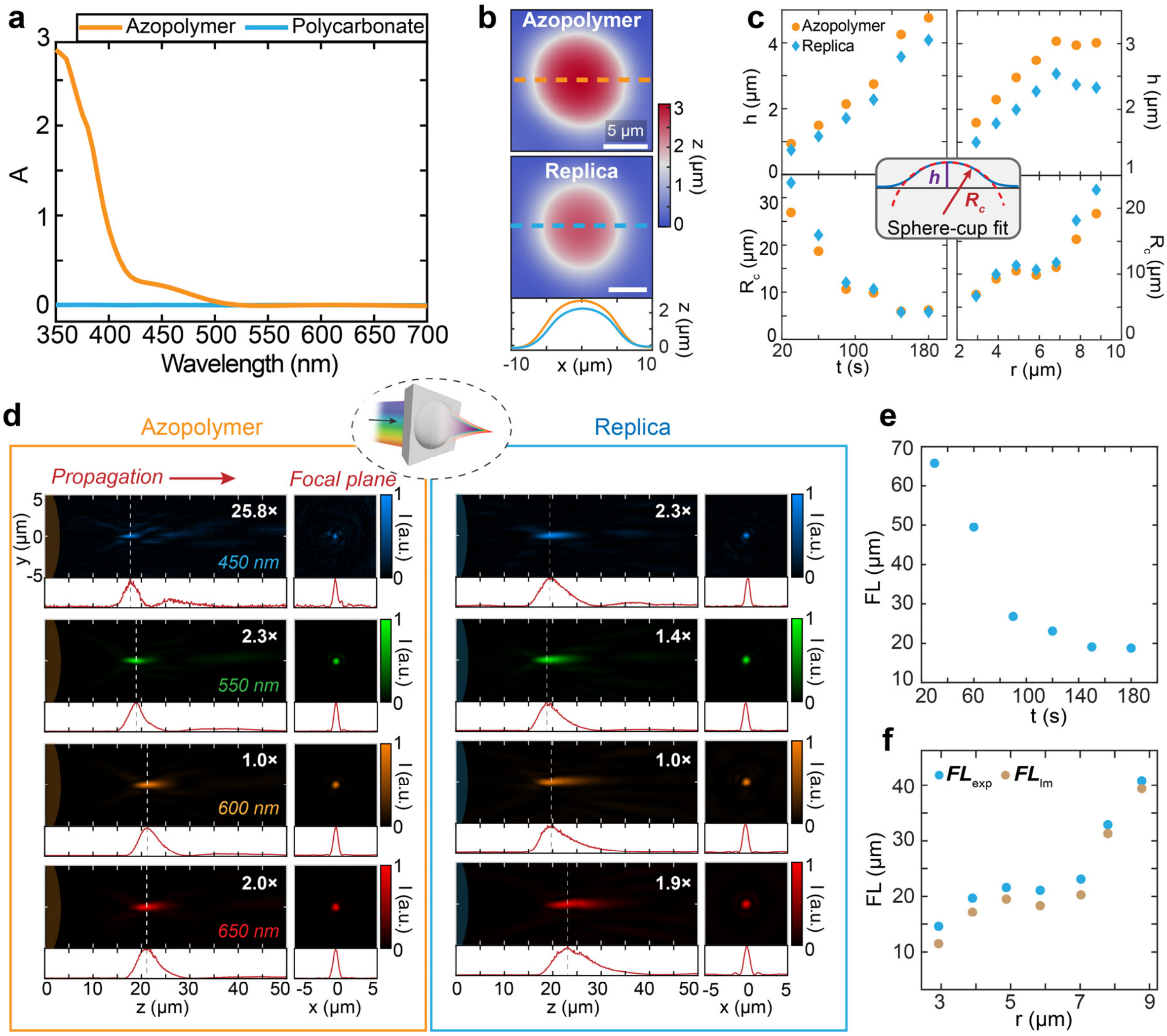

**Figure 9. Morphological control and optical characterization of azopolymer microlenses and their polycarbonate replicas. a)** Measured absorbance spectra of the azopolymer and polycarbonate samples. **b)** AFM topographies of a representative azopolymer microlens and its PC replica, and profiles extracted along the dashed lines. **c)** Fitted $h$ and $R_c$ for the azopolymer masters (orange circles) and PC replicas (blue diamonds). The left column shows the dependence on master-writing time $t$ at fixed dark-circle radius $r$ = 5.9 µm, and the right column shows the dependence on dark-circle radius $r$ at fixed writing time $t$ = 120s. The inset defines the fitted geometrical parameters. **d)** Measured longitudinal PSF maps in the y-z plane and focal-plane images for an azopolymer microlens (left) and its PC replica (right) at 450, 550, 600, and 650 nm. The nominal wavelengths correspond to the centers of 10 nm spectral bands selected from a supercontinuum source. Curves below the images show the corresponding axial and transverse intensity profiles and dashed vertical lines identify the focal planes. Camera settings were held constant within each material series. Signals were normalized to the respective focal peak at 600 nm and rescaled for visualization using the indicated multiplicative factors. **e)** Experimental focal distance FL of PC microlenses as a function of master-writing time at fixed dark-circle radius, measured at 633 nm. **f)** Focal distance as a function of dark-circle radius at fixed writing time and a probe wavelength of 633 nm. Experimental values ($FL_{exp}$, blue) are compared with estimates from the lensmaker's equation using the fitted spherical-cap geometry ($FL_{lm}$, brown).

Changing the dark-circle radius provides a complementary means of controlling the focal response. At a fixed writing $t$ = 120s, the measured focal distance spans approximately 14–41 µm over the investigated radius range (**Figure 9**f). The dependence is not uniform across this range, with a stronger increase at the larger radii. Figure 9f also compares the measurements with propagation calculations using the measured AFM topographies and estimates from the lensmaker equation based on the fitted spherical-cap geometry. The estimations well reproduce the overall increase in focal distance with dark-circle radius. An additional characterization of the multi-color PSF of the microlens with obtained from r = 6.8 µm and t = 120s is shown in Supplementary Fig. 26. However, the measurements and the reproduced trend from calculations and simulations establish a clear connection between the digitally prescribed illumination geometry, the resulting surface master morphology, and the optical response of the replica.

# 3. Discussion and Conclusions

Current holographic patterning of azopolymer films has largely operated in regimes in which the surface relief follows the geometry of the illuminated regions only, yielding engraving-like, smooth, and relatively shallow topographies. In this work, we show that co-engineering bright and dark regions of a circularly polarized holographic intensity pattern provides a route to generating isolated protruding microreliefs with several micrometers of modulation directly from pristine flat films. The process uses only light exposure, without a mask and pre- or post-lithographic processing. In our scheme, the engineered dark features of the illumination defines the shape of the region of material accumulation, while the surrounding bright region provides the gradients and material reservoir required for inward mass transport and confined accumulation. The experimental results, together with VPA modelling of stress-driven deformation in azopolymers, support this interpretation. The model provides a physically grounded description of the principal deformation regimes, while the experimental geometrical studies establish practical design relationships under the investigated exposure conditions.

The significance of this concept lies not only in the direct formation of isolated, positive, protruding microreliefs from pristine flat films, but also in the ability to generate multiple protruding microreliefs within a shared illuminated region. Their geometry and separation can directly be prescribed through the dimensions and spacing of dark features embedded in a shared bright illumination field, subject to the optical resolution of the actual setup and the spatially coupled material response. Within this framework, a broader class of structures becomes accessible to holographic lithography of azopolymers, as demonstrated here for regular and irregular arrays with mixed primary geometries, as well as continuous freeform reliefs.

Another central result is that the reconfigurable nature of azopolymer mass transport, combined with dynamically programmable holographic illumination, allows temporal control to be included in the patterning workflow. Sequential exposures reshape initially rounded protrusions into microreliefs with flatter upper regions and introduce secondary features to produce hierarchical topographies. The same general formation and reshaping strategy applies to discrete arrays and connected freeform structures, with the transition between these configurations encoded digitally in the illumination patterns. Optical erasure further enables the same film region to be reused for different designs. The five successive write-erase-write cycles demonstrate repeated high-modulation patterning in the same area and with the same optical system. After erasure, the remaining broad surface variations have an RMS roughness of approximately 0.11 - 0.19 µm, substantially below the initial micrometer-scale relief modulation. Longer cycling sequences under fixed writing conditions would further clarify the evolution of the material response and residual topography.

The replication demonstration extends the optical morphological programmability of the azopolymer film to structures in a second material, having selected properties for intended functions, through a straightforward molding process. Transfer of the structures into polycarbonate preserves recognizable primary geometries and secondary surface features, while the measured profiles document reduction in vertical modulation in the range of 13-18%. These results demonstrate that the high-modulation surfaces generated by bright–dark illumination can function as optically programmable masters. The replicated microlenses further show how the rounded profiles produced by the process across different master-writing conditions can be used in a broader wavelength range in the visible, with an improved relative focal signal in the blue and green consistent with reduced attenuation compared with the azopolymer master.

At the same time, although the present platform offers broad design flexibility, several aspects remain important for extending the demonstrated capabilities. The attainable morphology depends jointly on the illumination geometry, exposure conditions, and the volume of material available for redistribution. Consequently, the separation thresholds and shape-control relationships obtained here should be regarded as design values for the investigated configurations. The most reliable operation is obtained when microreliefs are written in the saturation regime, where morphology becomes only weakly dependent on further variations in exposure conditions. This regime enables fully developed microstructures, but it also restricts the range of aspect ratios that can be accessed with the same degree of control. Since saturation is expected to depend not only on exposure dose, but also on illumination intensity and film thickness, a systematic study of their coupled effects would help widen this operating window and establish predictive relationships for microreliefs with different heights and aspect ratios. Complementary measurements across independently prepared films and writing sessions would further characterize the reproducibility of these relationships.

In addition, the structures demonstrated here are in the micrometer range defined by the optical configuration used in this work, but the underlying concepts are not intrinsically tied to the scale or field of view investigated here. The same bright-dark co-design principles could be implemented with optical configurations aimed at either improving spatial resolution or at increasing the patterned area. The latter aspect is particularly relevant for applications. In this context, stitching and tiling provide potential strategies for extending the structured area, but their implementation requires explicit engineering of the boundaries of each patterned field. The peripheral relief and depletion observed here show that the surrounding deformation must be considered together with the targeted structures. Gradual intensity transitions at the field boundaries, including apodization profiles, could be investigated to control this deformation and the overlap between adjacent exposures. Real-time monitoring of surface evolution by digital holographic microscopy could help guide such adjustments even during patterning.[14,33] These extensions are compatible with digital CGH design and could incorporate optimization algorithms that explicitly target intensity fidelity, dark-region contrast, and boundary profiles. Gradient-based approaches[51] provide one possible direction to explore alongside the MRAF iterative Fourier-transform algorithm used here, while retaining the general optical workflow. Treating both the main structures and peripheral regions as part of an optimization problem[52] for the entire holographic pattern could therefore support improved control of local morphology and extension to larger patterned areas.

In conclusion, our results establish a spatiotemporal all-optical framework for programmable and reconfigurable surface microstructures based on the combined control of bright and dark holographic features and exposure history. The approach connects direct protrusion formation from flat films with sequential reshaping, hierarchical patterning, repeated optical rewriting, and transfer into another material. The demonstrated optical functionality of the replicas provides a concrete use of this morphological control, linking digitally programmable master fabrication to functional microstructures in materials selected for their intended application.

# 4. Methods

### 4.1 Azopolymer synthesis and film preparation

The azopolymer was synthesized, purified, and characterized as previously reported ($M_w$ = 27 000; phase sequence: Glass 67 °C Nematic 113 °C Isotropic; $\lambda_{max}$ = 350 nm).[53] The reagents were purchased from Merck and used without further purification. The chemical structure and properties of our azopolymer are reported in our previous works.[17,29,35,39,40] The azopolymer films with a thickness of ~ 2.7 µm were obtained by spin coating the solution on 24 x 60 mm cover slides. After film deposition was completed, the samples were kept under vacuum at room temperature for 24 h to remove solvent traces and then stored at room temperature before photopatterning experiments.

### 4.2 Optical configuration of computer-generated hologram system

The computer-generated holographic setup is schematized in **Figure 2**a and detailed in our previous works.[17,29,31,35,40] A 491nm laser beam (Cobolt Calypso laser) was expanded and collimated before being directed to the screen of a reflective spatial light modulator (SLM-phase only, Holoeye Pluto 2.1) using a telescopic configuration with $L_1$ ($f_1$ = -50 mm) and $L_2$ ($f_2$ = 1000 mm). The wavefront of the reflected beam was modulated according to the phase profile (kinoform) displayed on the SLM as an 8-bit gray level image. The modulated beam propagated through a *4f* lens system ($L_3$ with $f_3$ = 300 mm and $L_4$ with $f_4$ = 175 mm), producing a demagnified image of the SLM field in the back focal plane of the Mitutoyo 50× microscope objective (NA = 0.55). This configuration enabled the projection of the holographic light intensity pattern (which we simply refer as *hologram*) onto the azopolymer film surface. For the real-time surface observations, a CCD camera coupled with a tube lens ($L_T$ with $f_{Lt}$ = 200 mm) was used to record the white light from a LED transmitted through the sample. The same configuration allowed imaging the writing light reflected from the sample plane through a beam splitter (BS) with 90/10 splitting ratio, used for hologram focusing and alignment.

The hologram design and kinoform calculation followed the procedure described in our previous works.[29,30,35] In brief, we designed each hologram as an 8-bit grayscale image (1080 × 1080 pixels) using conventional image-processing software. We then used an iterative Fourier transform algorithm based on the mixed-region amplitude freedom (MRAF) extension[54] of the standard Gerchberg-Saxton phase-retrieval algorithm[55] to calculate a set of 1000 kinoforms with independent initializations of the algorithm. During the photopatterning experiments, the set of kinoforms was displayed sequentially on the SLM screen at a frame rate of 60 Hz, to reduce the speckle noise affecting each hologram and produce smooth and contrasted average intensity patterns.[29,30,35]

### 4.3 Holographic intensity patterns characterization

The hologram images were obtained by averaging at least 1000 frames in holographic movie sequences. These were recorded by placing a flat mirror in the surface plane and recording a movie using the CCD of the real-time observation system implemented in the optical scheme described in section 4.2. The contrast was calculated using the standard relative contrast: $C = (\langle I_{bright}\rangle - \langle I_{dark}\rangle) / (\langle I_{dark}\rangle + \langle I_{bright}\rangle)$, were $I_{bright}$ and $I_{dark}$ are the average intensity levels measured within the region of the bright and dark in the averaged hologram images.

### 4.4 Azopolymer film exposure

For photopatterning experiments, the azopolymer films were mounted on an XYZ motorized stage (PI M-111.1DG with C-863 Mercury Servo Controller, from Physik Instruments). The photopatterning processes were performed using circular polarization and a constant output power from the writing laser at $\lambda = 491$ nm. Because the illuminated area depends on the hologram geometry, the local writing intensity at the sample plane varies between patterns even when the laser output power is kept constant. The estimated local intensity and exposure time used for each experiment are reported in Supplementary Table 1.
In addition to the writing holograms, a spatially uniform collimated assisting beam (at $\lambda = 405$ nm, circularly polarized) was directed onto the structuring area from the substrate side with an intensity of ~ 0.4 $Wcm^{-2}$ (**Figure 2**a). This intensity level is referred to as the assisting mode and this beam was applied simultaneously with the writing holograms during the formation of single protruding microreliefs and arrays, including the first exposure of multi-step sequences. The assisting beam promotes faster relief formation and higher attainable modulation while also suppressing the surface roughness suppressing the roughness that would otherwise develop within uniformly illuminated bright regions of the holograms (see Supplementary Fig. 27). These effects are consistent with coupled photochemical and photothermal contributions. Changes in the azobenzene isomer populations and molecular alignment dynamics can modify absorption of the writing beam, while illumination-induced heating can increase polymer mobility. Its influence on the mechanical response is represented phenomenologically in the VPA model through a reduced effective yield stress, without explicitly resolving the coupled photochemical and thermal processes.
In multi-step exposure experiments, the second and third exposures were performed without the assisting beam. For all step-resolved AFM comparisons, the corresponding writing sequence was repeated on adjacent regions of the same sample and intentionally terminated after the exposure step of interest
For the optical erasing described in Section 2.4, the intensity of the same 405 nm laser was increased to ~ 0.6 $Wcm^{-2}$, corresponding to its use in erasing mode [REF].[17,29,35]

### 4.5 Surface characterization

*Atomic Force Microscopy*

The fabricated surfaces were analyzed by an atomic force microscope (AFM, WITec Alpha RS300) operating in tapping mode. The scans were performed with a pixel resolution of 100 nm using a high-aspect-ratio probe (ISC-225C3_0-R from Team Nanotec, aspect ratio >5:1, radius <10 nm, half cone angle <5°) mounted on a cantilever with resonance frequency of 75 kHz. The influence of probe geometry was assessed by imaging a lithographically fabricated silicon micropillar reference sample using both the high-aspect-ratio probe employed in this work and a conventional probe (ARROW-FMR from NanoWorld, resonance frequency 75 kHz), with complementary SEM characterization (Supplementary Fig. 28). For the profiles compared, the high-aspect-ratio probe reduced the apparent difference between bottom and top widths by approximately 58% along x and 69% along y relative to the conventional probe, while the measured pillar heights agreed within 0.01 μm. This comparison supports the suitability of the selected probe for resolving steep micrometre-scale protrusions. The topographic images and the corresponding surface profiles were processed using *Gwyddion* software (version 2.62)[56], and finally plotted in Matlab®.
For the two-peak fitting shown in **Figure 3**d and Supplementary Fig. 9, was performed using an open-source peak fitter tool[57] implemented in Matlab®.

Volume-balance analysis was performed on a representative branched microrelief after the first exposure of the patterning sequence described in Section 2.3. Four AFM scans were used to reconstruct the surface topography across the protruding structure, the surrounding depleted regions, and the peripheral relief. The original film surface, identified from adjacent unexposed regions, defined the reference plane at z = 0. The accumulated and depleted volumes were calculated by integrating the positive and negative height deviations from this plane, respectively, with the depleted volume expressed as a positive magnitude. The ratio of the total accumulated to depleted volume was 0.97, consistent with approximately volume-conserving redistribution for the analyzed structure. The scan locations, reference regions, and volume definitions are shown in Supplementary Fig. 30.

*Scanning electron microscope (SEM)*

Scanning electron microscopy (SEM) was used as a complementary technique to corroborate the morphological data obtained from AFM for the structures on the azopolymer master. Measurements were performed using a field emission gun SEM (FEG-SEM) FEI/ThermoFisher Nova NanoSem 450. Prior to imaging, the samples were sputter-coated with a nanometric layer of a gold/palladium (Au/Pd) alloy using a Denton Vacuum Desk V coating system. Images were acquired in top view using a magnification ranging from 1500x to 15000x and, where required, in side view at magnification ranging from 2500x to 15000x by tilting the sample by 55° around the x-axis and applying the instrument's tilt correction.

### 4.6 Viscoplastic Photoalignment (VPA) model

All modeling data were derived from the recently developed viscoplastic photoalignment (VPA) model, which treats the material as an incompressible viscoplastic medium and therefore inherently assumes volume conservation, with Poisson ratio ν = 0.5 (see also Supplementary Note 1 and Supplementary Fig.29-30). According to the VPA model, the light field with circular polarization in the $xy$-plane generates the stress field $\boldsymbol{\tau}$ in the polymer matrix:[42,58]

$$\tau = \tau_0 \begin{pmatrix} 1/6 & 0 & 0 \\ 0 & 1/6 & 0 \\ 0 & 0 & -1/3 \end{pmatrix} \quad \text{with} \quad \tau_0 = -nkTV_r \tag{1}$$

where the magnitude of the light-induced stress $\tau_{eq} = \tau_0/2$ is proportional to the number density $n$ of backbone segments and the reduced strength of effective orientation potential $V_r$<0, which aligns the backbones along the polarization direction.[59] For a fixed wavelength and negligible thermal effects, $\tau_0$ scales approximately linearly with the laser intensity. According to Equation (1), the stress tensor is represented by the compressive component $\tau_z =$ $-\tau_0/3$, which acts in the major principal direction, and two tensile components $\tau_\perp = \tau_0/6$, which act in the two perpendicular minor principal directions. Such a light-induced stress tensor leads to uniaxial compression of the irradiated azo-polyacrylate along the direction of light propagation along *z*-axes and to radial stretching in the *xy*-plane.

VPA modeling of topographical structures was carried out using the finite element software ANSYS. Specifically, the Perzyna model was employed to describe the relationship between the rate of plastic strain $\dot{\varepsilon}_{pl}$ and the magnitude of light-induced stress $\tau_{\mathrm{eq}}$:

$$\dot{\varepsilon}_{pl} = \gamma\left(\frac{\tau_{\mathrm{eq}}}{\tau_{yield}} - 1\right) \tag{2}$$

where $\tau_{yield}$ represents the yield stress and the parameter $\gamma = \tau_{yield}/(3\eta)$ defines the viscosity $\eta$ of the plastic flow. When the light-induced stress substantially exceeds the yield stress, the deformation rate becomes proportional to the laser intensity. In the present study, $\tau_{yield}$ = 2 MPa was chosen, which is lower than the value of 5 MPa used in the previous work to model the same azo-polyacrylate.[42] Since the assisting beam operates at a shorter wavelength, it is expected to generate a higher cis-isomer population while contributing only weakly to orientation-dependent effects.[60] We therefore interpret its primary role as a modification of the material state, which is represented in the current model through a reduction of the effective yield stress. To further account for the assistant beam's role, we introduced a small baseline light intensity across the entire field ($V_r = -4$). According to Equation (2), the dynamic evolution of photodeformations is governed by the rate of plastic strain $\dot{\varepsilon}_{pl}$, which determines the speed of the process. The parameter $\gamma = 2.5 \cdot 10^{-3}$ $s^{-1}$ in Equation (2) was adjusted to reproduce the experimental timescale of photodeformation. For **Figure 1**e, Supplementary Fig. 1, **Figure 2**g, and **Figure 3**f, the reduced strength of the light-induced potential $V_r = -120$ was chosen to reproduce the experimentally observed deformation profiles under illumination at 12.0 $Wcm^{-2}$ and 0.375 $Wcm^{-2}$ for the assisting beam. This determines the magnitude of light-induced stress $\tau_{\mathrm{eq}}$ = 62.2 MPa, when the number density of backbone segments is assumed to be $n = 2.5 \cdot 10^{26}$ $m^{-3}$.[59] The light-induced stress tensor was incorporated in ANSYS software using the custom subroutine Userthstrain.[45] The thermal strain was chosen in this subroutine with the help of the Perzyna model (2) in such a way that it produced the plastic strain prescribed locally by the light-induced stress. In the present model, the parameters adjusted relative to previous studies ($\tau_{yield}$, γ, $V_r$) primarily affect the kinetics of the deformation process, whereas the characteristic doughnut morphology generated by circularly polarized illumination is largely preserved.

### 4.7 Fabrication of replica into polycarbonate film

Replication of selected surface structures was carried out in polycarbonate (PC) using a two-step replica molding procedure involving an intermediate polydimethylsiloxane (PDMS) mold. PDMS was prepared by mixing the precursor and curing agent at a 10:1 weight ratio. The mixture was cast onto the patterned azopolymer master and maintained at 30 °C for 18 h before being peeled off, yielding a negative replica of the original surface relief. A few drops of a PC solution in tetrachloroethane were deposited onto a glass slide, and the PDMS mold was placed on top with its patterned surface facing the solution. The solvent was allowed to evaporate through the pores of the PDMS mold, which was removed after 2 h to obtain a PC film reproducing the original surface topography.

Absorption spectra of both the azopolymer and polycarbonate were recorded using a PerkinElmer LAMBDA 900 UV-Vis-NIR spectrophotometer. Measurements were performed in absorbance mode over the wavelength range of 350-700 nm, with a sampling interval of 5 nm.

**4.8 Multi-wavelength microlens characterization**

The azopolymer master microlens and the corresponding polycarbonate replica were characterized by multi-wavelength point-spread-function (PSF) measurements. A supercontinuum laser (SuperK Compact – NKT Photonics) was used as the illumination source. The laser was spectrally filtered using four narrowband-pass filters (FWHM ≈ 10 nm) centered at 450, 550, 600, and 650 nm. These wavelengths were selected to sample the microlens response across the visible spectral range. For each wavelength, a three-dimensional PSF dataset was acquired under normal incident illumination of the sample, from the substrate side. The sample was mounted on a motorized XYZ stage and the transmitted light was recorded by using the same microscope objective, tube lens, and CCD imaging described for the CGH setup in Section 4.2. A schematic of the PSF measurement configuration is provided in Supplementary Fig. 22.

The three-dimensional PSF was reconstructed from a sequence of intensity images acquired at successive axial positions. The image plane was initially positioned approximately at the microlens aperture plane, which was defined as $z = 0$ plane. To locate the axial position of the aperture plane, the hologram pattern was brought into focus on the depressed area adjacent to the microlens. The sample was then translated along the optical axis in steps of 0.1 µm, with one image recorded at each position. The resulting image stack was used to reconstruct the evolution of the focal intensity distribution along the propagation direction.

# Acknowledgements

This work was supported by an ERC grant (HyperMaSH, 101164874), funded by the European Union. Views and opinions expressed are however those of the authors only and do not necessarily reflect those of the European Union or the European Research Council Executive Agency. Neither the European Union nor the granting authority can be held responsible for them. M.Sap. and N.L. acknowledge support by the grant GR 3725/10-1 from Deutsche Forschungsgemeinschaft.

# Author Contributions

S.L.O. conceived the study. I.K.J., M.Sap., and S.L.O. developed the methodology. I.K.J., M.Sal., F.R., and F.B. carried out experiments. N.L.and M.Sap. carried out the VPA simulations. A.P., I.K.J., M.Sal, N.L., and S.L.O processed and analyzed the data. S.L.O., F.B., and M.Sal. provided resources. I.K.J., F.R., M.Sal., and N.L. prepared the figures and visualized the results. S.L.O. supervised the project. I.K.J., F.R., M.Sap., and S.L.O. wrote the original manuscript. M.Sal., M.Sap., and S.L.O. reviewed and edited the manuscript.

# Conflict Interests

Authors declare no competing interests.

## Data availability statement

The data supporting the findings of this study are available from the corresponding authors upon reasonable request.

**Supplementary Information for**

# All-optical Lithography for Spatiotemporal Patterning of High-Modulation Azopolymer Microreliefs

I Komang Januariyasa,[1] Francesco Reda,[1] Nikolai Liubimtsev,[2] Achu Purushothaman,[1] Marina Saphiannikova,[2,3] Fabio Borbone,[4] Marcella Salvatore,[1,*] and Stefano Luigi Oscurato[1,*]

[1]Physics Department "E. Pancini", University of Naples Federico II, Complesso Universitario di Monte Sant'Angelo, via Cinthia, 80126, Naples, Italy.

[2]Division Theory of Polymers, Leibniz Institute of Polymer Research Dresden, 01069 Dresden, Germany.

[3]Faculty of Mechanical Science and Engineering, Dresden University of Technology, 01062 Dresden, Germany.

[4]Department of Chemical Sciences, University of Naples "Federico II", Complesso Universitario di Monte Sant'Angelo, Via Cintia, 80126 Naples, Italy.

*Stefano Luigi Oscurato, e-mail: stefanoluigi.oscurato@unina.it

*Marcella Salvatore, e-mail: marcella.salvatore@unina.it

# 1. Supplementary Figures

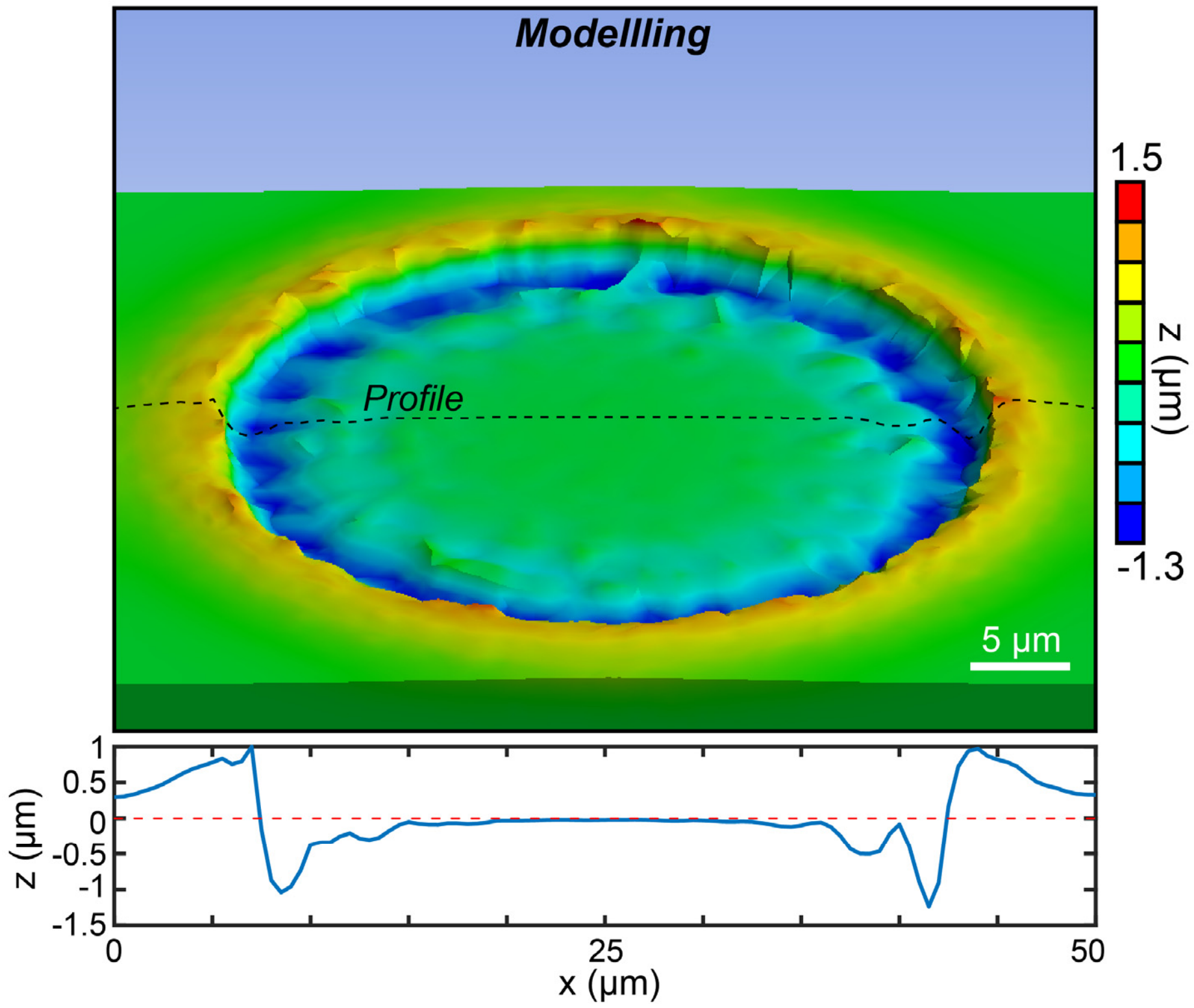

**Figure S1. Viscoplastic PhotoAlignment (VPA) modeling of azopolymer surface deformation.** VPA model prediction (three-dimensional view and cross-section) of azopolymer surface deformation under disk-shaped holographic illumination. The modelling parameters are detailed in the experimental section in the main text.

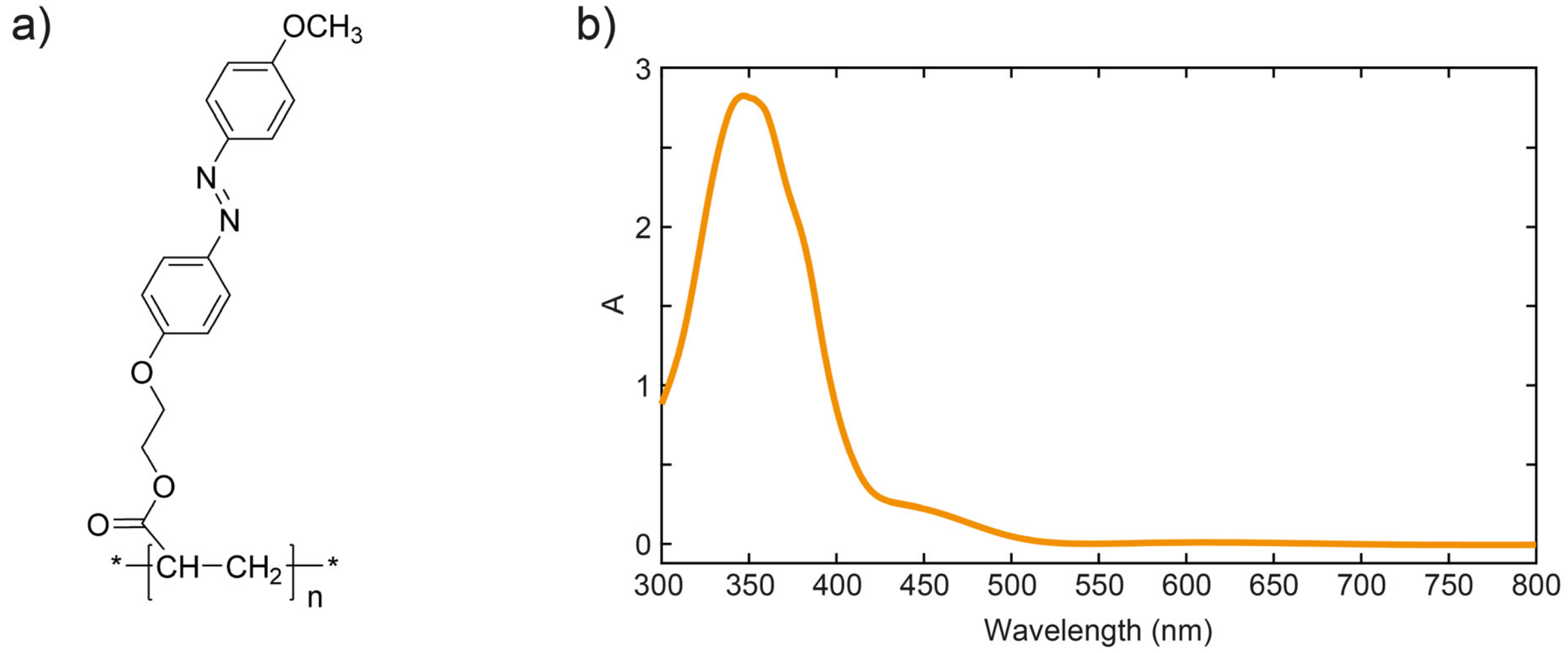

c)

| $M_w$ | $M_n$ | $M_w/M_n$ |
|---|---|---|
| 27000 Da | 21100 Da | 1.28 |

d)

| $T_g$ | $T_i$ |
|---|---|
| 67 °C | 113 °C |

**Figure S2. Physicochemical characterization of the azobenzene-functionalized acrylic polymer used in this work.** (a) Chemical structure of the acrylic polymer bearing photoresponsive azobenzene moieties as side chains of the polymer backbone. (b) UV-visible absorption spectrum of a transparent polymer thin film prepared by spin coating a solution of the polymer in 1,1,2,2-tetrachloroethane onto a 170 µm-thick microscope coverslip. (c) Molecular-weight characteristics determined by gel permeation chromatography (GPC): weight-average molecular weight, Mw = 27000 Da; number-average molecular weight, Mn = 21100 Da; and dispersity, $M_w/M_n$ = 1.28. (d) Thermal-transition temperatures measured by differential scanning calorimetry (DSC): glass-transition temperature Tg = 67°C and isotropization temperature Ti = 113°C. A liquid-crystalline phase develops upon annealing above Tg and becomes isotropic at Ti.

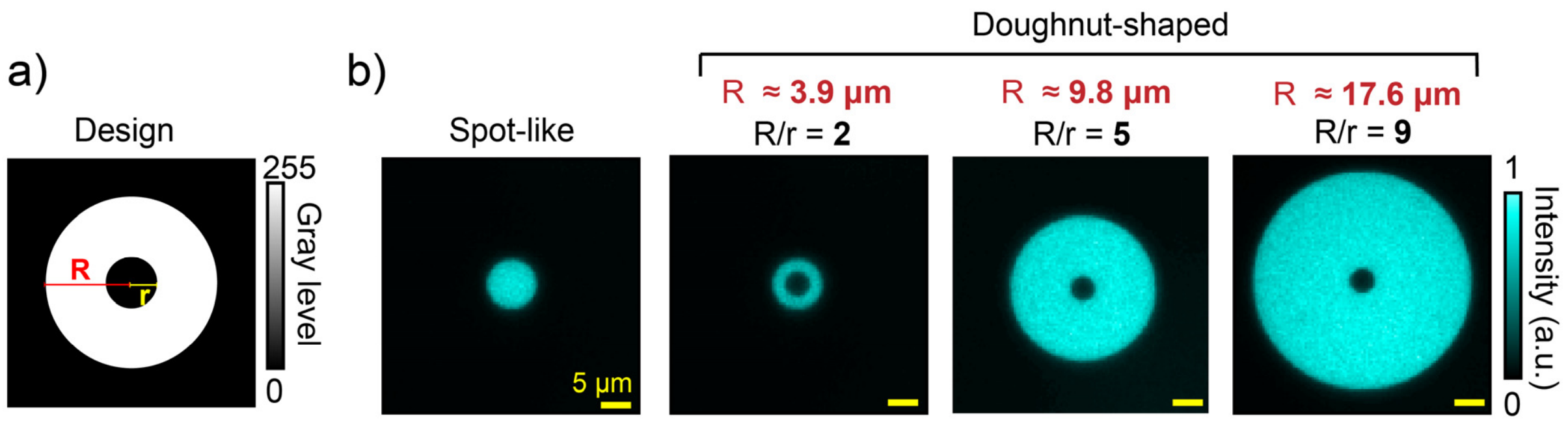

**Figure S3. Spot-like and doughnut-shaped holograms with different R/r values.** a) Grayscale design for a doughnut-shaped intensity pattern. *R* and *r* are the radii of the bright ring and the inner dark circle of the doughnut, respectively, both measured from the center of the doughnut pattern. b) Holographically generated patterns for increasing *R/r* ratios. The spot-like pattern has a radius of ~ 3.9 µm. The value of *r* ~ 1.95 µm was kept constant for all *R/r* ratios.

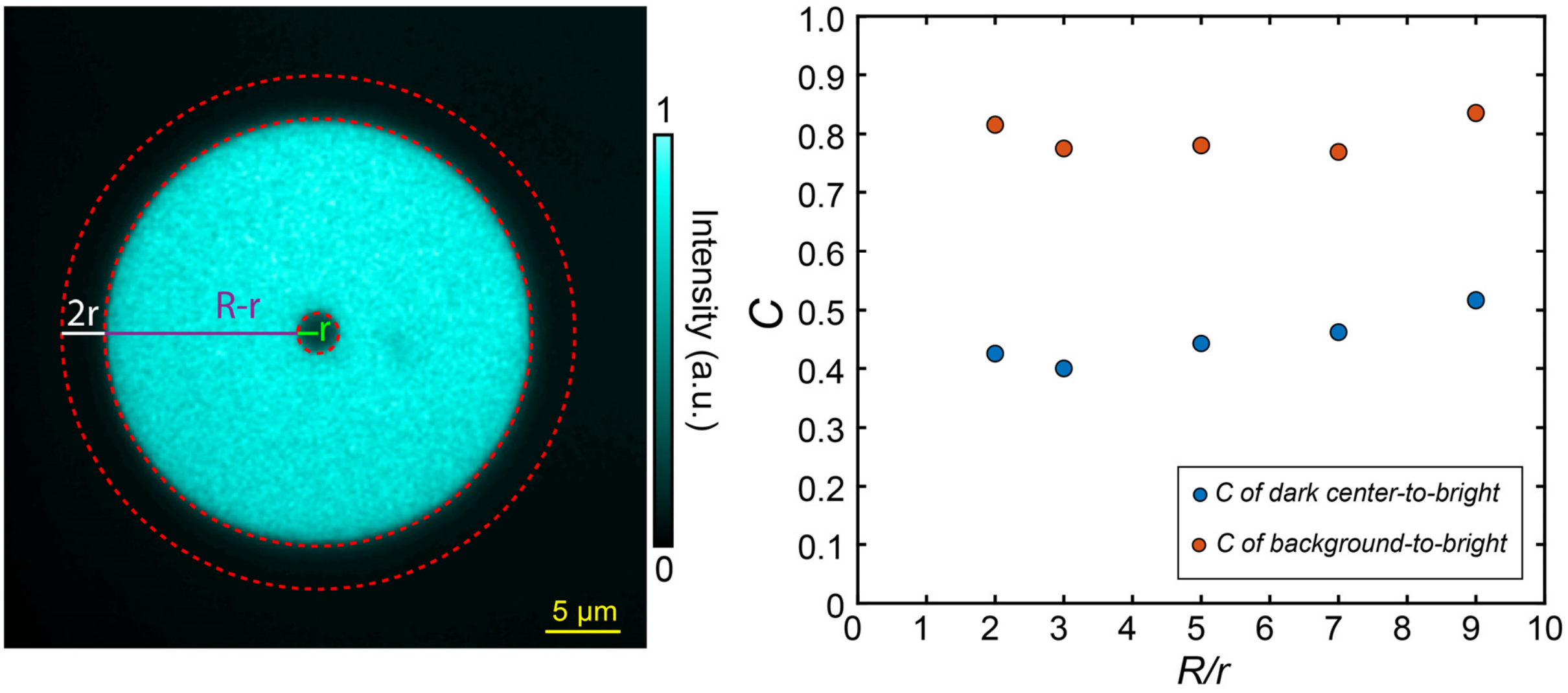

**Figure S4. Contrast of the isolated dark core and dark background relative to the bright ring for different *R/r* ratios.** The value of $r \sim 1.95$ μm was kept constant for all *R/r* ratios. The contrast was calculated using the standard relative contrast: $C = (\langle I_{bright} \rangle - \langle I_{dark} \rangle) / (\langle I_{dark} \rangle + \langle I_{bright} \rangle)$. Here, $I_{bright}$ is the average intensity within the bright ring, which has a radial size of R−r and an area of $\pi(R^2 - r^2)$. For the contrast between the dark center and the bright ring, $I_{dark}$ corresponds to the average intensity inside the central dark core, with radius r and area $\pi r^2$. For the contrast between the surrounding dark background and the bright ring, $I_{dark}$ is instead the average intensity within the dark annulus located at the outer perimeter of the bright ring. This annulus has a radial thickness of 2r and an area of $\pi[(R+2r)^2 - R^2]$. The transition boundaries between the dark core and bright ring, and between the bright ring and outer dark region, were defined as the radial positions corresponding to the half-height of the intensity profile, measured from the minimum (dark) intensity to the maximum (bright) intensity in each case. The average contrast *C* of dark center-to-bright ring = 0.45 ± 0.04 and contrast *C* background-to-bright ring = 0.80 ± 0.03. The uncertainty is the standard deviation.

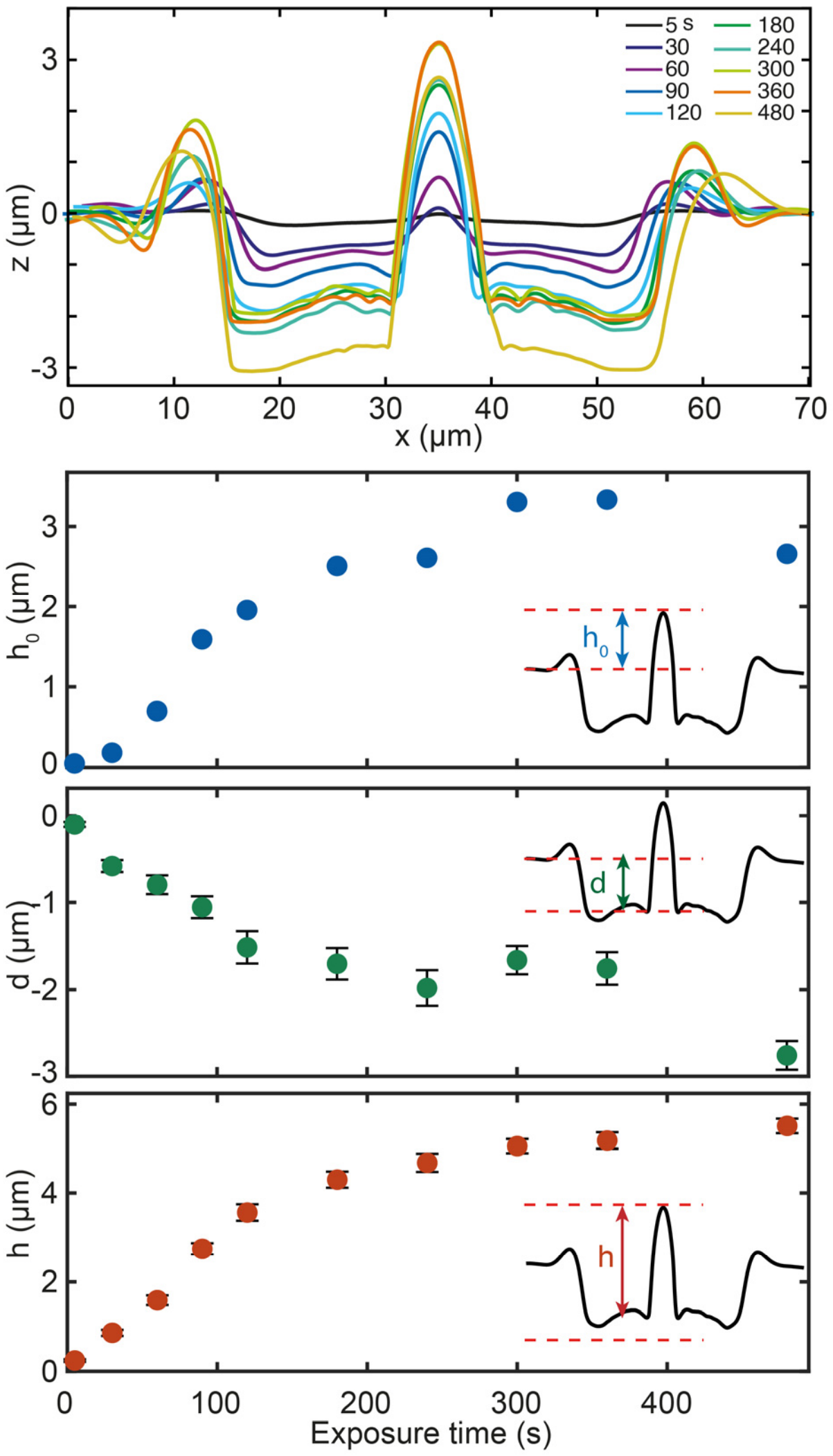

**Figure S5. The morphological evolution of an isolated protruding microstructure formation.** All micro-protrusions were produced with the same local hologram intensity (see Supplementary Table 1). The hologram pattern used in this experiment is the doughnut pattern with $R/r = 9$ and $r \sim 1.95$ μm. $h_0$, $d$, and $h$ are defined as the relative peak height to the non-exposed level, the depression depth relative to the non-exposed level, and the effective height of the micro-protrusion (peak-to-baseline), respectively.

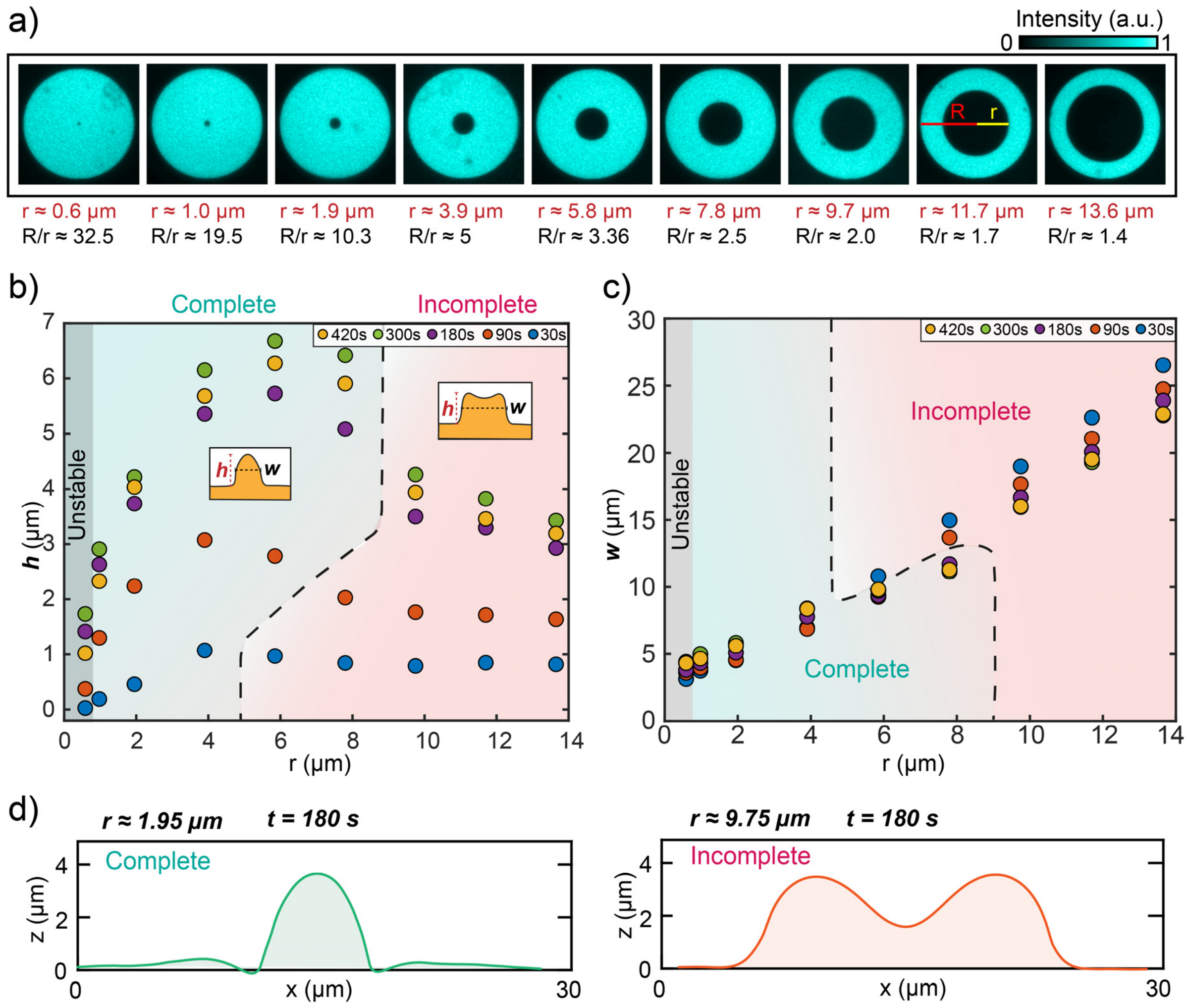

**Figure S6. Evolution of the shape and the stability of microprotrusions.** a) Doughnut-shaped intensity patterns were used to produce isolated protruding microstructures with different radii $r$ of the central dark region, while keeping the bright ring radius constant at $R \approx 19.5$ μm. b-c) Evolution of the height $h$ and the full-width-half-maximum $w$, respectively, of the isolated protruding microstructures as a function of $r$ for five exposure times. Each measurement of $h$ and $w$ refers to a single, individually analyzed micro-protrusion. The height $h$ corresponds to the maximum modulation relative to the adjacent depleted region, whereas the width $w$ is the FWHM of that same individual microrelief. All micro-protrusions were produced using the same local hologram intensity (see Supplementary Table 1). For very small $r$ (gray region) approaching the resolution limit of the projection optics, reduced hologram contrast was observed and unstable reliefs were obtained under our experimental conditions, with modulation comparable to the roughness of the depleted region; the dashed gray line separates a regime where a single, complete and well-developed protrusion forms (green area) from a regime where accumulation in the central region becomes incomplete and the topography is characterized by a double-peak feature (red area). d) AFM profiles illustrating the “complete” and “incomplete” microrelief from two different $r$ made at the same exposure time.

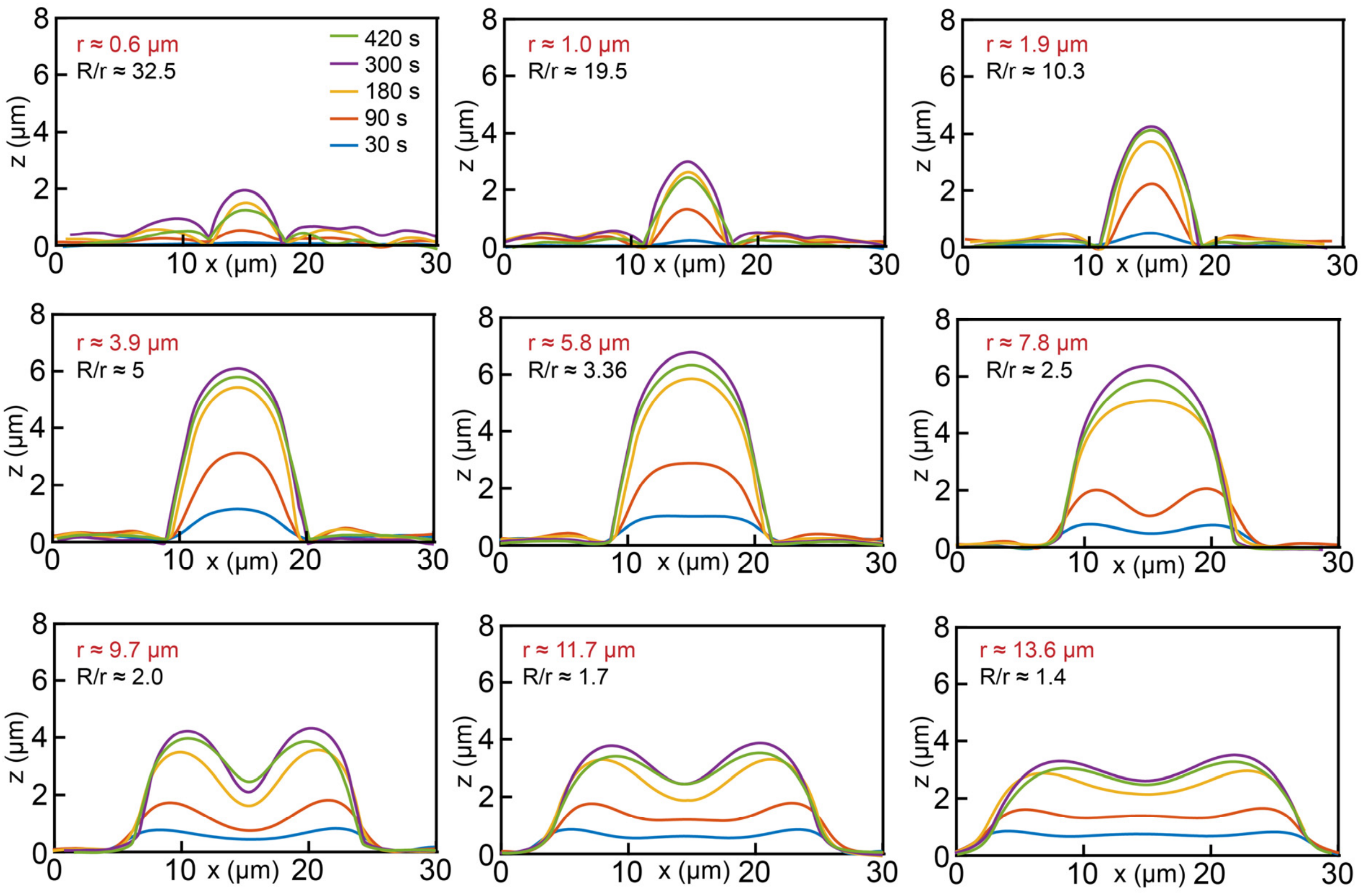

**Figure S7. Evolution of the profiles of isolated protruding microstructure for different *r* values for constant *R* ≈ 19.5 µm) and exposure times.** The profiles of the isolated protruding microstructure correspond to the data shown in Figure 2f and Figure S6. The morphological profiles were extracted from AFM images using a single line scan passing through the center of each protruding microreliefs.

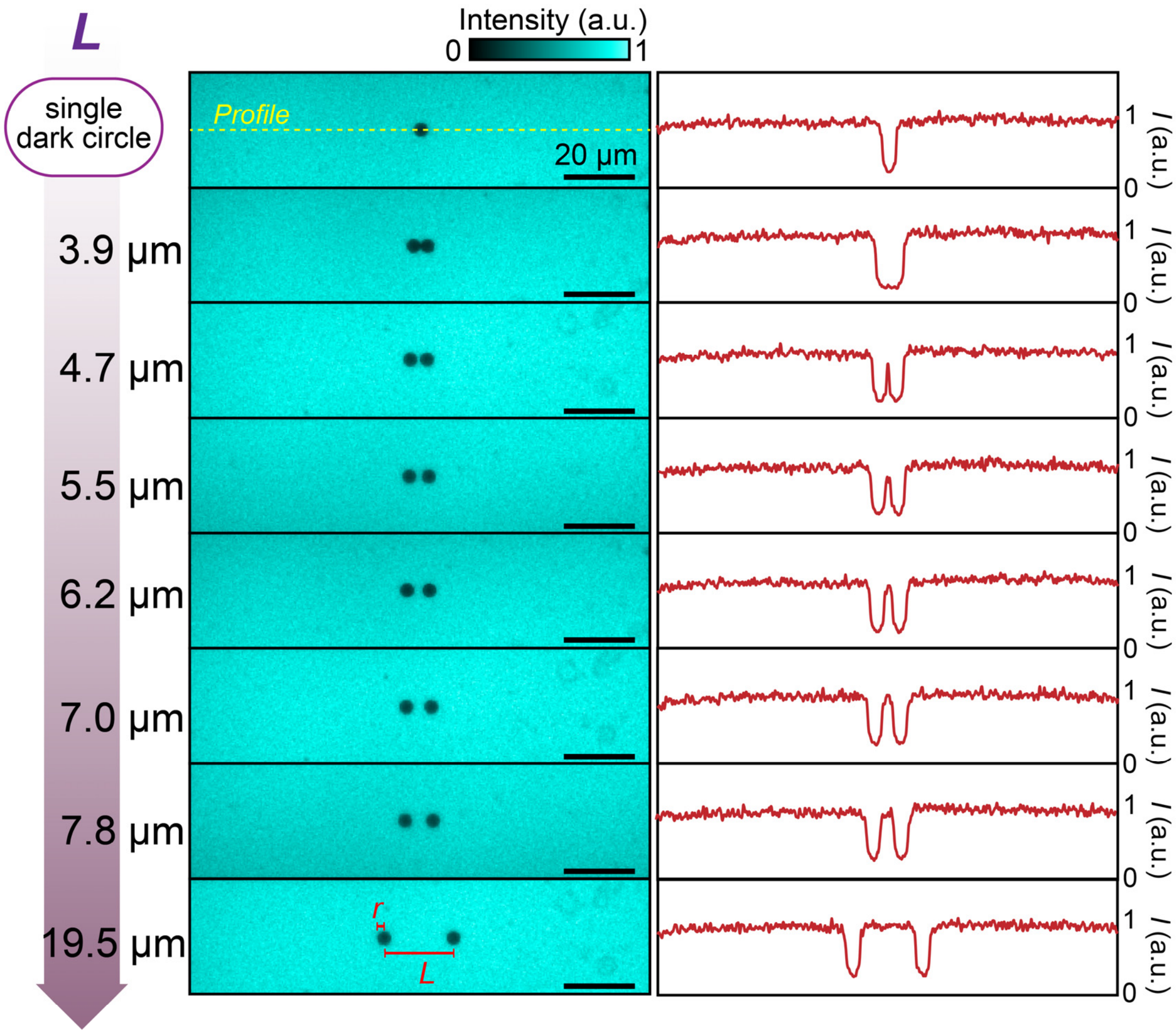

**Figure S8. Holographic intensity patterns used for resolution study and corresponding intensity profiles.** The radius $r$ of the dark circles was fixed at ~1.95 µm, while center-to-center distance $L$ between two dark circles was varied from $L = 0$ (single dark circle) to $L \approx 58.5$ µm. The hologram images shown here represent selected examples illustrating the gradual separation of the dark cores pair.

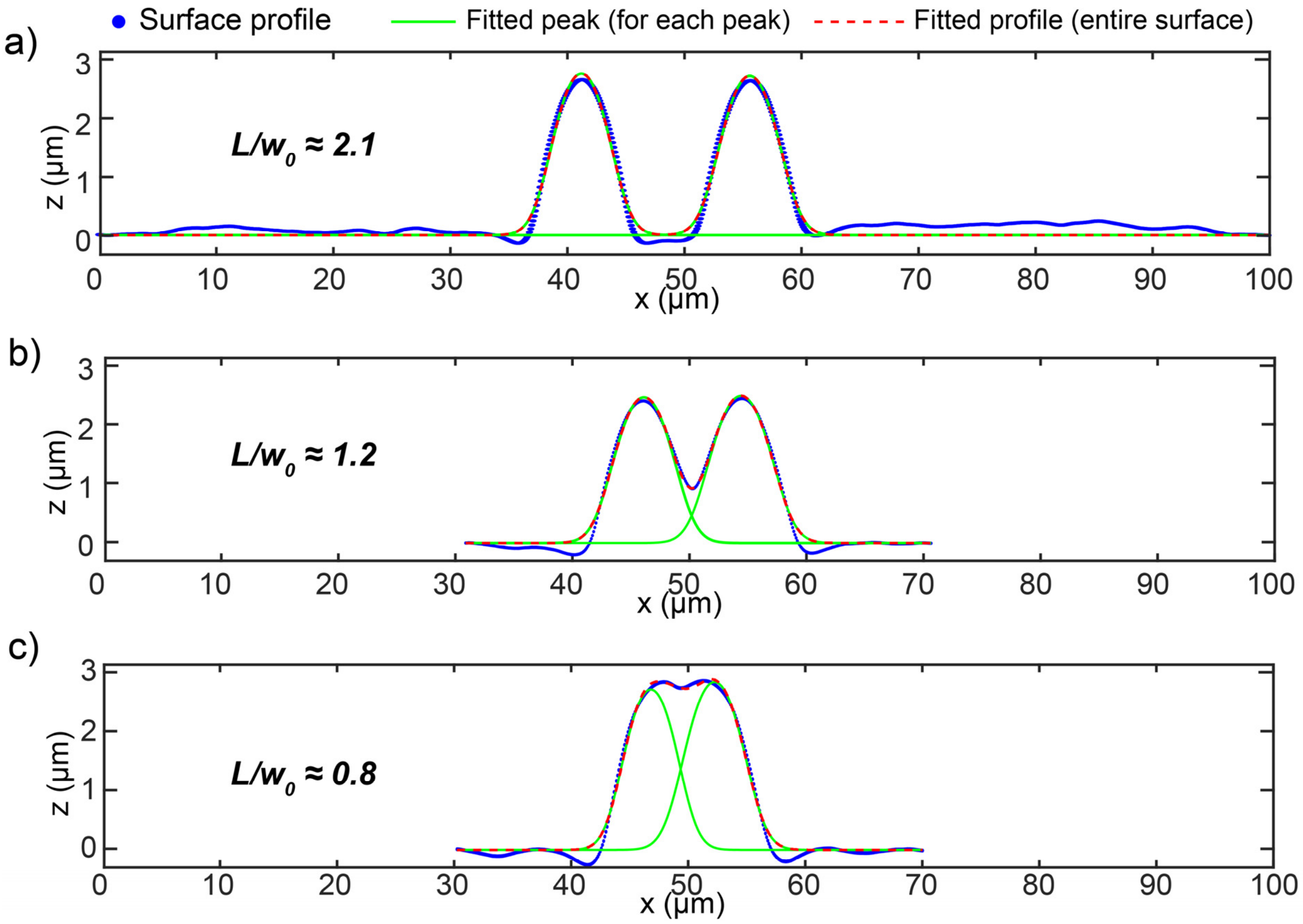

**Figure S9. Two-peak fitting used to determine peak separation for different *L/w₀* ratios**. *L*: center-to-center distance of two adjacent dark circles in the holographic pattern; $w_0$: full-width-half-maximum of the microrelief produced with single dark core (see Figure S8.). a) $L/w_0 \approx 2.1$, b) $L/w_0 \approx 1.2$ and c) $L/w_0 \approx 0.8$ exemplify the case of the pair of protruding microreliefs when it is “well-spaced” as they have base separation, “well-resolved” as they are separated at half-maximum of modulation, and “not resolved” as they exhibit non-separated feature at half-maximum of modulation, respectively. Fitting was performed in MATLAB® using the *peakfit.m* script,[54] employing a logistic distribution and a quadratic baseline (see Experimental section in the main text).

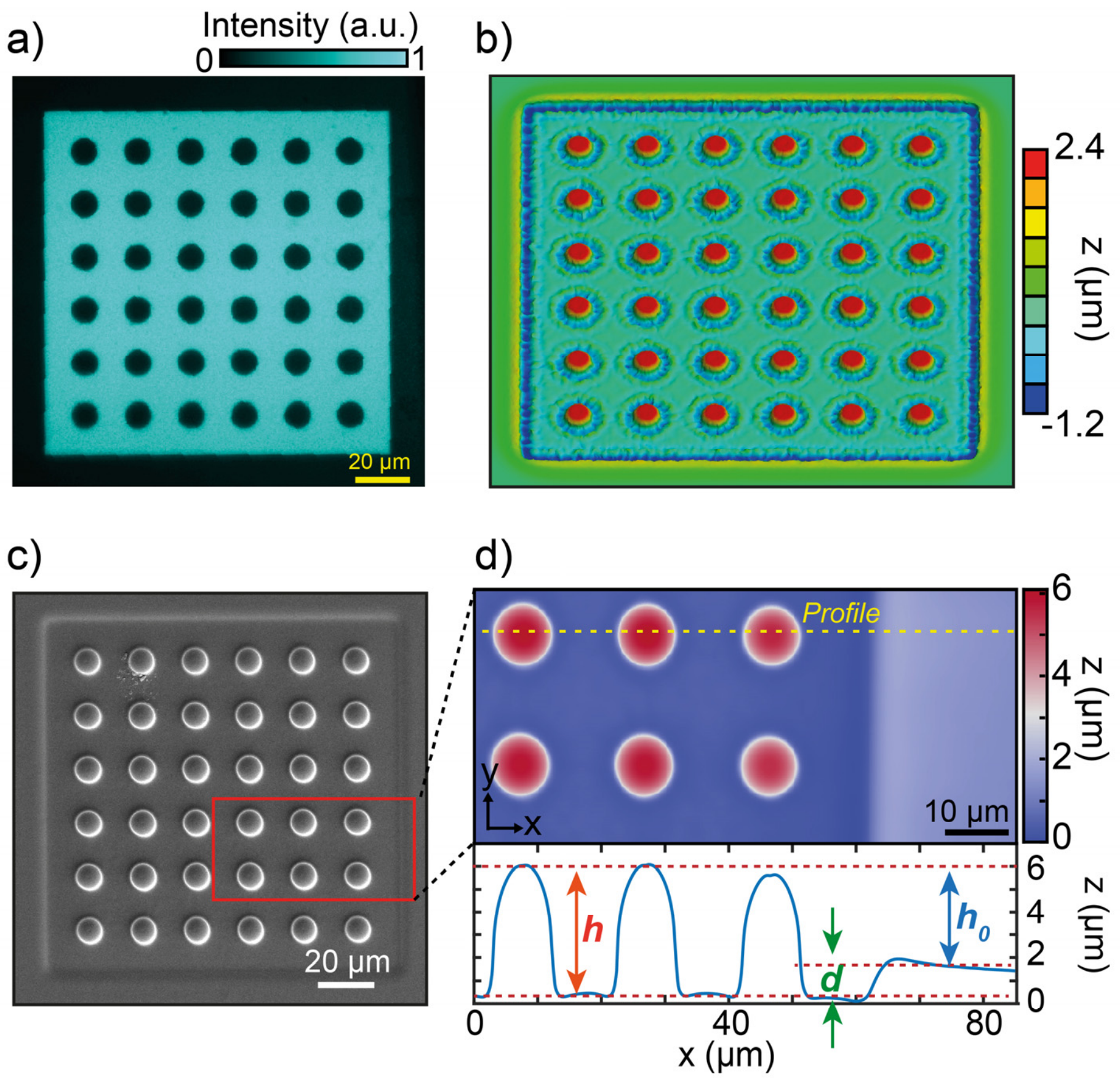

**Figure S10. Transition between illuminated and the non-illuminated areas during fabrication.** a) Holographic intensity pattern. Square area of 127 × 127 µm$^2$. b) The simulation prediction of the Viscoplastic photoalignment model of the produced surface from the exposure of the hologram shown in (a). c) The surface image of the resulting array of isolated protruding microstructures on the surface taken using scanning electron microscope. The holographic pattern, the modelling and the fabricated surface shown here are the same data set shown in Figure 3e, 3f and 3h, respectively, with additional view on the boundary and non-exposed area around the patterned surface. d) AFM image and its topographic profile (yellow dashed line) acquired at the boundary between the illuminated and non-illuminated regions. The profile shows the transition from the base of the isolated protruding microstructure array to the non-illuminated surface. The base level of the isolated protruding microstructure array lies below that of the non-illuminated area with relative height $d \approx 1.2$ µm, while the isolated protruding microstructures rise above the non-illuminated surface with a relative height of $h_0 \approx 4.6$ µm and an effective height of $h$ relative to its baseline level.

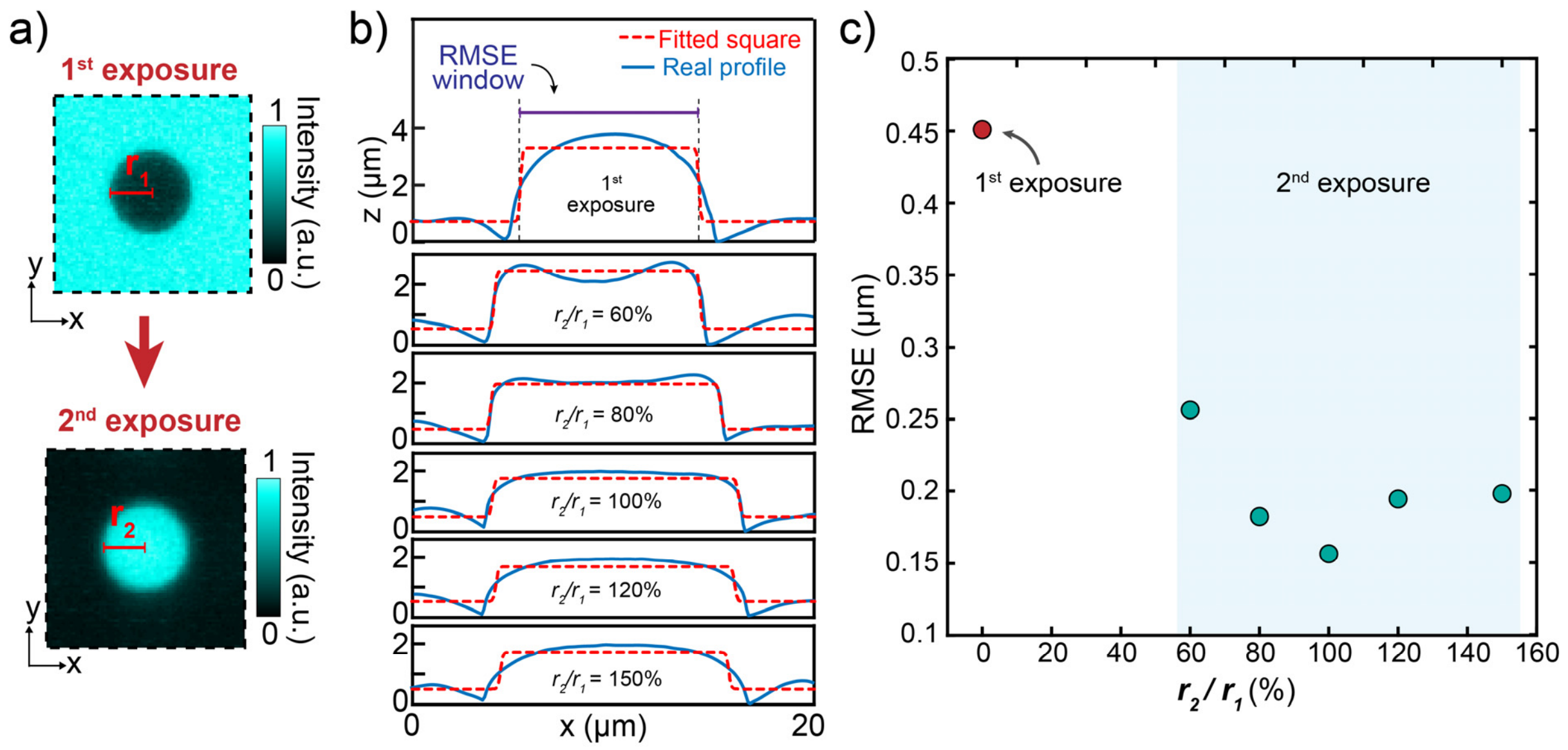

**Figure S11. Optimization of the 2-step exposure strategy.** a) Holographic patterns used for the 1st and 2nd exposures. b) AFM topographic profiles (traced along the microstructure diameters) after the 1st exposure (top panel) and after the 2nd exposure for increasing $r_2/r_1$ ratios obtained by keeping fixed the $r_1 \sim 4.9$ µm and increasing $r_2$ values. The time exposure was kept constants for all $r_2/r_1$ variations. c) RMSE values as a function of $r_2/r_1$ ratio measured from the isolated protruding microstructure topographic profile relative to the fitted square profile in the span of *RMSE window* (entire span of top plateau of each fitting as indicated in b). Each value corresponds to an average of 5 isolated protruding microstructures. The RMSE window was chosen to focus on how the double-exposure can form flatten top plateau. The fitting was done by first, estimating initial guesses by detecting individual plateau regions and extracting their average height, width, and periodic spacing, which provided physically meaningful starting values for the global fit. The micropillar array was then modeled using a uniform rectangular-pulse representation, where the AFM height profile ($z(x)$) is expressed as a baseline ($b$) offset plus the contribution of $N$ identical pulses with common height ($h$), width ($w$), period ($P$), and lateral shift ($\Delta S$). The model used was pulse square: $z(x) = b + h\sum_{k=0}^{N-1} T[\frac{x-(\Delta S+kP)}{w}]$. The smoothed rectangular function is defined as: $T(z) = 0.5(\tanh\big(E(0.5 - |z|)\big) + 1)$, where $E$ is a parameter to define the edge sharpness. The tanh-based smoothing preserves the rectangular shape ($E$ was set to 100 in all fitting, that gives a slope edge width $\approx 0.02w$ in each side) while ensuring numerical stability during nonlinear least-squares optimization by avoiding absolute rectangle edge. $r_2/r_1 = 1$ (100%) was selected to fabricate quasi-sharp microcylinders with the double-step exposure (minimum RMSE value).

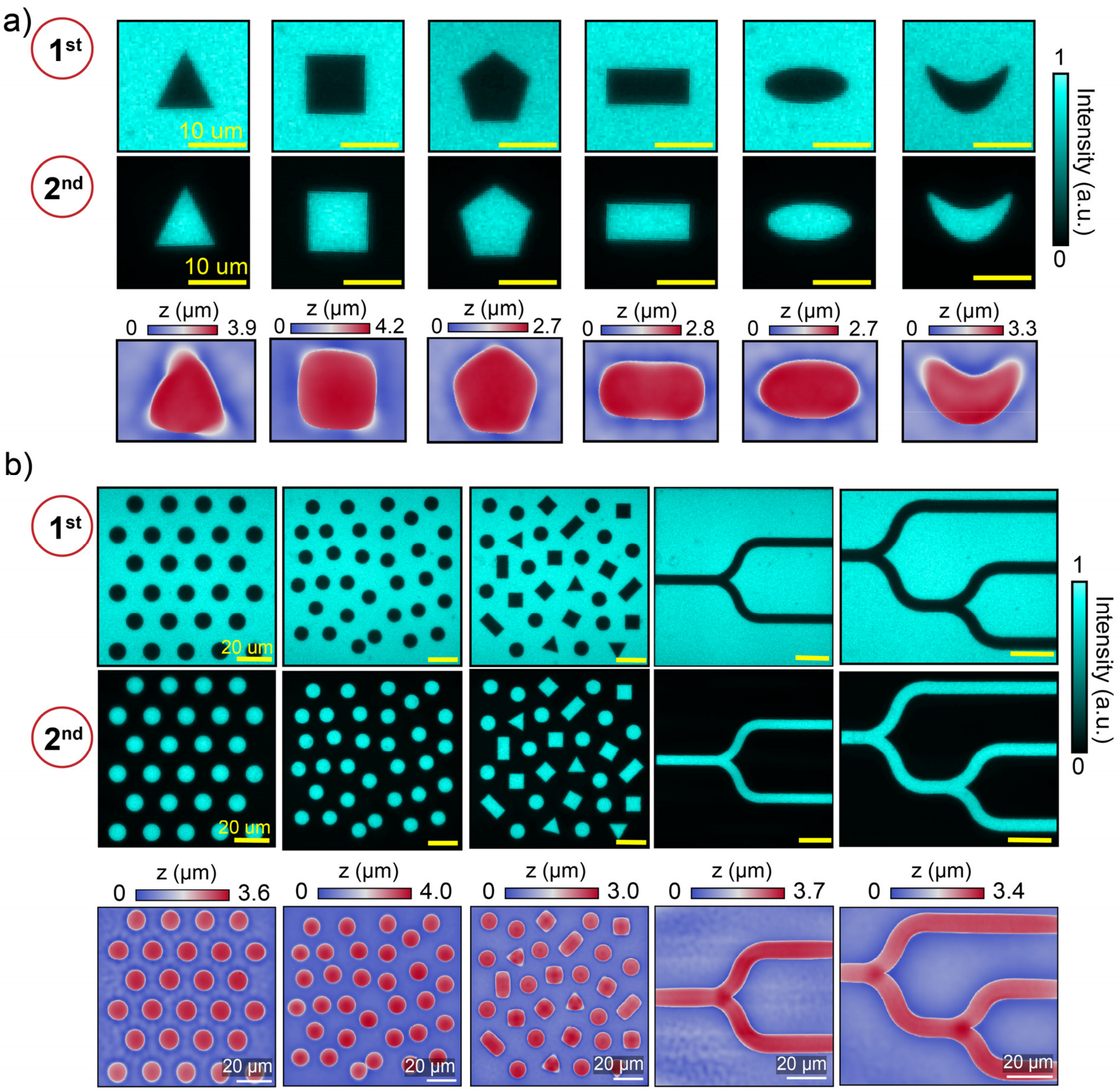

**Figure S12. Holographic intensity patterns for 2-step exposure with various shapes, array configurations, and free-form patterns.** a) Individual shape programmability in the 2-step exposure for creating micro-posts with diverse lateral shapes. b) The array arrangement and free-form design programmability for creating various microstructural array and complex micro-patterns.

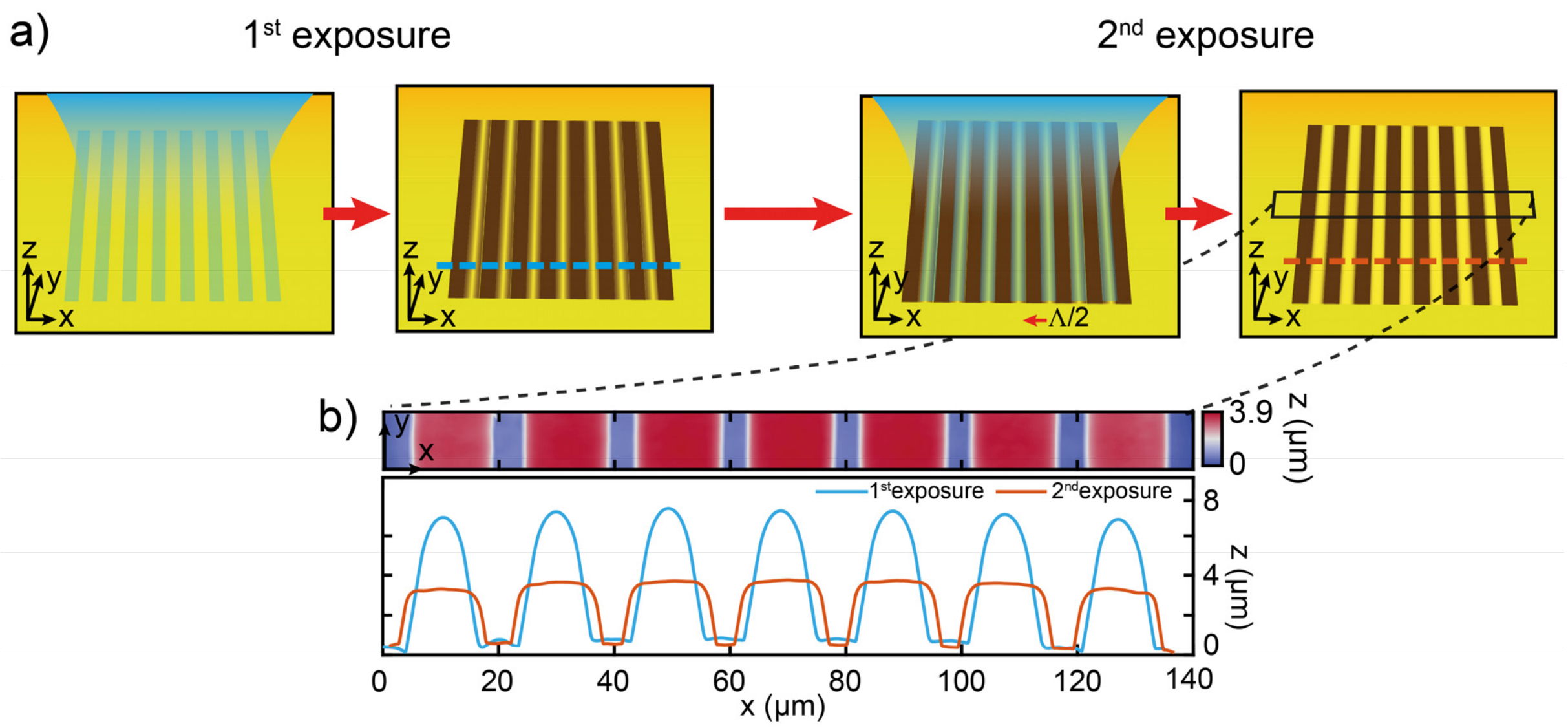

**Figure S13. 2-step strategy to produce 1D periodic continuous morphology.** a) Schematic of the consecutive 2-step exposure using periodic binary-line intensity pattern (line width and period ~ 9.75 μm). b) AFM topographic profile fabricated from 1st and 2nd exposure.

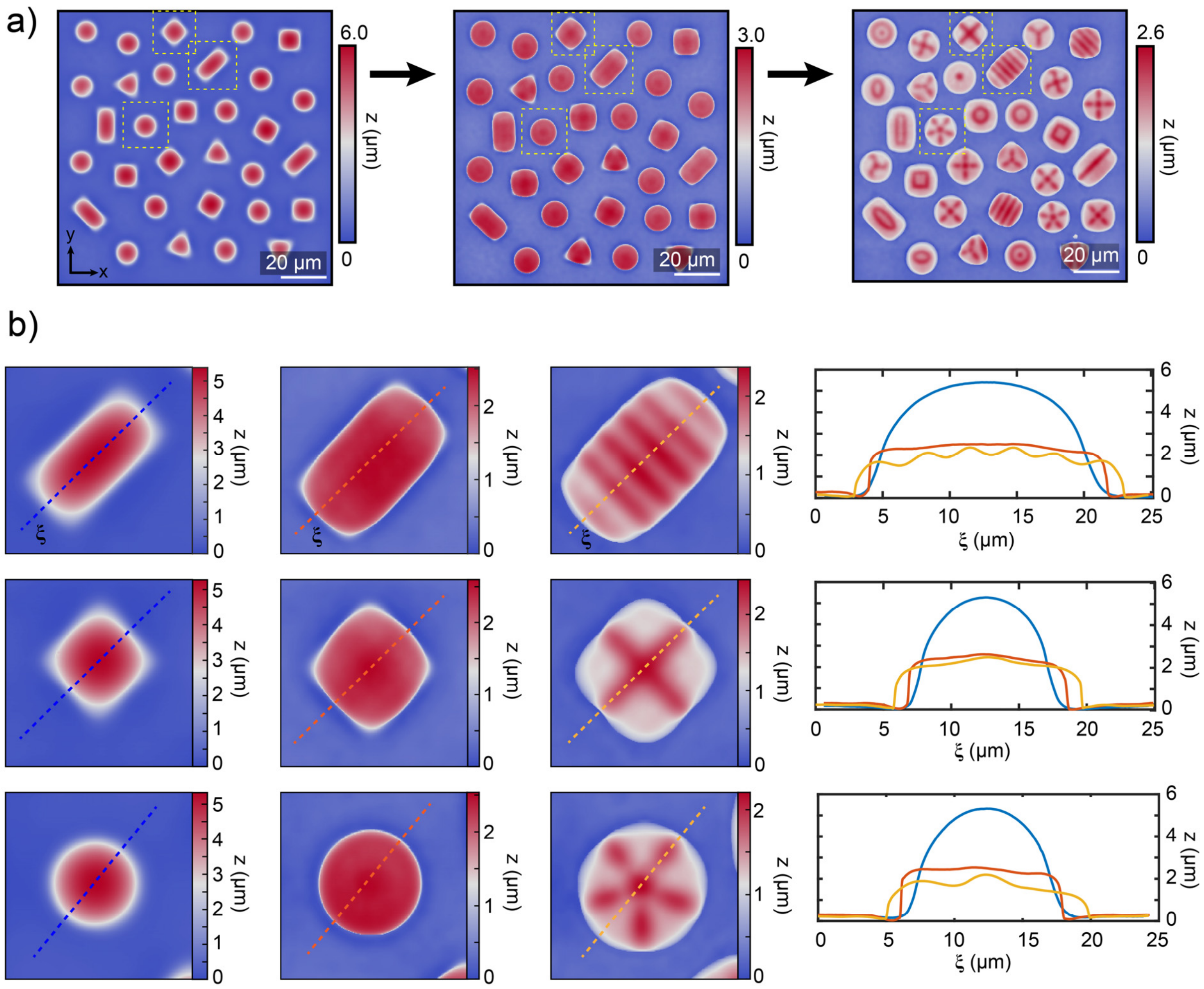

**Figure S14. The morphological evolution of the 3-step exposure strategy.** a) The surfaces produced from each stage of exposure, from left to right are 1st, 2nd, and 3rd exposure, respectively. The surface studied here is the surface shown and discussed in Figure 6c–e. b) The detailed morphological analysis of three selected individual microstructures. The profiles reported here were taken from a single pixel line crossing the apex of the microstructures.

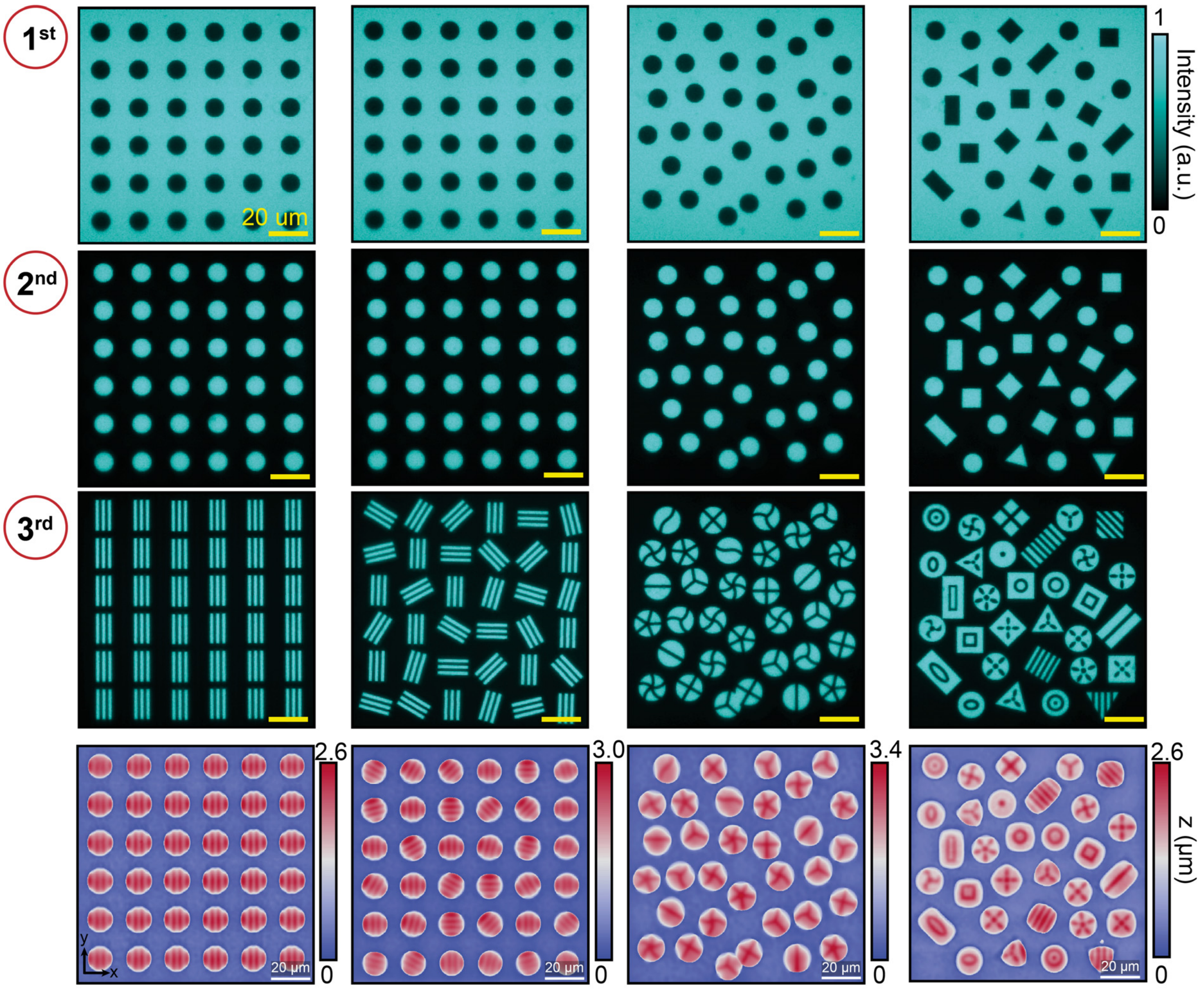

**Figure S15**. **Various morphological shapes and arrays produced using 3-step exposure approach.** Holographic intensity patterns used for 3-step exposure with different combinations of spatial designs of the primary structure array (1st and 2nd exposure) and the secondary structure (3rd exposure) and their corresponding produced surfaces. The rightest holograms set and the corresponding fabricated microstructures array are the same data shown in Figure 6.

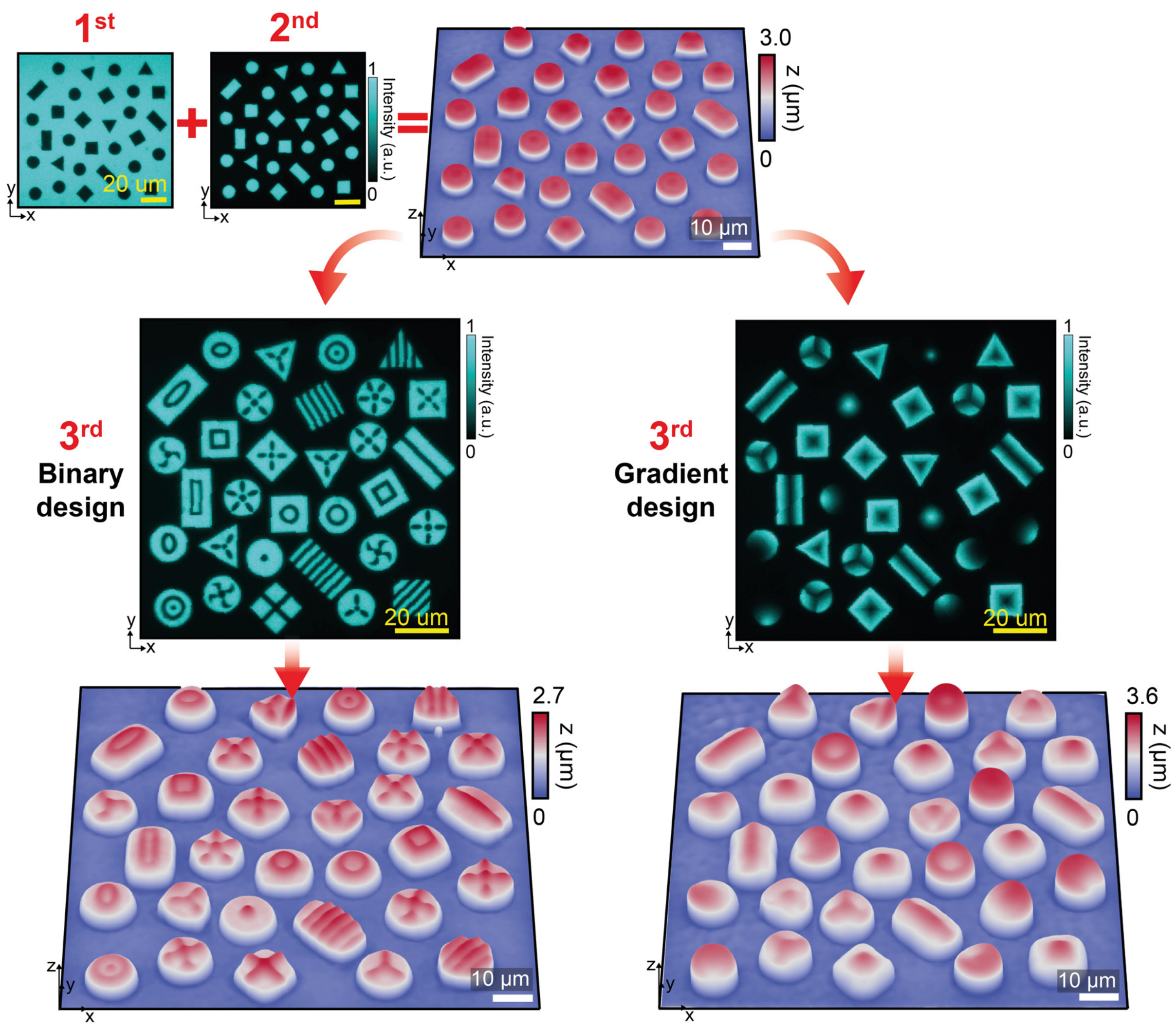

**Figure S16. Alternative fabrication paths for complex 3D microarchitectures using multi-step exposure.** The chart compares two distinct 3D microarchitectures that can be achieved from the same design. In the first design (left flow chart), a binary intensity pattern was used for the 3$^{rd}$ exposure, yielding microstructures with a 2-level hierarchy. In the second design (right flow chart), a gradient intensity pattern was used for the 3$^{rd}$ exposure to produce smoothly slanted features across the microstructures, thereby affecting the overall architectural design. The 1$^{st}$ and 2$^{nd}$ exposure correspond to the conditions shown in Figure 5c in the main text. The final microstructure obtained via binary approach corresponds to Figure 6c. The final microstructure obtained via the gradient approach was fabricated after 30 s for the 3$^{rd}$ exposure (see Supplementary Table 1 for details).

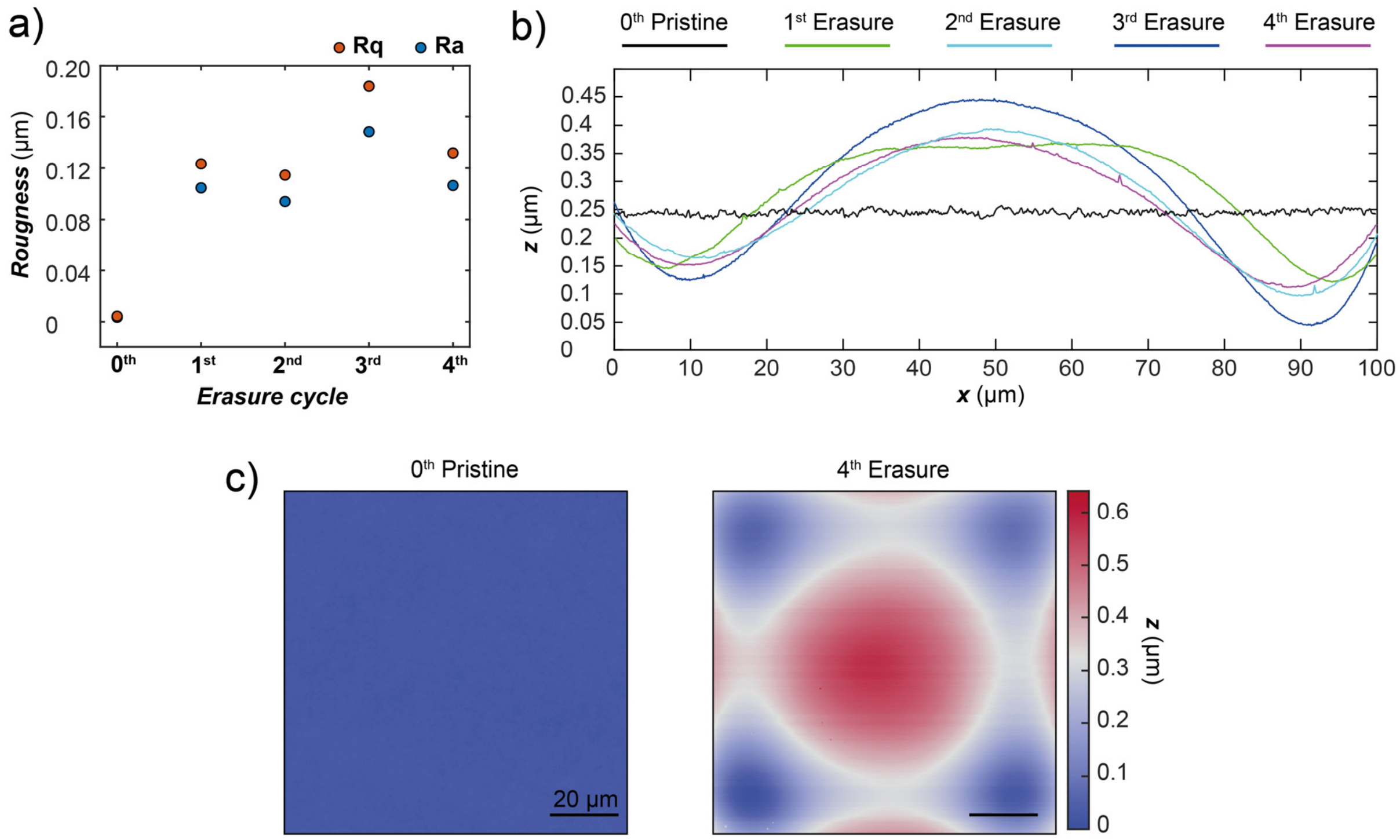

**Figure S17. The roughness of the erased surface in the write-erase-write strategy. a)** The roughness of the surface as a function of erasure cycle. The root-mean-square roughness (*Rq*) was derived from the deviation of the surface to its own average level using the standard root-mean-square method. The average roughness (*Ra*) was derived from the average of the absolute deviation of the surface to its own average level. Both roughness parameters were calculated using the entire surface. **b)** the 1D profiles of each erasure cycle (including the pristine surface) taken at the middle part across the horizontal direction on the corresponding surface. **c)** The top-down view of the pristine surface and the surface after the $4^{th}$ erasure.

The maximum number of rewriting cycles achievable with the present all-optical patterning configuration has not been explicitly determined. Repeated operation is expected to be ultimately limited by photobleaching, particularly under cumulative exposure to the higher-intensity 405 nm illumination used for erasure, which may progressively reduce the photoresponsive capability of the azopolymer. In our previous study on shallow surface-relief gratings, 15 cycles were achieved without substantial degradation in pattern quality or response speed [Ref. XX]. These results support the potential for repeated optical reconfiguration, although they do not establish the cycling endurance of the higher-modulation structures investigated here. The accessible number of cycles is expected to depend on the illumination conditions and the cumulative exposure required for writing and erasure.

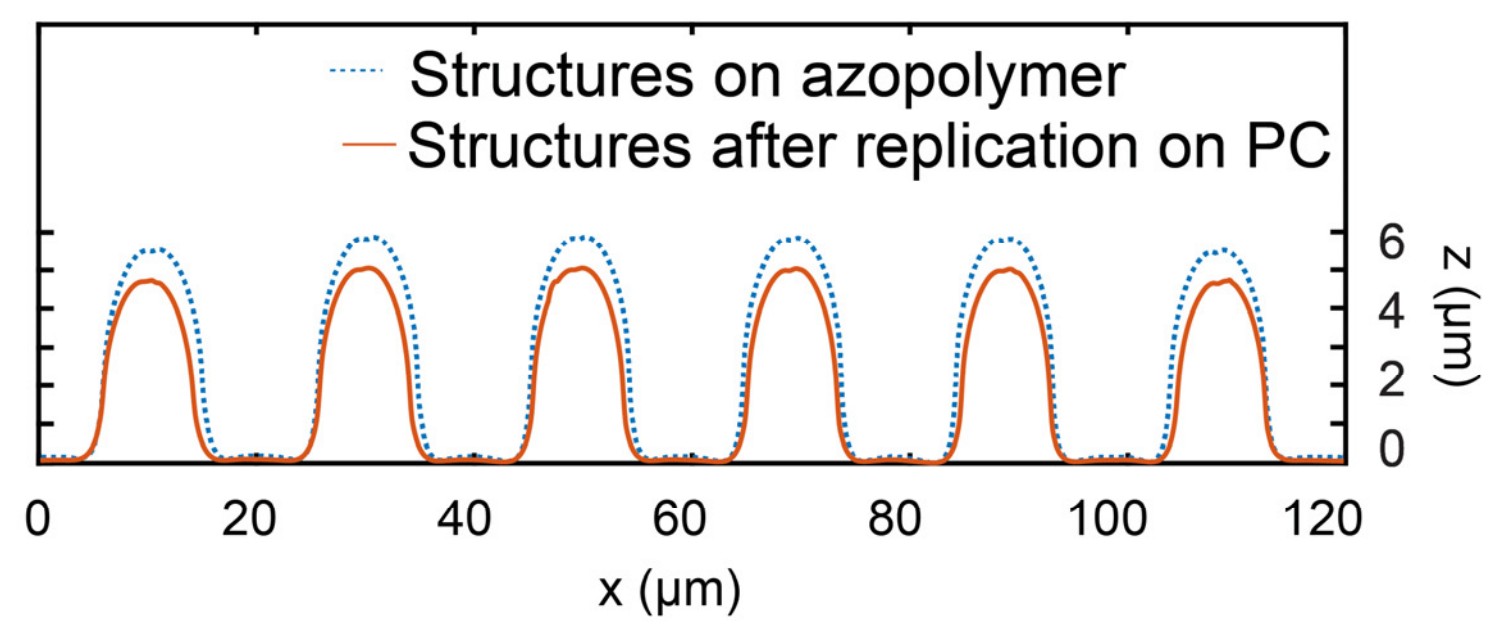

**Figure S18. The size reduction of the microstructure after replication.** The microprotrusions array shown and discussed in Figure 3g – h was used for quantification of the size reduction of the individual microstructure after the replication procedure. 36 microprotrusions were used to get the average of the initial height from master azopolymer microprotrusions array and the average of the height post-replication on the polycarbonate. The calculation yields approximately 12.8% height reduction.

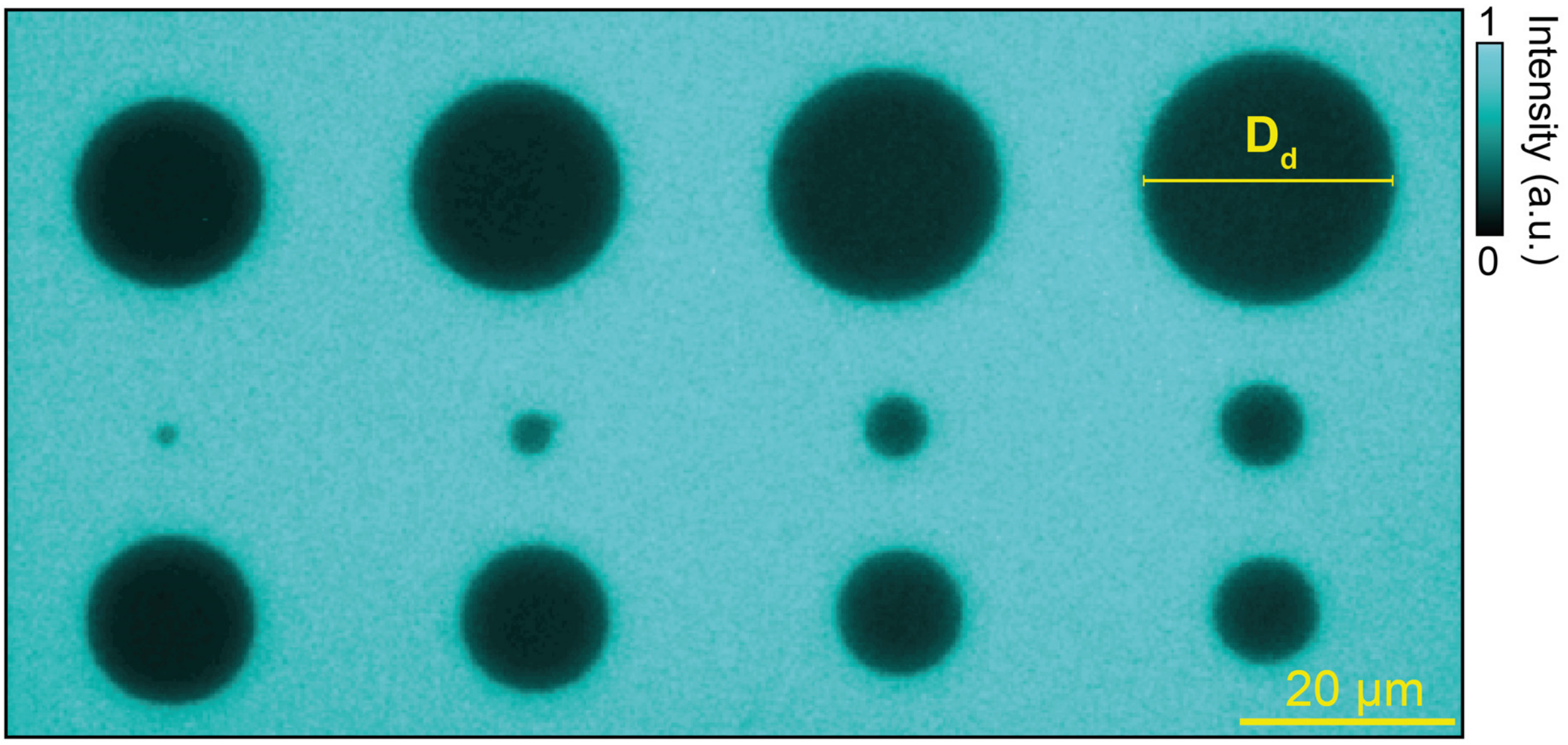

**Figure S19.** Hologram design of the dark circles to produce microlens with varying diameters ($D_d$) of approximately 3.9, 5.85, 7.80, 9.75, 11.7, 13.65, 15.60, 17.55, 19.5, 21.45, and 23.40 μm, ordered from the smallest to the biggest, respectively.

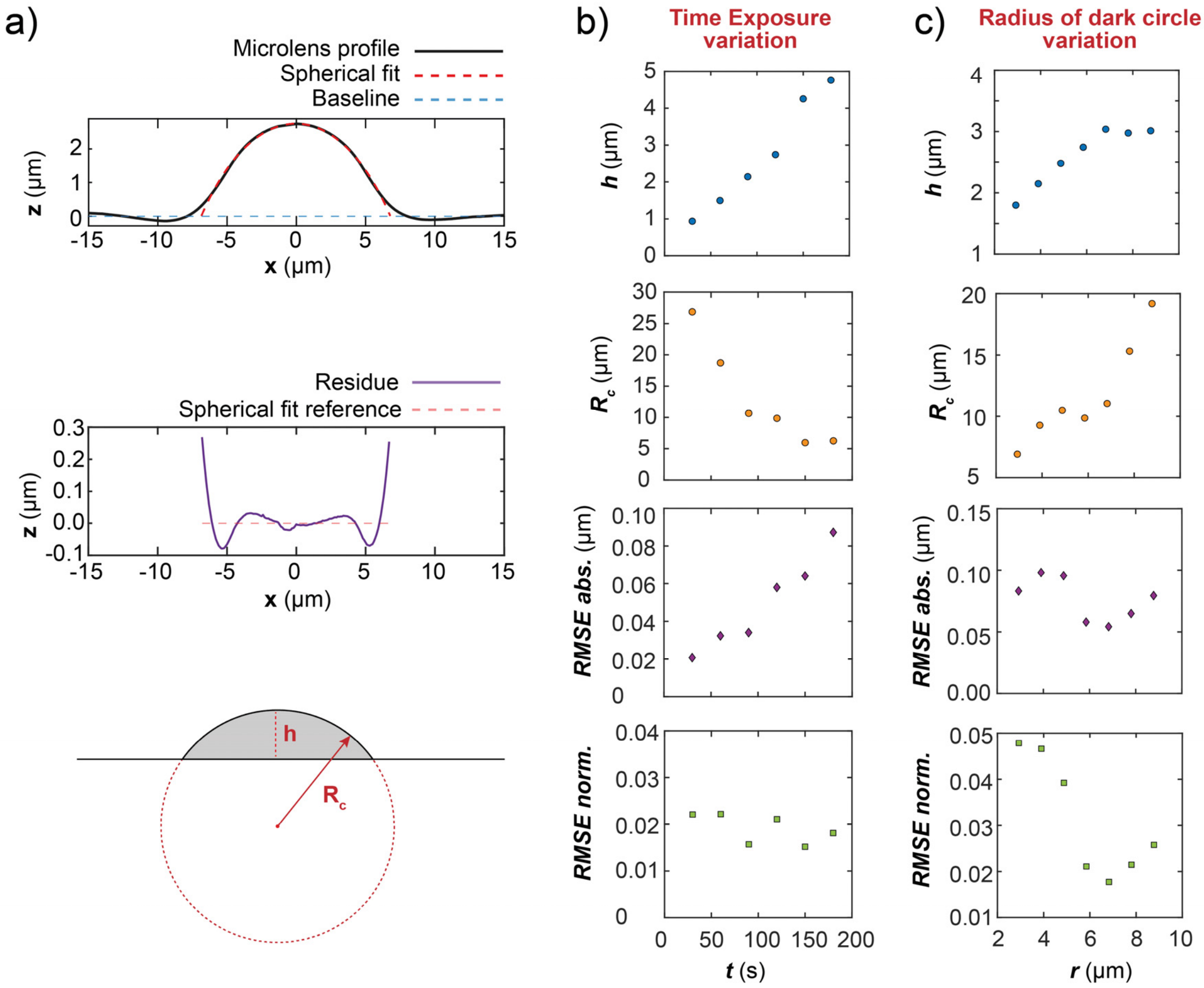

**Figure S20. The morphological study of azopolymer master microlens using spherical fit. a)** The spherical fit procedure to approximate the fabricated microlens with sphere cap geometry. The spherical fit of the microlens used geometrical spherical definition $(x-a)^2 + (y-b)^2 = R^2$ to describe 1D ideal spherical cap where ($a$, $b$) and R are the center in the cartesian coordinate and the radius curvature of the ideal sphere, respectively. In order to find the best spherical fit that represents the micro-bump produced in this study, an iterative of nonlinear least-squares procedure was performed using the equation $\min_{a,b,R} \sum_i (\sqrt{(x_i-a)^2+(y_i-b)^2} - R)^2$ which practically optimizing the data point of the experimental 1D microlens profile from AFM to the 1D ideal spherical profile. The fraction of the experimental 1D microlens profile varied in the iteration procedure, ranging from upper-half to total height of the microlens profile. **b)** and **c)** summarize the quantitative data of azopolymer microlens effective height ($h$), approximated radius curvature ($R_c$), root-mean-square-error (*RMSE*) both in absolute and normalized terms, for time exposure (***t***) and radius of dark circle in hologram design (***r***), respectively. The absolute RMSE (*RMSE abs.*) was derived directly from the microlens profile data point with the spherical fit curve. Normalized RMSE (*RMSE norm.*) was calculated using equation *RMSE abs.* / $h$. The time exposure variation was conducted by setting the $r$ = 5.9 µm. The $r$ variation was done by keeping the time exposure constant at 120 s.

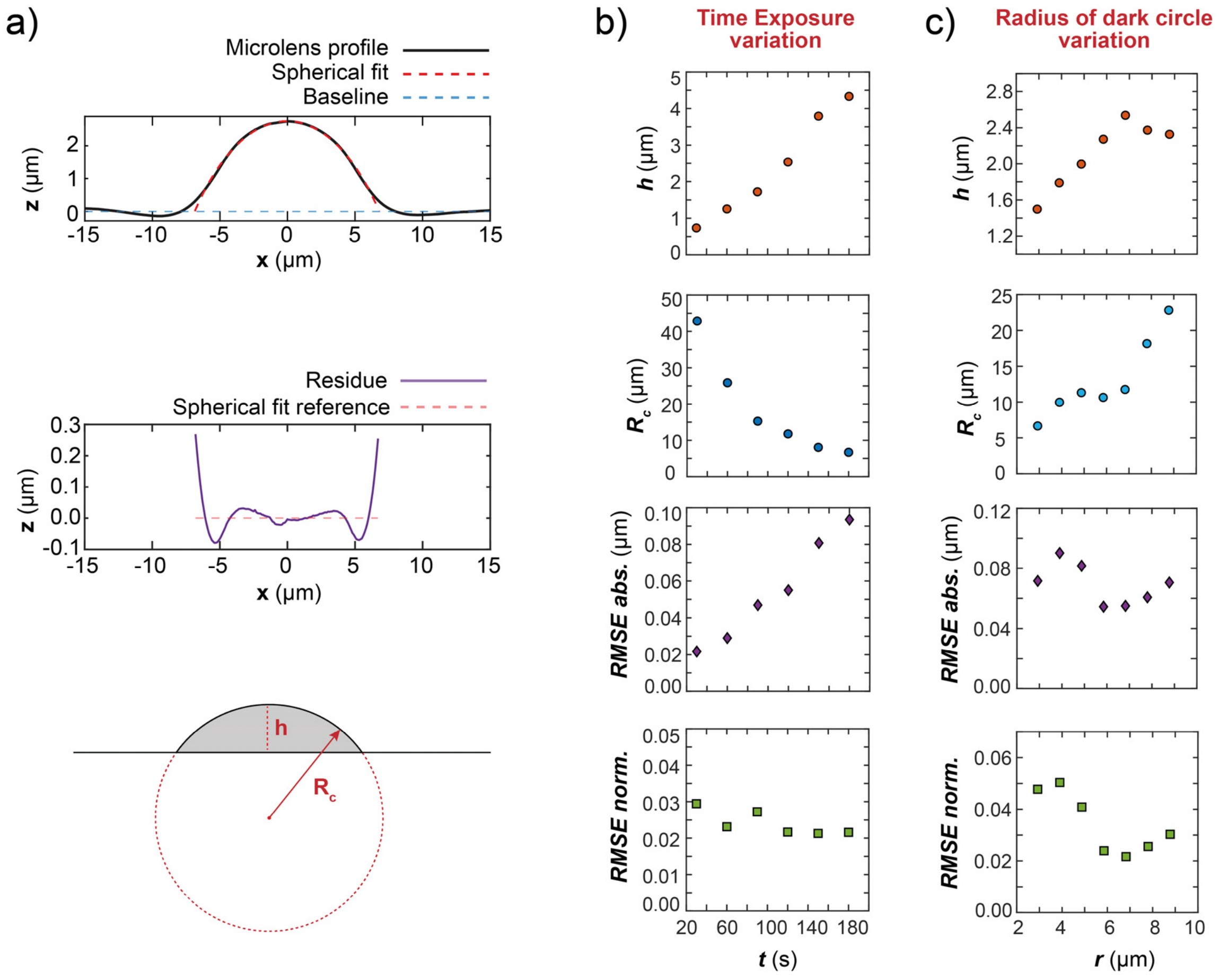

**Figure S21. The morphological study of replica microlens (on polycarbonate) profiles using spherical fit. a)** The spherical fit procedure to approximate the replicated microlens with sphere cap geometry. The spherical fit of the microlens used geometrical spherical definition $(x-a)^2+(y-b)^2=R^2$ to describe 1D ideal spherical cap where $(a, b)$ and R are the center in the cartesian coordinate and the radius curvature of the ideal sphere, respectively. In order to find the best spherical fit that represents the micro-bump produced in this study, an iterative of nonlinear least-squares procedure was performed using the equation $\min_{a,b,R} \sum_i (\sqrt{(x_i-a)^2+(y_i-b)^2}-R)^2$ which practically optimizing the data point of the experimental 1D microlens profile from AFM to the 1D ideal spherical profile. The fraction of the experimental 1D microlens profile varied in the iteration procedure, ranging from upper-half to total height of the microlens profile. **b)** and **c)** summarize the quantitative data of microlens effective height (*h*), approximated radius curvature ($R_c$), root-mean-square-error (*RMSE*) both in absolute and normalized terms, for time exposure (*t*) and radius of dark circle in hologram design (***r***), respectively. The absolute RMSE (*RMSE abs.*) was derived directly from the microlens profile data point with the spherical fit curve. Normalized RMSE (*RMSE norm.*) was calculated using equation *RMSE abs.* / *h*. The time exposure variation was conducted by setting the *r* = 6.8 µm. The *r* variation was done by keeping the time exposure constant at 120 s.

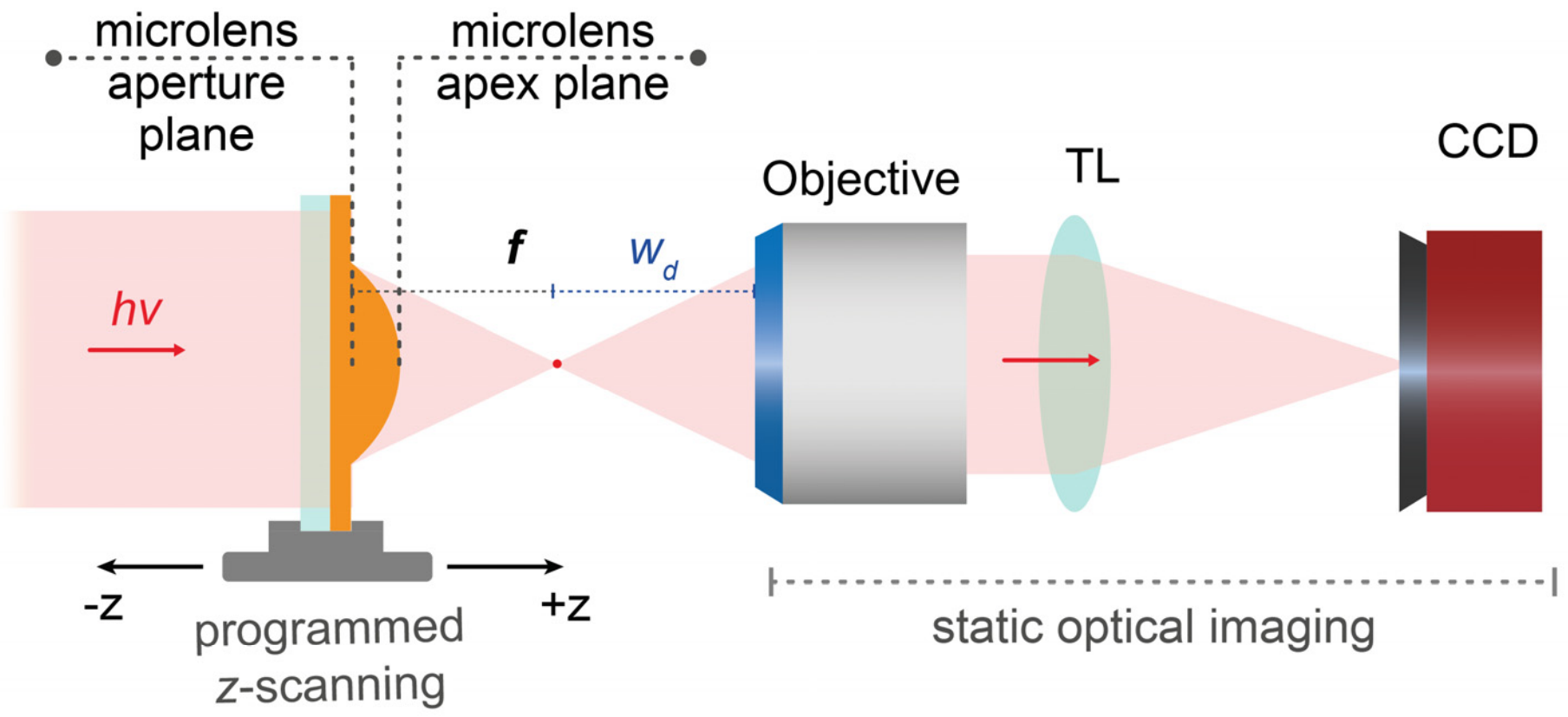

**Figure S22. The optical setup for point-spread-function (PSF) study of the microlens.** The static optical imaging system consists of an objective with working distance of $W_d$, tube lens (TL), and a charge-coupled device (CCD). The three-dimensional PSF was recorded by first positioning the image plane of the imaging system approximately at the aperture plane of the microlens. Then, the motorized z-stage and CCD were synchronized to move of a ~0.1 µm step further away relatively from the objective (to -z direction) and then to capture the image, respectively. With a proper number of steps, a 3D PSF produced by the microlens can be reconstructed from these stacked images.

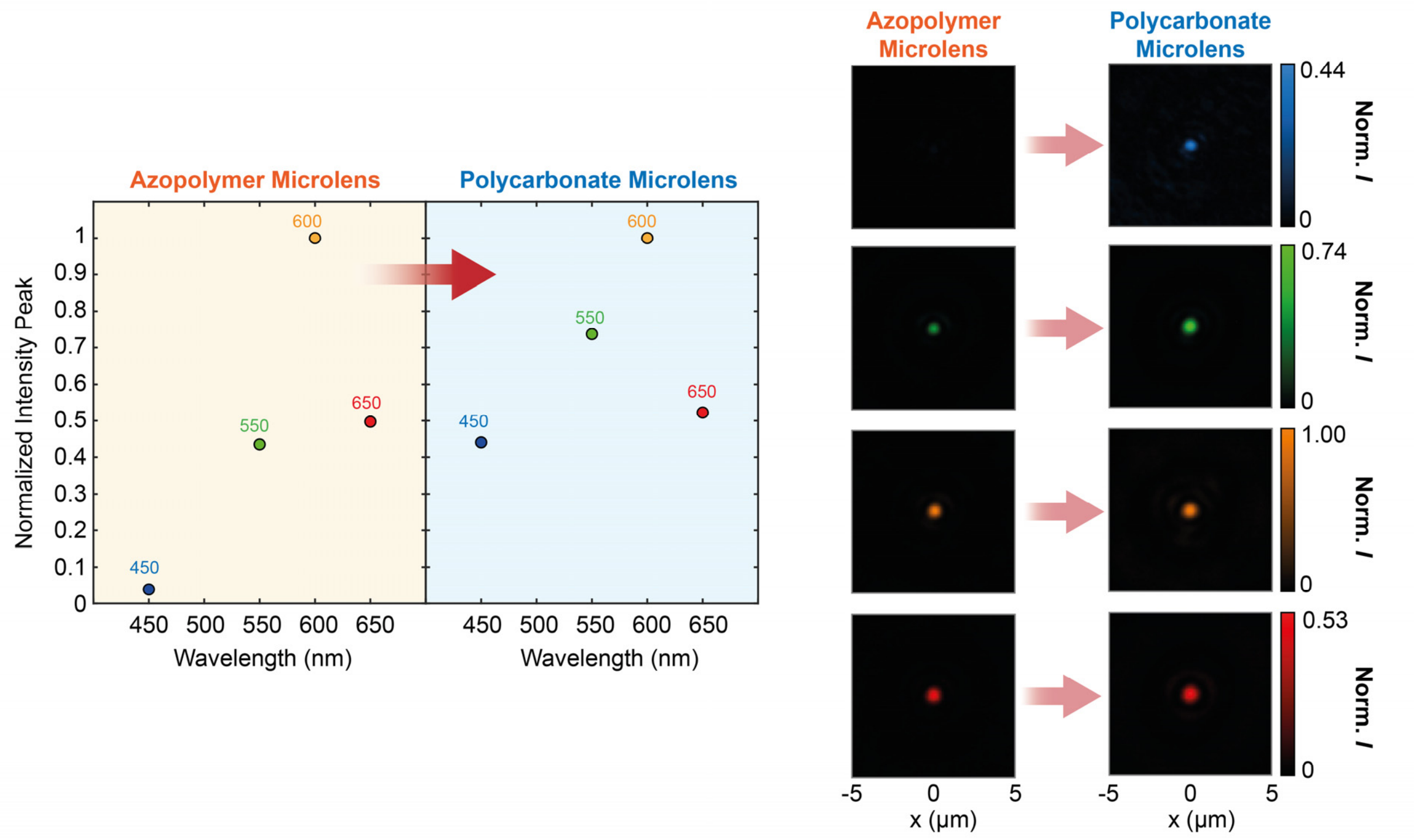

**Figure S23. Relative brightness analysis across visible wavelengths.** The relative brightness at several wavelengths in the visible spectrum was evaluated by imaging the focal spots formed at the focal plane of the microlens under study. For each wavelength, data acquisition was performed using identical camera settings (particularly exposure time and gain) to ensure quantitative comparability. The focal spot image and the corresponding incident beam power were recorded for every wavelength. To enable a meaningful comparison across the spectrum, the measured peak intensities were scaled according to their respective incident powers, effectively normalizing the data to a common incident beam power. The normalized peak intensity for each wavelength was then obtained by dividing its scaled peak value by the highest peak intensity among all wavelengths, yielding the relative brightness distribution for each material. This procedure was carried out independently for both the optically fabricated azopolymer microlens and its polycarbonate replica.

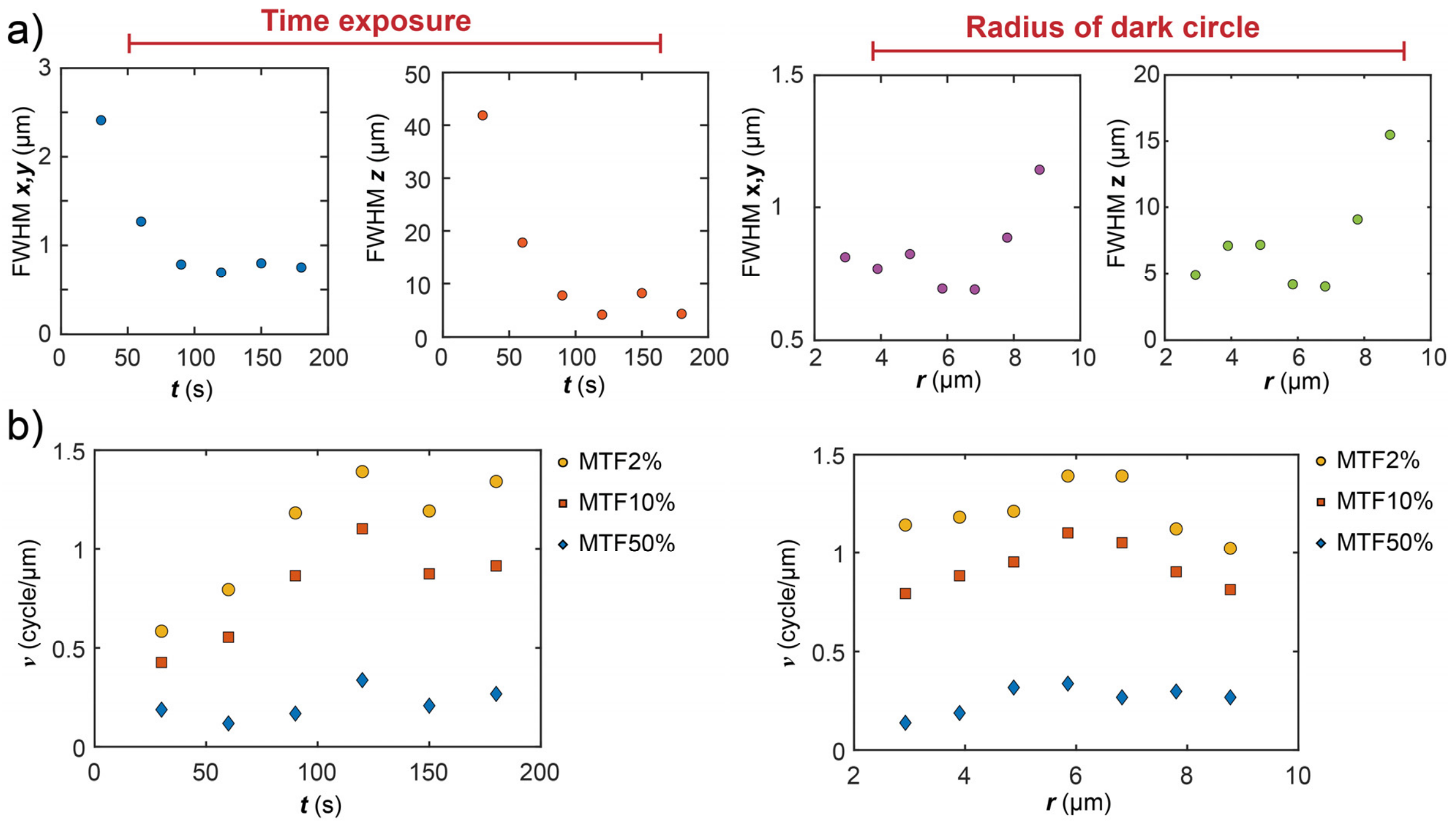

**Figure S24. The optical study of the azopolymer master microlenses. a)** The full-width-half-maximum (FWHM) across lateral axis (FWHM ***x,y***) of the focal spot and across axial axis (FWHM ***z***) along the beam propagation for both variations in ***t*** and ***r***. The FWHM ***x,y*** is the average of Gaussian-fitted FWHM along ***x***- and ***y***-axis of the focal spot. The FWHM ***z*** was derived from the experimental axial intensity profile by taking the effective FWHM of the peak. **b)** The calculated modulation-transfer-function (MTF) at different levels, including 50% (MTF50%), 10% (MTF10%), and 2% (MTF2%) which also acts as the effective cutoff frequency in this study. The MTF was calculated by computing the 2D Fourier transform of the focal spot image to obtain the optical-transfer-function and taking the absolute value for proper MTF, which then went through normalization. As the focal spot from a circularly designed microlens has circular symmetry, a radial averaging of the 2D MTF was done which yield 1D radial averaging profile of the MTF. The 1D radial average profile then provided the approximated MTF at different levels as the function of spatial frequency (***ν***). The experiment was done using a 633 nm laser as the probe beam. The time exposure variation was conducted by setting the $r$ = 5.9 µm. The $r$ variation was done by keeping the time exposure constant at 120 s.

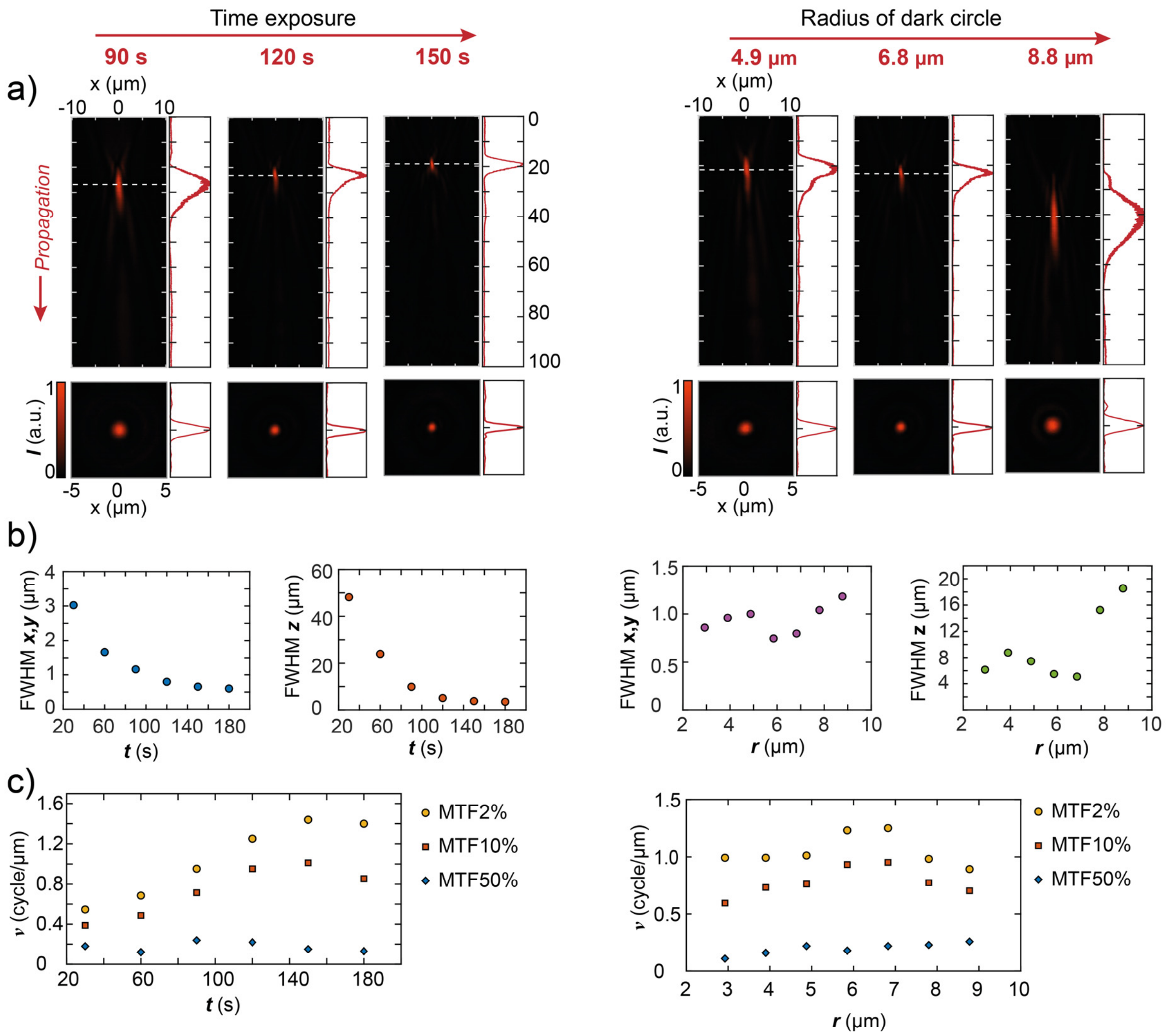

**Figure S25. The optical study of replicated microlenses. a)** The beam propagation and the focal plane image of different microlens for different time exposure (***t***) and radius of dark hole in the hologram design (***r***). **b)** The full-width-half-maximum (FWHM) across lateral axis (FWHM ***x,y***) of the focal spot and across axial axis (FWHM ***z***) along the beam propagation for both variations in ***t*** and ***r***. The FWHM ***x,y*** is the average of Gaussian-fitted FWHM along ***x***- and ***y***-axis of the focal spot. The FWHM ***z*** was derived from the experimental axial intensity profile by taking the effective FWHM of the peak. **c)** The calculated modulation-transfer-function (MTF) at different levels, including 50% (MTF50%), 10% (MTF10%), and 2% (MTF2%) which also acts as the effective cutoff frequency in this study. The MTF was calculated by computing the 2D Fourier transform of the focal spot image to obtain the optical-transfer-function and taking the absolute value for proper MTF, which then went through normalization. As the focal spot from a circularly designed microlens has circular symmetry, a radial averaging of the 2D MTF was done which yield 1D radial averaging profile of the MTF. The 1D radial average profile then provided the approximated MTF at different levels as the function of spatial frequency (***ν***). The experiment was done using a 633 nm laser as the probe beam. The time exposure variation was conducted by setting the $r$ = 6.8 µm. The $r$ variation was done by keeping the time exposure constant at 120 s.

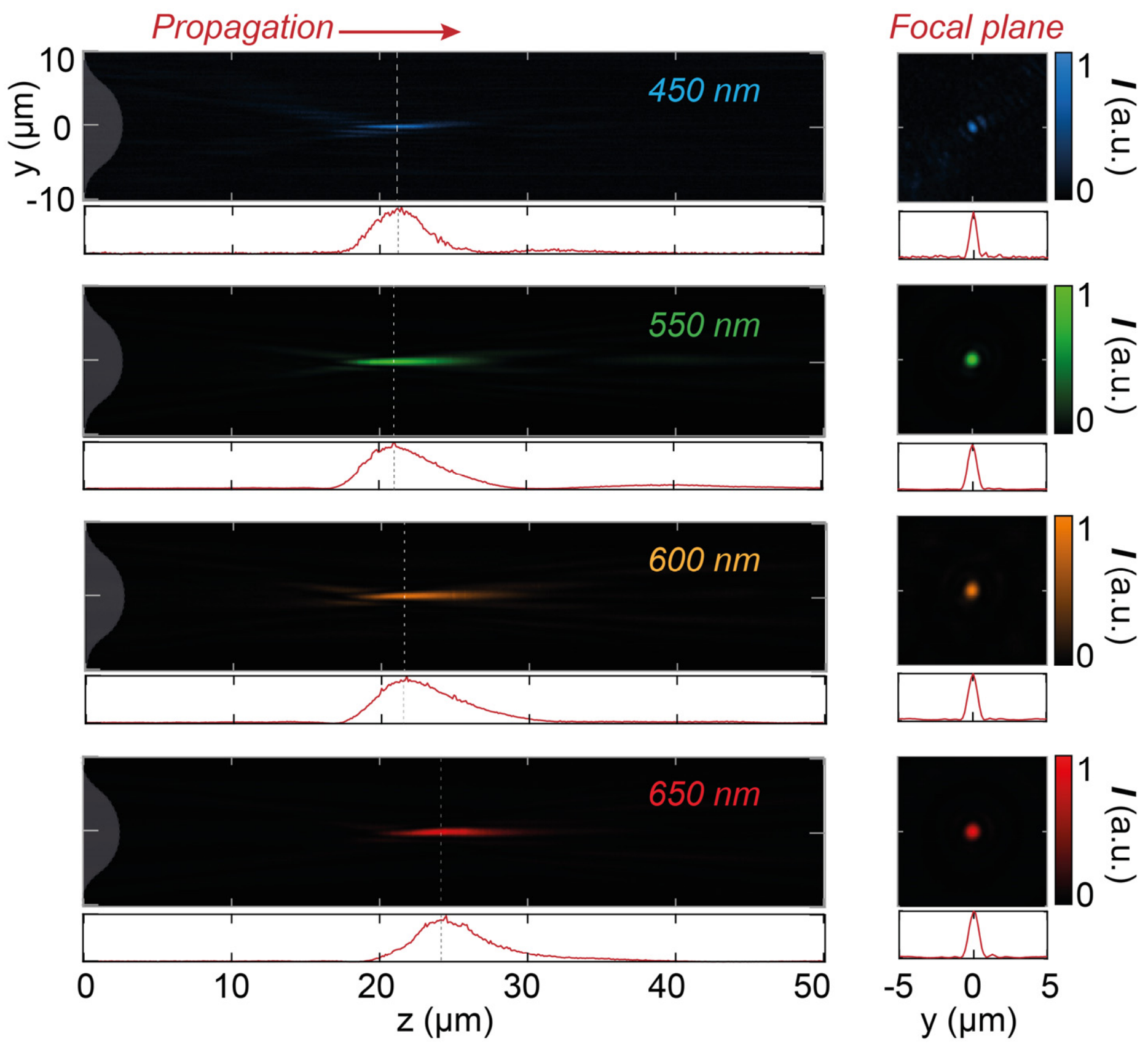

**Figure S26. The multi-color study of the replicated microlens.** The studied microlens is the replicated microlens produced from radius of dark circle of 6.8 μm and time exposure of 120 s.

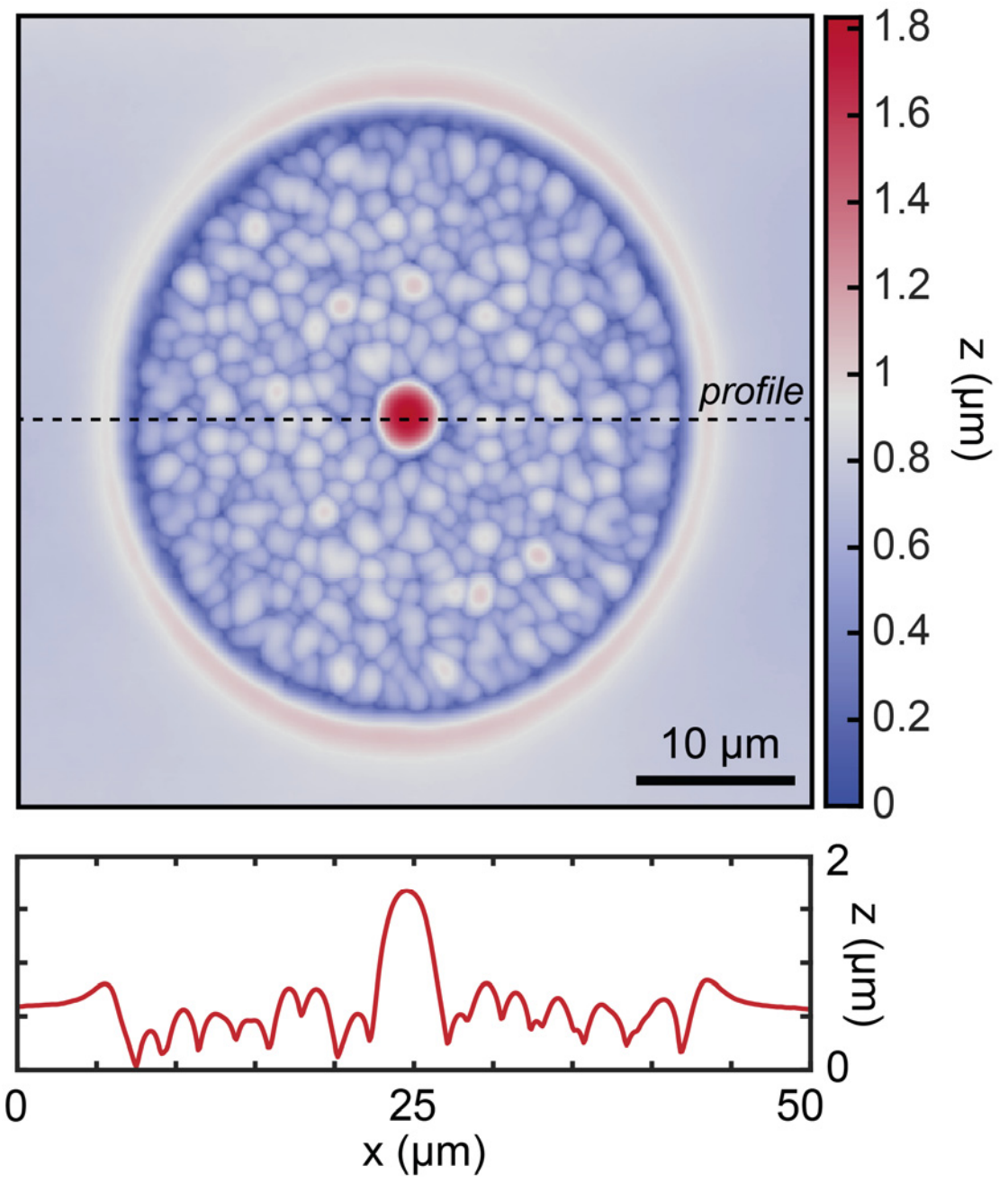

**Figure S27. Surface formation after illumination with a doughnut intensity pattern without the assisting beam (405 nm).** The hologram used was the doughnut-shaped pattern with *R/r* = 9 shown in Figure S3.. The illumination parameters (intensity of the writing beam at 491 nm and exposure time) were the same as those used to produce isolated protruding microstructures in Figure 2d (*R/r* = 9) (see Supplementary Table 1 for details).

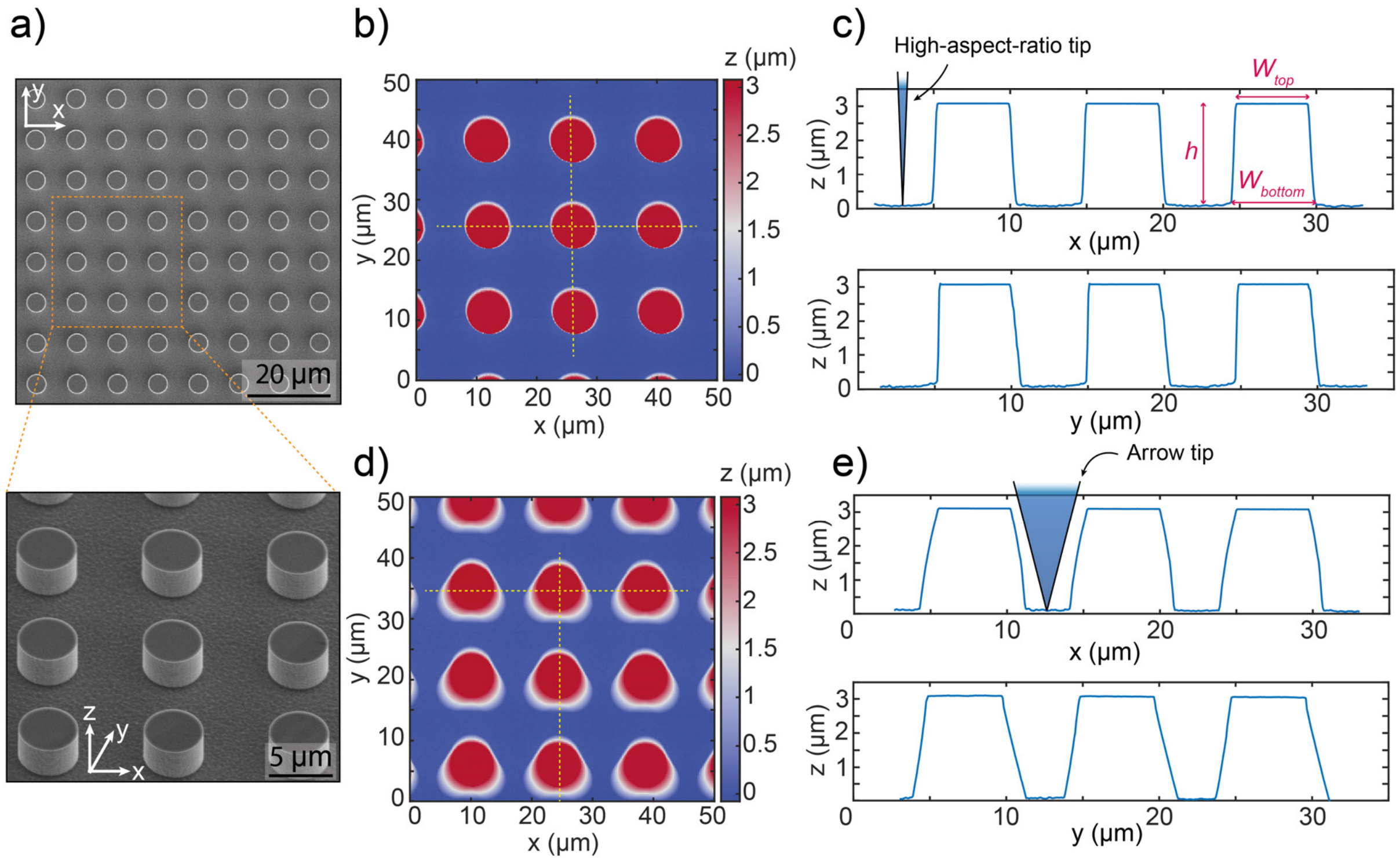

**Figure S28. The effect of the atomic force microscope tip geometry on the reconstruction of microstructure morphology.** a) The morphological image of the silicon master of micropillar array manufactured by MicroFabSolutions s.r.l. with nominal height = 3.1 ± 0.1 µm, diameter = 4.7 ± 0.1 µm, and pitch = 10.0 ± 0.1 µm, acquired with scanning electron microscopy. b) The top-down morphology image of 3×3 silicon micropillars array taken with atomic force microscope (AFM) using high-aspect-ratio tip (ISC 225C3.0–R). c) The silicon micropillar profiles scanned along $x$ and $y$-axis from the surface morphology in (b) and the scheme of the high-aspect-ratio tip with 1:1 scaling relative to the micropillar profile. Quantitative analysis of micropillar profile in $x$-axis yield $h \approx 3.02$ µm, $W_{top} \approx 4.77$ µm, and $W_{bottom} \approx 5.74$ µm, and in $y$-axis yield $h \approx 3.06$ µm, $W_{top} \approx 4.79$ µm, and $W_{bottom} \approx 5.63$ µm. (e) The morphology of the silicon micropillar array scanned using NanoWorld® "arrow" tip. The schematics of the tips in the panels c) and e) are in scale with the array profile, using the nominal geometry provided by the vendors. (f) The side profile of the silicone micropillar extracted from morphology in (e) and the scheme of the Arrow™ tip with the scaling of 1:1 relative to the micropillar profile. The quantitative analysis derived from side profile at $x$-axis provided $h \approx 3.02$ µm, $W_{top} \approx 4.81$ µm, and $W_{bottom} \approx 7.12$ µm, and in $y$-axis yield $h \approx 3.05$ µm, $W_{top} \approx 4.96$ µm, and $W_{bottom} \approx 7.67$ µm. The NanoWorld high-aspect-ratio provides a superior accuracy for side profiling of the silicone micropillar relative to the standard arrow tip.

| Number of Figure | Intensity of Hologram λ=491 nm (W/cm$^2$) | | | Time Exposure (s) | | |
|---|---|---|---|---|---|---|
| **2**c – d, **S27** | ~ 63 | | | 180 | | |
| **2**e, **S5** | ~ 63 | | | (**2**e) 60, 180, and 300<br>(**S5**) 5, 30, 60, 90, 120, 150, 180, 240, 300, 360, 420, and 600 | | |
| **2**f, **S6**, **S7** | ~ 38 | | | 30, 90, 180, 300, and 420 | | |
| **3**c – d | ~ 27 | | | 300 | | |
| **3**g – h, **4**d (panel 1st exp, **S10** | ~ 27 | | | 300 | | |
| | 1st exposure | 2nd exposure | | 1st exposure | 2nd exposure | |
| **4**c – d (panel 2nd exp) | ~ 27 | ~ 133 | | 300 | 30 | |
| **S12**a (triangle) | ~ 38 | ~ 319 | | 150 | 20 | |
| **S12**a (square) | ~ 42 | ~ 167 | | 150 | 25 | |
| **S12**a (pentagon) | ~ 41 | ~ 183 | | 150 | 15 | |
| **S12**a (rectangle) | ~ 41 | ~ 183 | | 150 | 15 | |
| **S12**a (ellipse) | ~ 41 | ~ 183 | | 150 | 15 | |
| **S12**a (boomerang) | ~ 41 | ~ 183 | | 150 | 20 | |
| **S12**b (circular in hexagonal array) | ~ 41 | ~ 190 | | 150 | 20 | |
| **5**b, **S12**b (circular in random array) | ~ 30 | ~ 153 | | 300 | 30 | |
| **5**c, **S12**b (random in random array) | ~ 28 | ~ 142 | | 300 | 30 | |
| **5**e, **S12**b (branched pattern) | ~ 38 | ~ 329 | | 360 | 15 | |
| **5**f, **S12**b (3-branched pattern) | ~ 40 | ~ 230 | | 300 | 20 | |
| **S13**b | ~ 26 | ~ 31 | | 240 | 180 | |
| | 1st exposure | 2nd exposure | 3rd exposure | 1st exposure | 2nd exposure | 3rd exposure |
| **S15** (1st from left) | ~ 27 | ~ 133 | ~ 104 | 300 | 30 | 30 |
| **S15** (2nd from left) | ~ 27 | ~ 133 | ~ 90 | 300 | 30 | 30 |
| **S15** (3rd from left) | ~ 28 | ~ 153 | ~ 97 | 300 | 30 | 30 |
| **6**c – e, **S14**, **S15** (4th from left) | ~ 28 | ~ 142 | ~ 90 | 300 | 30 | 45 |
| **S16** (gradient) | ~ 28 | ~ 142 | ~ 146 | 300 | 30 | 30 |
| **7** (1st cycle) | ~ 38 | | | 300 | | |
| **7** (2nd cycle) | ~ 36 | | | 180 | | |
| **7** (3rd cycle) | ~ 47 | | | 300 | | |
| **7** (4th cycle) | ~ 37 | | | 180 | | |
| **7** (5th cycle) | ~ 37 | | | 300 | | |
| **9**b, **9**c, **9**e, **9**f, **S20**, **S21**, **S24**, **S25** | ~ 37 | | | (**9**b) 120 s<br>(**9**c) 30, 60, 90, 120, 150, 180 s<br>(**9**e) 30, 60, 90, 120, 150, 180 s<br>(**9**f) 120 s<br>(**S20**, **S21**, **S24**, **S25**) 30, 60, 90, 120, 150 s for time exposure variation and 120 s for radius of dark circle variation | | |

**Supplementary Table 1. The intensity of holographic illumination and time exposure from different experiments.** The intensity of holographic illumination was calculated by using formula of *Intensity = (beam power) / (area of bright part of hologram)*. Power of beam was measured with a power meter at the plane before the Objective (see Experimental section for the CGH setup). The area of the bright part of the hologram was calculated by first binarizing the hologram image (see Experimental section for hologram recording procedure). In the binarization step, pixels with values below the threshold were assigned “0”, and those above the threshold were assigned “1”. The area corresponding to the “1” pixels was then extracted. It gives a reasonable approximation of the real local intensity of each hologram, since most of the holographic patterns (except hologram in Figure S16, gradient design) are binary patterns by design.

**Supplementary Table 2.** Comparison of the main processing characteristics and demonstrated capabilities of established microfabrication approaches, previous azopolymer patterning methods, and the all-optical strategy developed in this work.

| Criterion | Optical Lithography | Electron Beam Lithography | Interference Lithography | Direct Laser Writing / 2PP | UV-Nanoimprint Lithography | Previous Azopolymer Patterning | This Work |
|---|---|---|---|---|---|---|---|
| **Initial substrate / structure** | Flat resist-coated substrate | Flat resist-coated substrate | Flat photosensitive film | Photoresist / photopolymer resin | Pre-patterned physical mold | Flat azopolymer film | **Flat pristine azopolymer film** |
| **Physical mask / master required** | Often | No | No | No | Yes | No | **No** |
| **Chemical development / etching** | Usually required | Usually required | Often required | Development typically required | No development for replication | No | **No** |
| **Morphology defined by** | Mask + process | E-beam scan | Interference field | Laser trajectory | Physical master | Optical intensity / polarization field | **Bright/dark optical field + exposure history** |
| **Direct positive 3D relief from flat film** | Process-dependent | Yes, after processing | Limited | Yes | No - replicated from mold | Mostly shallow / engraving-like | **Yes** |
| **Vertical modulation** | Process-dependent | Process-dependent | Typically shallow | μm to >100 μm | Mold-defined | Typically <1 μm | **Several μm; 5.4 ± 0.3 μm demonstrated** |
| **Patterning parallelism** | High | Low | High | Low | High | High | **High - full-field holographic exposure** |
| **Design freedom** | High | High | Limited | High | Mold-limited | High | **High-discrete, freeform, polygonal, hierarchical** |
| **Post-writing reshaping** | No | No | No | No | No | Yes | **Yes** |
| **Erase / rewrite same area** | No | No | No | No | No | Possible | **Yes - 5 cycles demonstrated** |
| **Hierarchical / temporal control** | Process-intensive | Process-intensive | Limited | Sequential writing | New mold/process | Limited | **Digital multi-exposure control** |
| **Functional replication** | Yes | Yes | Yes | Yes | Intrinsic | Selected cases | **Yes - transfer into PC** |
| **Demonstrated optical function** | DOEs / micro-optics | Nanophotonics / nanostructures | Gratings / photonic structures | Micro-optics / 3D structures | Replicated optical elements | Optical encoding / diffractive structures | **Replicated PC microlenses** |

**Supplementary Table 3**. For studies based on pre-patterned structures, the height reported corresponds to the initial fabricated structure unless a larger light-generated vertical modulation is explicitly demonstrated.

| Study | Patterning strategy | Initial substrate / structure | Mold or pre-patterning required | Demonstrated height / modulation | Lateral feature size | Aspect ratio | Demonstrated array / area | Patterning mode / throughput | Reversibility demonstrated | Modeling approach/predicting capability | Demonstrated function |
|---|---|---|---|---|---|---|---|---|---|---|---|
| **This work** | CGH-based all-optical lithography using engineered dark regions; sequential spatiotemporal exposures | **Flat, pristine azopolymer film** | **No** | **5.4 ± 0.3 µm** average modulation for the 6 × 6 micro-bump array | **FWHM 8.6 ± 0.2 µm**; period 19.6 ± 0.3 µm | **≈0.63** (5.4/8.6) | **6 × 6 array**, P = 19.6 ± 0.3 µm; arbitrary/random arrays and free-form patterns also demonstrated | **Parallel full-field holographic exposure**; representative 1-step array: **300 s**; multistep structures: 300 s + 30–45 s | **Full optical erase and rewrite with a different pattern in the same surface region** | **VPA modeling used to reproduce and interpret the main deformation regimes and their dependence on illumination geometry** | Multi-focus microlenses; hierarchical structures; reprogrammable master templates |
| **Januariyasa et al., Light Sci. Appl. 2026** | Vectorial field-guided photo-reconfiguration using spatially structured polarization | Pre-patterned azopolymer micropillar arrays | **Yes – PDMS replica molding from master** | Initial $H_0$ = **8.4 µm** for homogeneous-field experiments; $H_0$ ≈ **3.0 µm** for vectorial-field experiments | $D_0$ ≈ **4.0 µm** or **6.7 µm**; pitch ≈10–12 µm | Up to **≈2.1** for initial $H_0/D_0$ = 8.4/4.0 | Projected addressable area ≈ **130 × 78 µm²** | Parallel SLM polarization projection; individual pillars independently addressable | Reconfiguration demonstrated; | **Quantitative predictive VPA model**, experimentally validated; target anisotropic/bent/chiral morphologies predicted | Programmable anisotropic, bent, S-shaped and chiral microstructures |
| **Gao et al., Nanoscale 2025** | Light-driven deformation, sliding and rotation of pDR1m pillars | Thermally imprinted pDR1m pillar array | **Yes – 2PP master → PDMS mold → thermal imprint** | Initial fabricated height **459 ± 66 nm**; height decreases during deformation | Diameter **1.14 ± 0.04 µm**; master pitch **4 µm** | Initial h/d ≈ **0.40** | Array; macroscopic area not quantitatively reported | Parallel illumination; characteristic deformation times **30–600 s** | Local morphological evolution/recovery reported | **Phenomenological/semi-quantitative three-regime model**; not general forward morphology prediction | Biointerfacing; elongation, sliding, rotation and nanobump formation |
| **Salvatore et al., RSC Appl.** | Digital holographic lithography of | **Flat spin-coated film,** | **No** | Submicrometric relief; | Micrometer-scale holographic | N/A for relief profiles | Pattern demonstrated within AFM-accessible | **Parallel full-field CGH exposure**, single exposure sequence | Azopolymer reconfigurability established; **full erase/rewrite not** | Calibration/LUT relating optical grayscale to relief height; **no predictive** | Multilevel QR-code / optical information encoding |

| Study | Patterning strategy | Initial substrate / structure | Mold or pre-patterning required | Demonstrated height / modulation | Lateral feature size | Aspect ratio | Demonstrated array / area | Patterning mode / throughput | Reversibility demonstrated | Modeling approach/predicting capability | Demonstrated function |
|---|---|---|---|---|---|---|---|---|---|---|---|
| **Interfaces 2025** | grayscale/multilevel reliefs | **1.7 µm thick** | | | resolution; multilevel QR-code pixels | | fields up to **100 × 100 µm²** | | **demonstrated in this paper** | **mechanical morphology model** | |
| **Wang et al., Adv. Photonics Res. 2025** | Polarization-controlled deformation of azo-molecular-glass pillar arrays | Hot-embossed IA-Chol submicron pillars | **Yes – PDMS mold / hot embossing** | Initial height **761 ± 9 nm** | Diameter **501 ± 7 nm**; interpillar distance **1001 ± 7 nm** | Initial h/d ≈ **1.52** | Large periodic array; exact processed area not reported | Parallel masked polarized exposure; representative exposures **40–100 s** | Different deformation states possible; **complete reset/rewrite not demonstrated** | **No quantitative morphology prediction** | Polarization image recording/reconstruction; diffraction; PDMS/PMMA replica transfer |
| **Januariyasa et al., ACS AMI 2023** | Wavelength-dependent 3D photo-reconfiguration | Square array of cylindrical azopolymer micropillars | **Yes – replica/soft molding** | Initial $H_0$ = **10.0 ± 0.2 µm**; | D = **5.0 ± 0.2 µm**, P = **10.0 ± 0.2 µm**; wettability array D = 8.0 ± 0.2 µm, P = 12.0 ± 0.2 µm | Initial H/D = **2.0** for main experiment | Periodic arrays; total area NR | Parallel homogeneous illumination; exposure-dependent deformation | Multiple optically induced morphologies from same initial structure; **not full erase-to-flat/rewrite** | Quantitative morphological analysis and wavelength/penetration-depth interpretation; **no mechanical morphology forward model** | Tunable 3D shape and hydrophobicity |
| **Wang et al., Adv. Optical Mater. 2022** | LPL-induced topographical transition of azo-molecular-glass pillar arrays | Pre-patterned IA-Chol submicron pillar array | **Yes** | Initial submicron pillar height; light transforms/merges structures | Initial diameter ≈ **0.50 µm** | ≈0.5 for the starting structures in the geometry reported | Periodic array; total area NR | Parallel CW 532-nm illumination | Morphology can evolve with dose; **full reset/rewrite not established** | Optical simulations used for local fields; **mechanistic interpretation and no quantitative morphology prediction** | Complex periodic surface patterns, fringes/ripples/wrinkles |
| **Salvatore et al., J. Phys.** | Polarization-driven reshaping of preformed | Cylindrical azopolymer | **Yes – pre-patterned/imprin** | Micron-scale preformed pillars; no direct | Micron-scale pillars / programma | Geometry-dependent | Large-area structuring demonstrated | **Parallel optical reconfiguration** | Reconfigurable morphology demonstrated; not equivalent to flat- | Experimental study **no quantitative forward morphology model** | Programmable anisotropy, diffraction gratings, tunable |

| Study | Patterning strategy | Initial substrate / structure | Mold or pre-patterning required | Demonstrated height / modulation | Lateral feature size | Aspect ratio | Demonstrated array / area | Patterning mode / throughput | Reversibility demonstrated | Modeling approach/predicting capability | Demonstrated function |
|---|---|---|---|---|---|---|---|---|---|---|---|
| **Photonics 2021** | micropillars into gratings | micropillar array | **ted surface** | generation of protrusions from flat film | ble grating periodicity | | | | film erase/rewrite of arbitrary structures | | orientation/periodicity |
| **Jo et al., Nano Lett. 2021** | Photoreconfiguration of molded pillars into shark-skin-like denticles | Molded azopolymer pillar array | **Yes** | Height inherited from preformed micropillars; reshaped rather than generated from flat film | Denticle lateral size ≈ **5 µm** reported for the fabricated structures | NR consistently from accessible primary text | Large patterned arrays | Parallel optical exposure | Morphological reconfiguration; full erase/rewrite not demonstrated | Phenomenological analys of illumination parameters **no quantitative forward morphology model** | Shark-skin-mimetic hydrophobic/antifouling/low-drag surfaces |
| **Wang et al., Sci. Rep. 2021** | CPL-driven self-organized topographical transition | Hot-embossed IA-Chol submicron pillar array | **Yes** | Submicron pre-existing topography; new pillars form during transition | Initial pillar diameter ≈ **0.38 µm** | Sub-unity | Periodic array; macroscopic area NR | Parallel CPL exposure | Exposure-induced transition; no arbitrary erase/rewrite | Mechanistic interpretation; **no quantitative forward morphology model** | Threefold pillar-density enhancement / ordered surface patterns |
| **De Martino et al., Nano Lett. 2020** | Light-driven reshaping of vertical pDR1m pillars into bars | Thermally molded square micropillars | **Yes – PDMS mold / thermal imprint** | **≈800 nm** | **≈1.0 µm** lateral size | **≈0.8** | Arrays; total area NR | Parallel green-light illumination | Dynamic reshaping demonstrated; **no full reset/rewrite workflow** | **no quantitative forward morphology model** | Dynamic manipulation of cell-membrane curvature / actin response |
| **Jo et al., ACS AMI 2020** | Slant-polarized-light bending of azopolymer pillars | Molded/pre-patterned pillar array | **Yes** | Pre-existing micron-scale pillars; light produces bending rather than height creation | **≈1 µm diameter** | NR from accessible primary text | Periodic array | Parallel slanted illumination | Shape controlled optically; full reset/rewrite not demonstrated | Mechanistic field-gradient interpretation; **no quantitative forward morphology model** | Bent pillars, anisotropic wetting and directional adhesion |

| Study | Patterning strategy | Initial substrate / structure | Mold or pre-patterning required | Demonstrated height / modulation | Lateral feature size | Aspect ratio | Demonstrated array / area | Patterning mode / throughput | Reversibility demonstrated | Modeling approach/predicting capability | Demonstrated function |
|---|---|---|---|---|---|---|---|---|---|---|---|
| **Oscurato et al., ACS AMI 2017** | Polarization/intensity-driven reshaping of imprinted 3D posts | Imprinted azopolymer micropillar array | **Yes – imprinting** | Micron-scale preformed pillars; height not newly generated by optical patterning | Micron-scale pillars | NR here unless taken directly from SI | Large-area parallel reconfiguration | Parallel illumination | **Reversible reshaping stated and demonstrated** | Phenomenological; **no quantitative forward morphology model** | Wetting anisotropy, unidirectional spreading, programmable wetting paths |
| **Pirani et al., Sci. Rep. 2016** | Polarization-controlled reversible reshaping of PAZO/PMMA pillars | Soft-lithographed circular micropillar array | **Yes – soft lithography** | Initial H = **1.5 µm** | D = **3 µm**, pitch = **5 µm** | Initial h/d = **0.50** | Large-area square lattice; analysis over multiple pillars | Parallel illumination; reversible cycle ≈ **400 s** | **Yes: repeated elongation/restoration cycles**, with gradual degradation | **no quantitative forward morphology model** | Reversible pillar elongation/orientation |
| **Lee et al., ACS AMI 2015** | Directional superficial photofluidization for deterministic reshaping of pre-existing 3D features | Preformed 3D azopolymer architectures | **Yes – prestructured objects** | Geometry-dependent | Micron-scale | Geometry-dependent | Parallel optical reshaping | Parallel polarized illumination | Reconfiguration demonstrated; not flat-film reset/rewrite | Phenomenological model; **no quantitative forward morphology model** | Deterministic shaping of complex 3D architectures |

# 2. Supplementary Notes

### Supplementary Note 1. Simplified volume conservation analysis on the isolated microrelief formation

When the ring is irradiated with circularly polarized light, the azopolymer inside it undergoes vertical contraction accompanied by radial expansion. The resulting surface relief can be divided into three regions: a micro-protrusion with the height $h_M$, a surrounding doughnut-shaped depression with the height $h_D$, and a circular hill at the outer boundary with the height $h_B$. For simplicity, we assume that the hill has the same lateral extension as the micro-protrusion radius $r$. Previous studies on the azopolymer used in this work have demonstrated that photoinduced deformations occur without measurable volume change. This observation is consistent volume balance measurements in Figure S30 and with the VPA model, which treats the material as an incompressible medium and therefore inherently assumes volume conservation. Because the azopolymer conserves volume during irradiation, the redistribution of material among these three regions must satisfy the following condition:

$$\pi(R+r)^2 h_0 = \pi r^2 h_M + \pi(R^2 - r^2)h_D + \pi((R+r)^2 - R^2)h_B$$

Dividing this relation by the initial film thickness $h_0$ and by $r^2$, we arrive at the following dimensionless equation:

$$\left(\frac{R}{r}+1\right)^2 = \lambda_M + \left(\left(\frac{R}{r}\right)^2 - 1\right)\lambda_D + \left(\left(\frac{R}{r}+1\right)^2 - \left(\frac{R}{r}\right)^2\right)\lambda_B,$$

where the stretch ratios $\lambda_i = h_i/h_0$ are introduced, $i = M, D, B$. Defining $k = R/r$, the stretch ratio of the micro-protrusion becomes:

$$\lambda_M = (k+1)^2 - (k^2 - 1)\lambda_D - (2k+1)\lambda_B$$

To estimate the micro-protrusion deformation, we fix the vertical contraction in the doughnut region to $\lambda_D = 1 - \varepsilon_D$ and the hill elongation to $\lambda_B = 1 + \varepsilon_D$, and vary the strain $\varepsilon_D$. The micro-protrusion increases monotonically with the ratio $k = R/r$ (see Figure S29). Despite its simplicity, the predicted stretch ratios are in reasonable agreement with both the experimental observations and the more sophisticated VPA simulations, which naturally capture the full spatial complexity of the deformation profile.

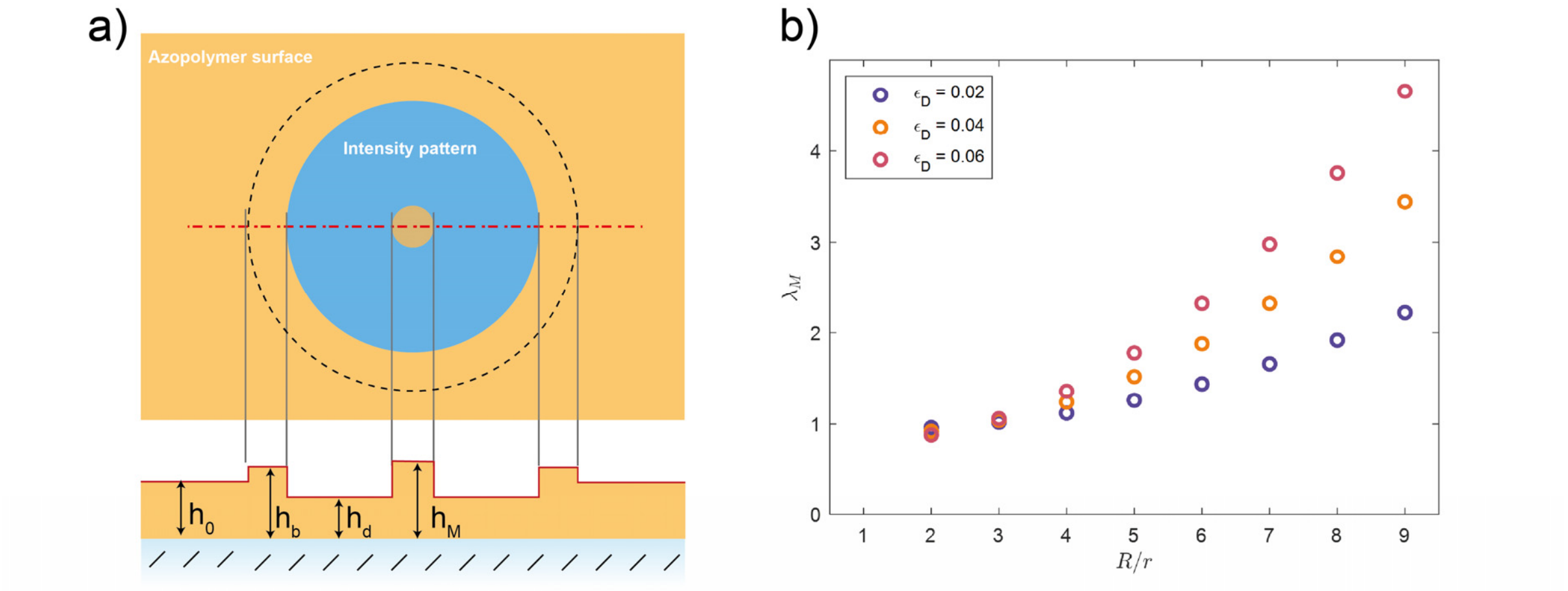

**Figure S29. Estimation of micro-protrusion stretching for different doughnut shapes.** a) Illustration of the modelled configuration. b) Stretch ratio $\lambda_M$ increases monotonically with increasing value of $R/r$.

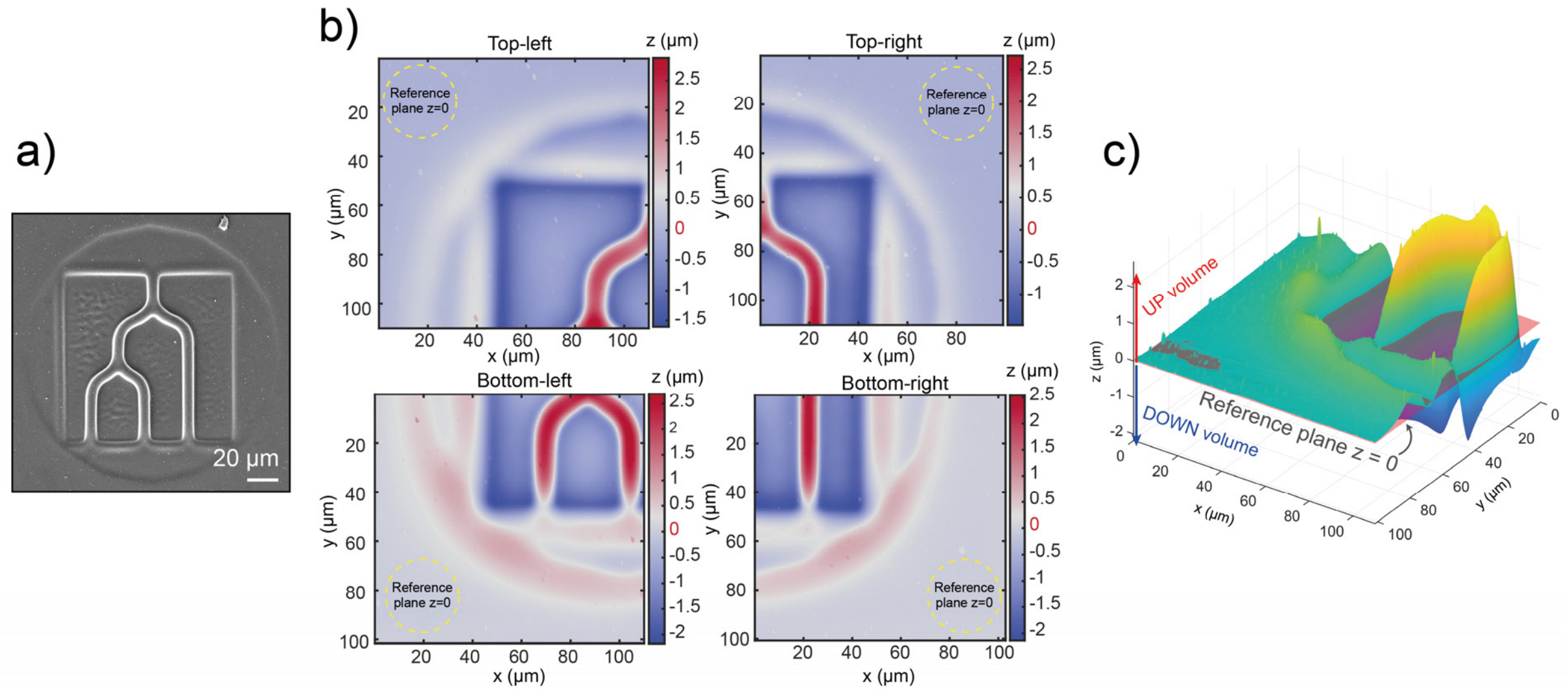

**Figure S30. AFM-based volume-balance analysis of a micropatterned surface.** (a) SEM image of the structure analyzed after the first exposure of the sequence used to fabricate the branched design shown in Figure 5f of the main text. (b) Four AFM scans used to reconstruct the surface topography over an area exceeding the range accessible in a single scan, covering the protruding structure, the surrounding depleted regions, the peripheral relief, and portions of unexposed regions. The original film surface was identified from the unexposed regions indicated by dashed circles and used to establish the reference plane at z = 0. (c) Example of calculation of the accumulated volume above the reference plane (UP volume) and the depleted volume below it (DOWN volume) for one AFM segment. These quantities were obtained by integrating the positive and negative surface-height deviations, respectively, with the DOWN volume expressed as a positive magnitude. Summing the contributions from the four segments gives a total UP-to-DOWN volume ratio of 0.97. The close balance between accumulated and depleted volumes supports approximately volume-conserving redistribution for the analyzed structure.